\documentclass[pre,aps,floats,superscriptaddress,floatfix,twocolumn,longbibliography]{revtex4-1}
\usepackage{amssymb,amsmath}
\usepackage{amsmath,amssymb}
\usepackage{enumitem,kantlipsum}
\usepackage{lipsum}
\usepackage{soul}
\usepackage{cancel}
\usepackage{gensymb}
\usepackage{graphicx}
\usepackage{psfrag}
\usepackage{color}
\usepackage{dcolumn}
\usepackage{bm}
\usepackage[normalem]{ulem}

\def\beq{\begin{equation}}
\def\eeq{\end{equation}}
\def\bea{\begin{eqnarray}}
\def\eea{\end{eqnarray}}

\usepackage{hyperref}
\hypersetup{
  colorlinks=true,
  citecolor=magenta,
  linkcolor=cyan,
}

\begin{document}
\title{Active XY model on a substrate: Density fluctuations and phase ordering}
\author{Astik Haldar}\email{astik.haldar@gmail.com}
\affiliation{Theory Division, Saha Institute of Nuclear Physics, HBNI, 1/AF Bidhannagar, Calcutta 700064, West Bengal, India}
\author{Apurba Sarkar}\email{apurbaphysics391@gmail.com}
\affiliation{School of Mathematical \& Computational Sciences, Indian Association for the Cultivation of Science, Kolkata-700032, West Bengal, India}
\author{Swarnajit Chatterjee}\email{swarnajitchatterjee@gmail.com}
\affiliation{Center for Biophysics \& Department for Theoretical Physics, Saarland University, 66123 Saarbr\"ucken, Germany}
\author{Abhik Basu}\email{abhik.123@gmail.com, abhik.basu@saha.ac.in}
\affiliation{Theory Division, Saha Institute of Nuclear Physics, HBNI, 1/AF Bidhannagar, Calcutta 700064, West Bengal, India}
\begin{abstract}
We explore the generic long wavelength properties of an active XY model on a substrate, consisting of collection of nearly phase-ordered {\em active} XY spins in contact with a diffusing, conserved species, { as a representative system of {\em active spinners with a conservation law}}. The spins rotate {\em actively} in response to the local density fluctuations and local phase differences, on a solid substrate. We investigate this system by Monte-Carlo simulations of an agent-based model, which we set up, complemented by the hydrodynamic theory for the system.  We demonstrate that this system can phase-synchronize {\em without} any hydrodynamic interactions. Our combined numerical and analytical studies show that this model, when stable,  displays hitherto unstudied scaling behavior:  As a consequence of the interplay between the mobility, active rotation and number conservation, such a system can be stable over a wide range of the model parameters characterized by a novel correspondence between the phase and density fluctuations. In different regions of the phase space where the phase-ordered system is stable, it shows phase ordering which is generically either {\em logarithmically} stronger than the conventional quasi long range order (QLRO) found in its equilibrium limit, together with ``miniscule number fluctuations'', or {\em logarithmically} weaker than QLRO along with ``giant number fluctuations'', showing a novel one-to-one correspondence between phase ordering  and density fluctuations in the ordered states.   Intriguingly, these scaling exponents are found to depend explicitly on the model parameters.  We further show that in other parameter regimes there are no stable, ordered phases. Instead, two distinct types of disordered states with short range phase-order are found,  characterized by the presence or absence of stable clusters of finite sizes. In a surprising connection, the hydrodynamic theory for this model also describes the fluctuations in a Kardar-Parisi-Zhang (KPZ) surface with a conserved species on it, or  an active fluid membrane with a finite tension, without momentum conservation and a conserved species living on it.  This implies the existence of stable fluctuating surfaces that are only logarithmically smoother or rougher than the Edward-Wilkinson surface at two dimensions ($2d$) can exist, in contrast to the $2d$ pure KPZ-like ``rough'' surfaces.

\end{abstract}

\maketitle

\section{Introduction}



Recent studies on nonequilibrium systems reveal a fundamental phenomenological difference with equilibrium systems that 
nonequilibrium steady states (NESS) can display orders, which are prohibited in equilibrium systems due to the basic laws of equilibrium statistical mechanics. A well-known example is a two-dimensional ($2d$) flock of self-propelled particles (e.g., a flock of birds, a school of fish and their artificial imitations). These systems can show true orientational long range order~\cite{2d-polar}, in contravention with $2d$ equilibrium systems with short range interactions and continuous symmetries (e.g., the equilibrium XY model) that can only show quasi-long range order (QLRO) but never a long range order~\cite{mwt}. Nonequlibrium systems can also display instabilities and phase separations not found in equilibrium systems. For instance, active nematic liquid crystals are found to be intrinsically phase separated~\cite{active-nematics}. Another example of such contrasting behavior between an equilibrium system and its nonequilibrium counterpart is a $2d$ superfluid: an equilibrium $2d$ superfluid displays algebraic order or QLRO~\cite{chaikin}, whereas its active analogue can show such an order only if the system is anisotropic~\cite{john-superfluid}.  Yet another example is that of a fluctuating surface: An ``Edward-Wilkinson'' (EW)~\cite{stanley} surface, that follows an equilibrium dynamics,  is ``logarithmically rough''  at $2d$, i.e., the variance of the height fluctuations grows logarithmically with the system size, in an exact correspondence with the QLRO of the $2d$ XY model at low temperature. In contrast, a surface that satisfies the well-known Kardar-Parisi-Zhang (KPZ) equation~\cite{kpz,stanley}, a paradigmatic nonequilibrium model, is ``algebraically rough'', i.e., the height fluctuation variance grows as a power law in the system size at $2d$.  Exploration of the natures of order, universality and steady states are a  primary goal in the general studies of nonequilibrium systems.

Universal scaling in driven nonequilibrium systems has been a topic of many theoretical studies. Studying the active or nonequilibrium versions of models of equilibrium statistical mechanics has been a useful and rewarding approach that has yielded unexpectedly rich nonequilibrium physics. A particularly fascinating example of this kind is the driven or active $2d$ XY model.   Wide ranging natural systems of physical, chemical or biological origin are described in terms of driven or active XY spins. This includes, e.g., active superfluids, which though generalize equilibrium superfluidity in ultra low-temperature condensed matter systems~\cite{chaikin}, can also  occur in very different systems like live bacterial suspensions~\cite{bacteria-super1,bacteria-super2,bacteria-super3},  often marked by resistanceless flows. Yet another class of systems where (active) XY-like phenomenologies is expected to be important includes many natural systems of diverse origin are described in terms of  collections of interacting oscillators~\cite{strogatz} that have same internal symmetry as the XY spin. There are many such examples, e.g., chemical oscillators~\cite{miyakawa-pre2002,toiya1,toiya2, 
fukuda-jphys, 
kiss-prl,migliorini, strogatz-physicad,BZ-vortical}, synthetic genetic 
oscillators~\cite{gen-osc}, and biologically relevant 
systems~\cite{frank22,frank-chiral,sebastian,liverpool,
frank-exp,somitogenesis,deve,quorum,Laskar}.  Another  physically different but related example is the KPZ equation for surface growth, which is intimately connected to the $2d$ driven XY model and does not admit a ``smooth'' phase at $2d$~\cite{stanley}.

The paramount issue in  $2d$ driven XY model is the existence of a stable orientationally ordered phase: What are the effects of   conservation laws and finite mobility on  phase ordering, i.e., do mobility and conservation laws together facilitate or hinder phase-ordering?  Apart from its theoretical motivation, it assumes importance as a prototype active model with diffusive motion having rotational invariance that displays QLRO phase order and normal number fluctuations in the equilibrium limit. Thus it offers an 
intuitive understanding of the impact of drive and mobility on the equilibrium states of $2d$ models with rotational invariance. For instance,  isotropic active superfluids without number conservation on substrates, e.g., driven fluids of exciton polaritons, excitations of two-dimensional quantum wells in optical cavities \cite{polariton}, have no algebraic order~\cite{john-superfluid}, in contrast to its equilibrium analog; see also Refs.~\cite{dest1,dest2,dest3,dest4,dest5}. Can number conservation with mobility restore order, and if so, at what conditions and what is the nature of this ``order''?   Further, biologically relevant rather ubiquitous experimentally realizable examples with possible ordered phases, which motivate our study, include ``active carpets'' of cilia or bacterial flagella, in which 
active carpets of cilia~\cite{cilia1} or bacteria~\cite{ramin,bact1},   move chemical
substances or other materials, such as sperm in fallopian
tubes~\cite{fallo1} { and mucous in the respiratory tract~\cite{volvox}, trapped
near the carpet.} {\em In vitro} magnetic cilia carpets~\cite{mag-cilia} is another related example that inspires the present study. We are also motivated by some recent studies~\cite{uriu1,uriu2}, which suggest that cell movements can indeed promote
synchronization (i.e., phase-ordering) of coupled genetic oscillators. Similarly,  a collection of interacting mobile oscillators reveal  the significant effects of the oscillator mobility on the steady states, with the possibility of synchronization at $2d$~\cite{frasca,peruani,demian1,demian2}. 
Although it is generally believed that elasto-hydrodynamic interactions are crucial in synchronization of biological flagella~\cite{lauga1,yeomans1,saint1}, recently,  other mechanisms like ``cell-rocking''~\cite{geyer}  and phase-dependent forces~\cite{ramin-bennet}  are argued to be crucial in the synchronization of swimming unicellular green alga {\em Chlamydomonas}, whereas direct hydrodynamic interactions are shown to contribute insignificantly.  
In a recent study on a linearized hydrodynamic model of diffusive oscillators~\cite{tirtha-oscii}, it has been found by using a linearized treatment that a very large oscillator diffusivity can either promote or destroy synchronization.  Lastly, it remains a paramount theoretically unresolved issue as to how to suppress the instability of the $2d$ KPZ equation, that is formally related to the $2d$ driven XY model. All these results call for systematic, generic study on the order and steady states of a collection of  active XY model with number conservation in $2d$. 

Finding generic conditions either favorable or disadvantageous to  the phase-ordering in a $2d$ active XY model with a conservation law remain theoretically challenging, but poorly understood till now. The fundamental issue here is the existence (or lack of it) of ordered states of the active $2d$ XY model: Given the applications of the active $2d$ XY model as a possible basis for theoretical understanding of wide-ranging phenomena of diverse origin, it would be clearly desirable to explore the concept of ordering and the extent of universality in the active $2d$ XY model, and the role of drive on them, and find the scaling laws for such a system. We do so below by focusing on the specific case of a $2d$ collection of  active phase-ordered XY spins in contact with a conserved density. We show such a system can synchronize {\em without} any long-range hydrodynamic interactions, and both the local rotation of the spins, and phase difference-dependent current of the conserved species play a crucial role in synchronization.

\begin{figure}[htb]
 \includegraphics[width=8cm]{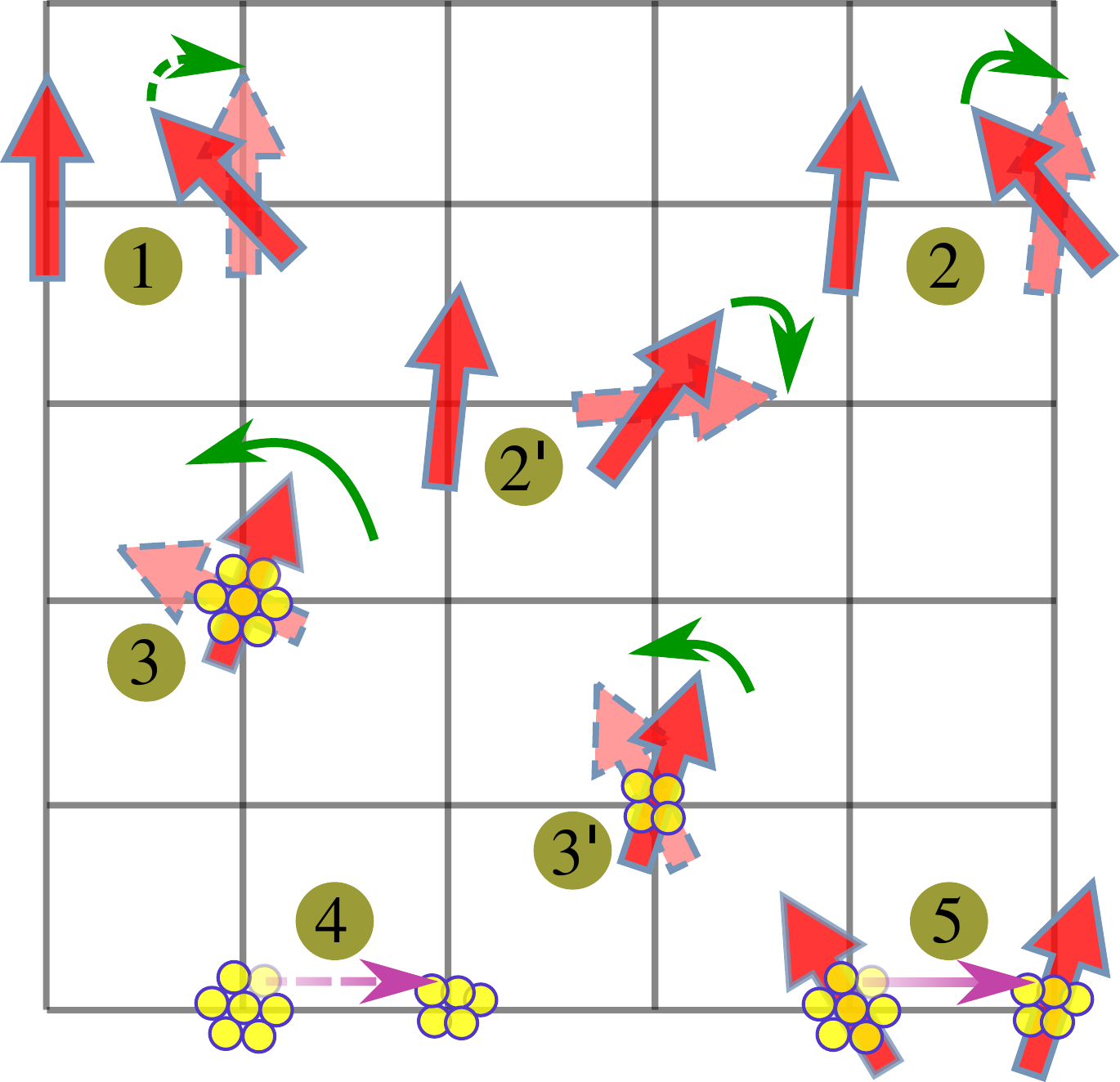}
\caption{
A schematic diagram demonstrating the different local microscopic dynamical update rules in an active model of 2D XY spins interacting with a conserved, mobile species on a substrate. The microscopic processes involve five distinct local processes. The
phase changes by local alignments due to the spin stiffness (process marked $1$), and active rotations proportional to functions of the local phase differences ($2$ \& $2^\prime$) and local concentrations ($3$ \& $3^\prime$). Unlike the local alignments in $1$, which always try to minimize the phase differences between neighbouring spins, the phase updates caused by local phase inhomogeneities in $2$ \& $2^\prime$ introduce a net rotation of the spins, which either decrease ($2$) or increase ($2^\prime$) the phase differences between neighboring spins. The dashed-outlined arrows represent the updated phase of the spins in response to the aforementioned factors. The mobile species moves by diffusion ($4$), and also in response to local phase heterogeneities ($5$).
}
\label{fig:modefigure}
\end{figure}

In this article, we consider a nearly phase-ordered active XY model interacting with a conserved, diffusively mobile species on a $2D$ substrate. We study the stability of small fluctuations around uniform states in this model, and the scaling properties of these fluctuations in the stable states. Inspired by the phenomenologies of bacterial quorum sensing and synchronization in response to complex and dynamic environments~\cite{quorum1,quorum2,quorum3}, we assume the microscopic phase and density dynamics 
 to depend on the local phase and density inhomogeneities in the model. The activity of the XY spins arises from their tendency to rotate {\em actively} in response to local number density of the diffusive species and magnitude of the nearest-neighbor phase differences, processes which have no equilibrium analogs, and by local alignments with neighbors. Further, the conserved current depends on the local phase differences.  From the stand point of active matter systems, our model forms a new class of ``polar'' active matter, known as {\em active spinners without directed walking}. These are distinct from the celebrated ``moving XY model'' for polar ordered active fluids on substrates~\cite{2d-polar,john-flock,vicsek}. The latter model is characterized by ``self-propulsion'' of the particles. This is absent in our model; instead, the particles can rotate {\em actively}, resulting into stable phase-ordered NESS with novel parameter-dependent scaling behavior, or lack of phase-ordered states with distinct density morphologies. 
See Fig.~\ref{fig:modefigure} for a schematic model diagram. 
  We study our active XY model by Monte-Carlo simulations (MCS). To understand the results from the MCS studies theoretically, we write down the hydrodynamic theory for this system based on symmetry principles and conservation laws, and analyze and solve the dynamical equations for the hydrodynamic variables. 
 
  In interesting connections, our hydrodynamic theory also formally applies to a generalized KPZ-like surface growth model given by a single-valued {\em non-compact} height field, coupled to the conserved species living on the surface, or an active fluid membrane with a finite surface tension but without momentum conservation having a conserved species living on it.

Our MCS studies of the agent-based model, backed by our hydrodynamic theory, predict that 
this system of  active  XY spins can be in stable phase-ordered states over wide-ranging model parameters. We attribute the ordering to the interplay between the mobility and the number conservation of  the  mobile species, and therefore not found in its immobile analog,  or without number conservation. For wide-ranging choices of the model parameters, the variance of the phase fluctuations is found to { grow} {\em slower} or {\em faster} logarithmically with the system size {compared to the} $2d$ equilibrium XY model. The density fluctuations concomitantly are either miniscule (i.e., hyperuniform) or giant, compared to their equilibrium counterparts in a gas with short range interactions at a non-critical temperature, in the steady states. We also find breakdown of conventional dynamic scaling by logarithmic modulations of the diffusive dynamic scaling.   We further find that depending upon the model parameters, the hydrodynamic theory can get unstable by two distinct ways, {\em viz.} linear instability, and nonlinear effects, corresponding to the existence of two distinct types of disordered states in the agent-based model, characterized by the presence or absence of stable density clusters.

A short account of this problem and the principal results are available in Ref.~\cite{asp}. Detailed derivations and many additional results are presented here. 

The rest of this article is organized as follows. We start with a short summary of the technical results, which include a brief overview of the simulations of the agent-based model, and also the governing hydrodynamic equations themselves, in Sec.~\ref{sh-summ}.  We set up the agent-based discrete model in Sec.~\ref{modeling}. Next, in Sec.~\ref{num-res}, we present detailed numerical results on the phase-ordered states and the disordered states from our simulations of the agent-based model. In Sec.~\ref{hydro}, we derive the hydrodynamic equations. Then in Sec.~\ref{lin-theo}, we study the linearized equations, after expanding about the mean density $c_0$, and obtain the scaling properties of the correlation functions. Next in Sec.~\ref{rg}, we set up the perturbative renormalization group (RG) calculations to study the nonlinear effects going beyond the linearized approximations. In Sec.~\ref{scal}, we extract the scaling properties and discuss the nature of order in the nonlinear theory by using perturbative RG methods, and  calculate the various correlation functions in real space in the equal-time limit. In Sec.~\ref{dest-order}, we discuss the possible routes to destabilization of the ordered states in the hydrodynamic theory. We further discuss the phase diagram of the hydrodynamic model in Sec.~\ref{phase-sec}. In Sec.~\ref{tp-def}, we briefly discuss and speculate on the role of the topological defects in the steady states.  Finally in Sec.~\ref{summ}, we conclude this paper with a summary and discussion of the results. Some of the technical details and several additional results are given in Appendix for the interested readers.

\section{Summary of the technical results}\label{sh-summ}

Before we summarize our main technical results, it is instructive to first briefly recall the $2d$ XY model in equilibrium, which is used as a reference to our results on the $2d$ active XY model.
At $2d$, the classical XY model in thermal equilibrium is quite remarkable in not having a low temperature phase with a long-range order~\cite{mwt}. Instead it has a low temperature phase with QLRO, that undergoes a phase transition to a disordered state or a state with short-range order (SRO) via unbinding of topological defects~\cite{kt,chaikin}. In the QLRO phase, the variance of the fluctuations in the local phase $\theta({\bf x},t)$ in the XY model at $2d$ is
\begin{equation}
 \langle\theta^2 ({\bf x},t)\rangle_{eq} = \frac{T}{2\pi \kappa_0}\ln\left(\frac{L}{a_0}\right),\label{qlro}
\end{equation}
concomitant with an algebraic decay of the spin correlation functions~\cite{chaikin}. 
Here, $T$ is the temperature, $\kappa_0$ the spin stiffness, $L$ the system size, and $a_0$ the small scale microscopic cutoff, and we have set the Boltzmann constant $k_B=1$; $\langle...\rangle_{eq}$ implies averages over thermal noises or thermodynamic averages.  Equation~(\ref{qlro}) also holds for a fluctuating Edward-Wilkinson (EW) surface~\cite{stanley}, an equilibrium dynamics of a surface. Furthermore, in an equilibrium gas at $2d$ away from any critical points, the standard deviation $\sigma (N_0)$ of the number fluctuations in an open area having $N_0$ particles on average follow the well-known ``square root $N$'' law~\cite{pathria} 
\begin{equation}
 \sigma (N_0)\propto \sqrt{N_0}.\label{nnf-eq}
\end{equation}
Equations~(\ref{qlro}) and (\ref{nnf-eq}) serve as the benchmark of the results that we derive here from our model.

In this work, we introduce an agent-based model with novel microscopic dynamics to systematically study phase ordering, disorder and clustering. We consider a $2d$ collections of  active rotating XY spins of fixed length (or rotors) on a  square lattice {with periodic boundary conditions} and a mean number density $c_0\equiv N_\text{tot}/L^2$ of a diffusively mobile species with local concentration $c_\ell$ at site $\ell$. The microscopic dynamics of the XY spins and the conserved particles consist of simple rules on the phase updates of the spins together with random fluctuations, and on the positions of the particles with random fluctuations. The general forms of the update rules of density $c_\ell$ and phase $\theta_\ell$ {at site $\ell$} must be invariant under spatial translation and rotation, and individual rotations of the XY spins by $2\pi$.  Motivated by bacterial quorum sensing and synchronization in complex and dynamic environments~\cite{quorum1,quorum2,quorum3}  we assume the phase updates to depend upon the environmental inhomogeneities, e.g., local phase inhomogeneities and local concentration of the conserved species, and {also on the lack of clockwise-anticlockwise symmetry due to individual spin rotations}. Position update rules of the conserved species depend upon the local density inhomogeneities as well as local phase inhomogeneities.

Our simulations clearly reveal the existence of distinct ordered and disordered states. 

To quantitatively characterize the steady states, we calculate (a) the variance of the orientation fluctuations $\langle \theta^2\rangle \equiv \langle \sum_\ell (\theta_\ell -\overline \theta(t))^2\rangle/L^2$, $\overline \theta(t)\equiv \sum_\ell\theta_\ell(t)/L^2$,  and (b) the standard deviation $\sigma (N_0)\equiv \sqrt{\langle N^2\rangle - \langle N\rangle^2}$, where $N$ is the  total number of { particles} in an open square box of linear length $<L$, and $N_0\equiv \langle N\rangle$ is its average. Remarkably, the stable, phase-ordered states show a {\em variety of scaling}:

(i) Our systematic MCS studies show that depending upon the model parameters, the variance of the phase fluctuations $\langle \theta^2\rangle$ can grow with the system size $L$ {\em faster} or {\em slower} than $\ln L$, i.e., faster or slower than QLRO or the equilibrium limit result.  Our subsequent analytical approaches to the problem suggests $\langle \theta^2\rangle_{L^2}\sim (\ln L)^{\gamma_1}$. Here, the exponent $\gamma_1$ is found to be positive,  and vary with the model parameters that determine the various active dynamical processes. The dependence of the scaling exponents on the model parameters is truly novel and unique feature of this model.
 Qualitatively speaking, although like QLRO [see Eq.~(\ref{qlro}) above], $\langle\theta^2\rangle$   grows with the system size $L$, it does so either  logarithmically {\em slower} than QLRO (i.e., stronger order than QLRO or ``SQLRO'') for $0<\gamma_1<1$, or {\em faster} than QLRO (i.e. weaker order than QLRO or ``WQLRO'') for $\gamma_1>1$, depending upon the model parameters.  
 
 (ii) Similarly, the standard deviation of the density fluctuations $\sigma(N_0)$ can be bigger or smaller than $\sqrt{N_0}$, which holds in the equilibrium limit. Again, our analytical theory reveals that $\sigma(N_0)\sim \sqrt{N_0/(\ln N_0)^{\gamma_2}}$. The exponent $\gamma_2$, that can be positive or negative, varies with the same model parameters as for $\gamma_1$. Thus qualitatively speaking, like the equilibrium ``square root $N$-law'' or normal number fluctuations (NNF) $\sigma(N_0)$ grows with $N_0$, it does so {either} logarithmically {\em slower} than NNF (i.e., fluctuations suppressed in comparison to the equilibrium limit, known as miniscule number fluctuations (MNF) or hyperuniformity), or logarithmically {\em faster} than NNF (i.e., enhanced fluctuations in comparison to the equilibrium limit, known as giant number fluctuations (GNF)).
 
 (iii) In a  surprising correspondence between the phase and density fluctuations, our agent-based model shows simultaneous occurrence of SQLRO and MNF (WQLRO and GNF). 
 
 (iv) Our simulations further detect two distinct types of disordered states, i.e., states with short-range phase order (SRO), distinguished by the presence or absence of stable density clusters of typical  sizes.

 In order to understand and rationalize the results from MCS studies of the agent-based model, we construct the hydrodynamic theory of the nearly phase-ordered states of an active XY model coupled with a diffusively mobile conserved density. 
One of the hydrodynamic variables reflects the ``orientational order'' in the order parameter space, or ``phase-order'' of the XY spins.  This is described by the non-conserved, broken symmetry field or Goldstone mode $\theta({\bf r},t)$~\cite{chaikin}, a suitable coarse-grained version of individual spin (particle) based orientation of the spins with respect to some (arbitrary) reference axis.  Our second hydrodynamic variable is the conserved density fluctuations $\delta c({\bf r},t)\equiv c({\bf r},t)-c_0$ of the surface density $c({\bf r},t)$ of a diffusing species about its mean $c_0$. In our hydrodynamic theory, we study small fluctuations around a uniform reference state with $\theta({\bf r},t)=\theta_0$, measured with respect to an arbitrary reference axis, and density $c({\bf r},t)=c_0$. The hydrodynamic equations for these variables will be systematically derived in Sec.~\ref{hydro} below. These are obtained by gradient expansions around the uniform reference state, retaining the lowest order symmetry-permitted nonlinear terms. These equations are 
\begin{subequations}
 \begin{align}
 & \frac{\partial\theta}{\partial t}=\kappa\nabla^2\theta + \Omega_1\delta c+\frac{\lambda}{2} ({\boldsymbol\nabla}\theta)^2+ \Omega_2 (\delta c)^2 + f_\theta,\label{hydro1}\\
 &\frac{\partial\delta c}{\partial t}=\lambda_0\tilde\Omega_0 \nabla^2\theta + D\nabla^2\delta c + \lambda_0 \tilde \Omega_1 {\boldsymbol\nabla}\cdot(\delta c{\boldsymbol\nabla}\theta) + f_c.\label{hydro2}
 \end{align}
\end{subequations}
Here, $\kappa>0, D>0, \Omega_1, \lambda, \Omega_2, \lambda_0, \tilde\Omega_0, \tilde\Omega_1$ are phenomenological parameters of our model. Physical interpretations of the different terms in (\ref{hydro1}) and (\ref{hydro2}) are discussed in Sec.~\ref{hydro} below.  Equations~(\ref{hydro1}) and (\ref{hydro2}) may also be interpreted as the coupled hydrodynamic equations for a local single-valued height field $\theta$, measured with respect to an arbitrary base plane, and a conserved density $\delta c$ that exists on it; see also below.

The Gaussian noises $f_\theta$ and $f_c$  have zero mean and variances 
\begin{eqnarray}
 &&\langle f_\theta({\bf x},t) f_\theta(0,0) \rangle = 2D_\theta \delta^2({\bf x}) \delta(t),\label{f-theta} \label{noise-theta}\\
 &&\langle f_c({\bf x},t) f_c(0,0) \rangle = 2D_c (-\nabla^2) \delta^2({\bf x}) \delta(t). \label{f-c}\label{noise-c}
\end{eqnarray}
{ The hydrodynamic equations (\ref{hydro1}) and (\ref{hydro2}) together with (\ref{f-theta}) and (\ref{noise-c}) form the active spinner analog of the celebrated Toner-Tu model~\cite{2d-polar} for dry active flocks.}

We now list the principal results obtained by using (\ref{hydro1}) and (\ref{hydro2}) together with (\ref{f-theta}) and (\ref{noise-c}). 
It predicts either stable uniform NESS with diverse scaling properties of small fluctuations around them, or novel instabilities signalling the loss of any order. We show that the hydrodynamic equations that we derive here when linearized about $c_0$, the mean { particle} number density, can be linearly stable or unstable. We further show that in the linearly stable case, the NESS of such a system admits underdamped propagating waves having linear dispersion with the wavespeed being proportional to  $c_0$. If linearly unstable, then the growth rate of the modes is proportional to the wavevector. Focusing on the NESS for large system sizes  when the propagating modes clearly dominate over damping in the long wavelength limit, we show that the NESS  can be characterized by the variances of the phase fluctuations, density fluctuations and the associated correlation functions.  The stable states in the linear theory have scaling laws for the phase and density fluctuations indistinguishable from its equilibrium counterparts. Nonetheless, the system remains out of equilibrium even in the linear theory as can be seen from the non-zero phase-density cross-correlator in the linear theory.   The nonlinear effects either leave scaling of the linearly stable states essentially unchanged modulating only by introducing logarithmic corrections, or make the linearly stable states unstable, destroying any ordered states. Nonetheless, these logarithmic corrections in the stable case actually make the scaling regime rich and diverse, either by strengthening or weakening the phase order vis-\`a-vis the QLRO phase order in equilibrium or as predicted by the linear theory, and at the same time  suppressing or enhancing standard deviations of the density fluctuations vis-\`a-vis again the equilibrium limit result of NNF, or as predicted by the linear theory. We obtain these results by applying one-loop dynamic renormalization group methods on the hydrodynamic equations.  These results are consistent with those obtained from the MCS studies { on} the agent-based model.


The NESS of the nonlinearly stable states are characterized by a variety of quantities. For instance, the variance of the phase fluctuations 
\begin{eqnarray}
 \Delta_\theta^R\equiv\langle\theta^2 ({\bf x},t)\rangle_R\approx\frac{\overline D(0)}{4\pi\Gamma(0)}\left[\ln\left(\frac{L}{a_0}\right)\right]^{1-\eta},\label{eta-intro}
\end{eqnarray}
where $\eta$ is a scaling exponent that surprisingly depends on the dimensionless ratios $\mu_1,\,\mu_2$ of the ``bare'' or unrenormalized nonlinear coupling constants in the hydrodynamic equations of motion  that we derive later; $R$ refers to renormalized quantities. Further, $\eta$ can be positive or negative, with $\eta<1/3$ always.  Furthermore, $\overline D(0)$ and $\Gamma(0)$ are the unrenormalized (i.e., the microscopic or bare) values of the effective noise strength and damping coefficient. 
Since $\eta$ can be positive or negative, $\Delta_\theta$ can grow, respectively, {\em slower} or {\em faster} than $\ln (L/a_0)$, its form in QLRO or the equilibrium limit.  When $\theta({\bf x},t)$ interpreted as a single-valued height field, (\ref{eta-intro}) implies a fluctuating surface that is logarithmically {\em smoother} or {\em rougher} than an EW surface, respectively, for $\eta>0$ and $\eta <0$.
Similarly, 
the equal-time renormalized  phase-difference correlation function $C_{\theta\theta}^R(r)$ for the XY spins, an exact analogue of the phase-difference correlation function in the XY model, scales as
\begin{equation}
 C_{\theta\theta}^R(r)\equiv \langle[\theta({\bf x+r},t)-\theta({\bf x},t)]^2\rangle_R\approx\frac{\overline D(0)}{2\pi\Gamma(0)} [\ln(r/{a_0})]^{1-\eta},\label{theta-corr-intro1}
\end{equation}
in the limit of large $r$. This means, the renormalized spin correlation function $C_{ZZ}^R(r)$, defined analogously with the spin-spin correlation function for the usual (equilibrium) XY model (see later for a formal definition), decays {\em slower} or {\em faster} for $\eta >0$ or $\eta<0$, respectively, than its equilibrium limit (i.e., $\eta=0$), as we show later in details.  Once again, for a fluctuating surface, Eq.~(\ref{theta-corr-intro1}) means the equal-time height difference correlation function decays {\em logarithmically slower} or {\em faster} than in an EW surface as the separation $r$ grows.
%
%
 The above results on phase fluctuations clearly indicate order logarithmically stronger than QLRO for $\eta>0$ and logarithmically weaker than QLRO for $\eta<0$. We call these SQLRO and WQLRO phase order, respectively. Clearly, for $\eta=0$,   QLRO is retrieved, which is identical to the equilibrium results.

We further obtain the equal-time renormalized correlation function of the density difference $\delta c({\bf x},t)\equiv c({\bf x},t)-c_0$, where $c_0$ is the mean density. We find in the Fourier space, 
$C_{cc}^R(k)\equiv \langle |\delta c ({\bf k},t)|^2\rangle_R $ picks up a weak $k$-dependence displaying unusual scaling behavior:
\begin{equation}
C_{cc}^R(k)= \frac{\overline D(0)}{2\Gamma(0)}\left[\ln \left(\frac{\Lambda}{k}\right)\right]^{-\eta}
\end{equation}
in the hydrodynamic limit $k\rightarrow 0$; where $\delta c({\bf k},t)$ is the Fourier transform of $\delta c({\bf x},t)$, and $\Lambda=2\pi/a_0$ is an upper wavevector cutoff. 
 Clearly, for $\eta>0$, the density fluctuations are strongly suppressed in the long wavelength limit $k\rightarrow 0$ vis-\`a-vis the equilibrium result for a gas with short range interactions away from any critical points. This is a manifestation of {\em miniscule number fluctuations} (MNF) (also known as hyperuniformity), an exotic state of matter~\cite{hyper,hyper-rev}  rarely encountered in ordered active matter~\cite{shelley}. On the other hand, if $\eta<0$ the density fluctuations are hugely enhanced in the long wavelength limit, in comparison with NNF. These are {\em giant number fluctuations} (GNF), often encountered in the context of orientationally ordered active fluids~\cite{sriram-RMP}, and also in equilibrium superfluids~\cite{chaikin}; $\eta=0$ corresponds to the equilibrium result, as before.  Furthermore, the equal-time renormalized density autocorrelation function $C^R_{cc}(r)\equiv \langle \delta c({\bf x+r},t)\delta c({\bf x},t)\rangle_R$, which is just the inverse Fourier transform of $C_{cc}^R(k)$, has the  form
\begin{align}
 C^R_{cc}(r) &\equiv \langle \delta c({\bf x+r},t)\delta c({\bf x},t)\rangle_R \nonumber\\
  &\approx \frac{\overline D(0)}{4\pi\Gamma (0)} \frac{-1}{r^2} \eta[\ln(\Lambda r)]^{(-1-\eta)},\label{c-corr-intro}
\end{align}
for large $r$, $r\Lambda \gg 1$.
Thus for $\eta>(<)0$, $C^R_{cc}(r)$ falls of  relatively faster (slower).
This further implies that the standard deviation  of the number fluctuations $\sigma(N_0)$ in an area of linear size $\tilde L$ containing $N_0$ particles on average (i.e., $N_0 = \langle N\rangle$ as measured in the area $\tilde L\times \tilde L$) is
\begin{equation}
 \sigma(N_0)\propto \sqrt{N_0/(\ln N_0)^\eta }<(>) \sqrt{N_0}, \label{vari-den-intro}
\end{equation}
for $\eta> (<)0$.  Noting that $\sigma(N_0)\propto \sqrt{N_0}$ is the well-known ``square root mean'' law of fluctuations of  equilibrium statistical mechanics, which are  ``normal number fluctuations'' (NNF), expected in equilibrium systems with short range interactions away from any critical point, (\ref{vari-den-intro}) again implies  MNF, i.e., hyperuniform density and GNF, respectively, for $\eta> (<)0$. This is consistent with the discussions on MNF and GNF in this model above. NNF is retrieved when $\eta=0$.
We thus show that   SQLRO phase-ordering is associated with  MNF, whereas   WQLRO phase-ordering comes with  GNF.  This establishes a novel one-to-one correspondence between phase and density fluctuations in the ordered states, and forms a principal conclusion from this study.

We also find breakdown of conventional dynamic scaling in the stable ordered states, where time-scale $t$ no longer shows simple scaling with length-scale $r=|{\bf r}|$ via $t\propto r^z$, but is modulated by logarithmic corrections 
\begin{equation}
t\propto r^2 /[\ln (r/a_0)]^{(1-\eta)/2},
\end{equation}
for large $r$. Therefore, the dynamics is necessarily faster than ordinary diffusion, albeit only logarithmically.  This remains true even when $\eta=0$, i.e., when QLRO is observed. These results define the scaling behavior of this model. The model parameter-dependence of the scaling exponents is a key novel aspect, that reflects breakdown of the conventional notion of universality here.

 In other regions of the phase space, no uniformly ordered states are found.  The hydrodynamic theory predicts that in the parameter regimes corresponding to no stable uniform order, uniformly ordered states can get destabilized by two distinct mechanism - (a) due to linear instability, and (b) nonlinear instability of the linearly stable states. Results from {our} numerical simulations of the agent-based model also reveal the existence of two distinct kinds of  the disordered states: (a) one kind has an intrinsic length-scale that should control the morphology of the eventual steady states, whereas (b) disordered states of the second type are visually  similar to the rough phase of the $2d$ KPZ equation, lacking any intrinsic length-scale.
 
 In the immobile limit with uniform density,   or in the absence of number conservation, e.g., due to birth-death events,  the number density fluctuations relax {\em fast}, and is not a hydrodynamic variable. Therefore, the hydrodynamic equation for $\theta ({\bf x},t)$ reduces to the well-known Kardar-Parisi-Zhang (KPZ) equation for surface growth phenomena~\cite{john-superfluid}. The $2d$ KPZ equation does not have a stable ``smooth phase'' (equivalent to a phase ordered state in the spin language), but has only an algebraically rough phase~\cite{stanley}, which for phase $\theta$ should correspond to short range order (SRO), in agreement with the results from the agent-based model. 
 
 

The hydrodynamic theory predictions provide a generic theoretical framework to understand the numerical results from the agent-based model. In particular, both the agent-based model and the continuum hydrodynamic theory confirm the variety of scaling of the numerical analogs of $\Delta_\theta$ with $\ln (L/a_0)$, and  the standard deviation $\sigma(N_0)$ with $N_0$, demonstrating 
NESS with order both stronger and weaker than QLRO, together with MNF and GNF, respectively.  This reinforces the hydrodynamic theory predictions on the occurrence of SQLRO with MNF and WQLRO with GNF,  the principal qualitative feature of scaling in the phase-ordered states of this model. We further claim that the existence of two types of the disordered states of the agent-based model can be argued by considering the instabilities of the phase-ordered states in the hydrodynamic theory.


Our theory should form the basis of understanding of a wide class of driven systems, e.g., active superfluidity in bacterial suspensions on substrates, active carpets of cilia and their artificial imitations, and mobile oscillator synchronization. { Our hydrodynamic equations (\ref{hydro1}) and (\ref{hydro2}) also describe an active fluid membrane with a finite tension and a conserved species living on it and without momentum conservation~\cite{act-mem}. }  It is also intriguing from nonequilibrium statistical mechanics perspectives by showing how the well-known instability of the $2d$ KPZ equation can be suppressed by coupling it to a conserved density.










\section{Agent-based model}
\label{modeling}
In this Section, we construct an {\em ab initio} agent-based lattice-gas type model and study it by Monte-Carlo simulations. We use these MCS studies to characterize the phase-ordered states, and extract the scaling exponents for both density and phase fluctuations. The agent-based model is further used to systematically study the disordered phases.

We consider a collection of XY spins, each grafted on a lattice site of a two-dimensional square lattice of size $L^2$, interacting with a conserve, mobile particles of mean density $c_0=N_{tot}/L^2$; where $L$ is the linear system size along the X and Y directions, and $N_{tot}$ is the total number of mobile particles. On the lattice, a site with $x$- and $y$-coordinates is conveniently labeled by a single index $\ell \in [1,L^2]$ defined as
\begin{equation}
\ell=(y-1)L+x,
\end{equation}
where 
$x\in [1,L]$, $y\in[1,L]$. Each site has four nearest neighbors and we impose periodic boundary conditions along the $x$ and $y$ directions.

The initial condition of the system is then prepared by assigning each site $\ell$ a fixed number of $c_0$ particles  and a unit spin vector ${Z}_\ell=(\cos \theta_\ell, \sin \theta_\ell)$, where $\theta_\ell=\phi \pm r_\ell\Delta\phi$, {and $\phi \in (-\pi,\pi)$ is same for all sites}, $r_\ell\in (0,1)$ is a {uniform} random number and $\Delta\phi=\pi/36$. Therefore, the initial configuration of the system is nearly phase ordered with a given uniform density of the mobile species.

{\em Simulation Details:-} 
Now let us discuss the principal facets of the update rules that govern the system dynamics; a comprehensive details of the same are presented in Appendix~\ref{MC-update}. We use a square lattice for simplicity although the results should be independent of the underlying lattice structure in the long wavelength limit, as corroborated by our hydrodynamic theory later. We impose periodic boundary condition. To simulate the model, we make use of the canonical sampling Monte Carlo (MC) method where in a single MC step, we sequentially update both the phase and density of the $L^2$ sites where the sites to be updated are chosen in a stochastic manner. Therefore, certain sites may be chosen more than once, and certain sites may not be updated in a single MC step.

{ As mentioned above, inspired by the phenomena of quorum sensing by bacteria~\cite{quorum1,quorum2,quorum3}, in which bacteria may try to synchronize and rotate collectively in response to dynamically changing environmental inhomogeneities (which in the present model should be just local phase and density inhomogeneities), we define the update rules of the phases.} 

(i) Local alignment: This is present in the equilibrium limit as well, and involves the relaxation of the spins to align themselves locally with the neighbors. We model this term, by updating the local phase of site along the average orientation of the nearest neighbors including the site and then adding a small noise to it that describes the stochastic nature of the microscopic dynamics ({see also Ref.~\cite{vicsek} in the context of active polar flocks on a substrate}).

{  (ii) Phase difference-dependent local active rotation:  This is the local phase difference-dependent rotation of the spins, in response to the phase inhomogeneities with the nearest neighbor; see also e.g. Ref.~\cite{quorum2}. In the absence of any self-propulsion, we take it to be proportional to a function of the local phase differences, that is generally arbitrary, constrained only by the symmetry of the model under spatial translation, rotation and rotation of any spin by $2\pi$, and should break the clockwise-anticlockwise symmetry for (active) spin rotations. For simplicity, we choose them to be $g_1\cos (\sum_{\ell'}|\theta_\ell - \theta_{\ell'}|/4)$, where $\ell'$ is a nearest neighbor site of the randomly chosen site $\ell$; $g_1$ measures the relative amplitudes of this active processes.  The parameter $g_1$, a simulation parameter, controls the (relative) strength of this process in the overall update of the phase.

}


(iii) Number density dependent local active rotation:  Lastly, the spins can undergo local number-dependent rotation; see, e.g., Ref.~\cite{quorum1} for a similar mechanism. This term is modeled by assuming the local rotation to be proportional to $g_2\Omega(c_\ell)$, taken as a quadratic function of its argument for simplicity; where $c_\ell$ represents the instantaneous number of interacting mobile particles at site $\ell$. Simulation parameter $g_2$ controls the (relative) amplitude of this particular process in the overall update of the phase.



Finally, the phase of the random site is updated probabilistically where with probability $p_\theta$, we have a new phase having contributions from the local alignment (discussed in (i)) and from the phase and density fluctuations discussed in (ii) and (iii). With probability $1-p_\theta$, we only allow update of $\theta$ from the equilibrium relaxation due to the nearest neighbor orientations as described in (i). By varying $p_\theta$, we can control the relative strength of the microscopic active processes in the update of the phases. The probabilistic update is used to encode the inherent stochasticity of the noisy dynamics. 

For both the above mentioned update scenarios, the updating of the phase is also supplemented by a random noise $\xi$ drawn from a uniform distribution $[-\xi/2,\xi/2]$, which has an effect analogous to thermal fluctuations measured by the temperature in equilibrium systems. Here, let us emphasize the well known analogy between the flocking transition described by the Vicsek~\cite{vicsek} model and the ferromagnetic transition in the Ising model. The role played by the random noise $\xi$ associated with the alignment in the Vicsek model is equivalent to the role of temperature in the ferromagnetic transition of the Ising model; for less $\xi$ the system is ordered, while the system is disordered for higher $\xi$. 
The analogy between the noise $\xi$ and temperature $T$ in equilibrium could also be drawn from Eq.~\eqref{qlro}. In the QLRO phase of the equilibrium XY model,  the variance of fluctuations of the local phase $\langle\theta^2 ({\bf x},t)\rangle$ scales with $T$. It may be noted that our agent-based model reduces to the equilibrium XY model in the passive limit (i.e., when the ``active'' or nonequilibrium ingredients in the phase update rules are dropped); see  Fig.~\ref{fig:PureXY} in Appendix~\ref{equilibriumXYmodel} and the discussions there.


Likewise, the time-evolution of the local number density occurs via two different processes: 

(i) Random hopping in response to local density inhomogeneities: Particles hop randomly to the nearest neighbor sites with a lower density, which models ordinary diffusion and

(ii) By hopping in response to the  phase difference between the originating site $\ell$ and a randomly chosen nearest neighbor target site $v$.  This is reminiscent of the phase-dependent force introduced in a microscopic model of synchronization in {\em Chlamydomonas}~\cite{ramin-bennet}. Again, this function is arbitrary, being constrained only the above symmetries. Further, allowing the phase difference-dependent current to depend on the sense of rotation,  we take this function to be $c_{\rm sum}=\Omega^\prime (\langle c\rangle_v)\sin (\theta_\ell-\theta_v)$, for simplicity, where $\langle c\rangle_v$ is the mean density of sites $\ell$ and $v$. Function $\Omega^\prime(\langle c_v\rangle)$ is assumed to be a quadratic function of its arguments for simplicity that contains two parameters $c_s$ and $c_{s2}$;  see Eq.~\ref{omega_prime_cv} in Appendix~\ref{MC-update}. At each MC step, we calculate $c_{\rm sum}$: For $c_\text{sum}=\pm 1$, particles move from (to) the originating site  to (from) the target site.

In the same spirit of $p_\theta$, we define another probability $p_c$, which dictates whether the particle hopping to its nearest neighbor be determined by pure diffusion with probability $1- p_c$ ({as described in (i)}) or by the phase difference-induced current of the particles with probability $p_c$ ({as discussed in (ii)}).


It is instructive to compare our agent-based model with the Vicsek model for flocking~\cite{vicsek,ginelli}. In the latter model, the particles or agents are ``self-propelled', i.e., locally align themselves with the neighbors, and move along that direction, resulting into flocking at low noises and high densities. In contrast, in the present model, the spins are {\em immobile} and locally align with the neighbors, {\em but obviously do not move in that direction}. Instead, they {\em actively} rotate locally in response to the local phase difference in a way that breaks the clockwise-anticlockwise symmetry, and to the local number density as well. 
 


{

\section{Monte-Carlo simulations of the agent-based model}\label{num-res}

\subsection{General features}
The model which we have defined in Sec.~\ref{modeling}, is simulated on a square lattice with linear sizes $L=32, 64, 128, 192$ and $256$ with $c_0=5$. We have fixed the lattice spacing $a_0=1$. Starting from a nearly phase-ordered spin configurations, the Monte Carlo algorithm (see Appendix~\ref{MC-update}) evolves the system under various control parameters ($e.g.,$ $g_1$, $g_2$, $c_s$, $c_{s2}$, and $\xi$) until steady states are reached. In our simulations, $\xi=0.1$ (unless otherwise mentioned). 

{During the simulations, we first let the system to evolve} for $t_{ss} = 10^{4}$ MC step, so that steady states are reached, and following this, measurements are carried out in the steady states and noise-averaged data of the phase fluctuation, $\langle\theta^2\rangle$, are recorded up to some maximum time $t_m=10^6$ MC step. To achieve better statistics, data obtained are further averaged over $13$ independent initial spin configurations. The numerical code is implemented to measure the variance of the average spin fluctuations $\langle\theta^2\rangle$ in the following manner: 
\begin{equation}
\langle\theta^2\rangle=\langle(\theta-\bar{\theta}_{L^2})^2\rangle_{t,L^2},\label{thetasq-def-num}
\end{equation}
where $\bar{\theta}_{L^2}$ is calculated using 
\begin{equation} \label{theta_sysav}
 \bar{\theta}_{L^2} = \arctan[\langle \sin(\theta) \rangle_{L^2}/\langle \cos(\theta) \rangle_{L^2}],
\end{equation} 
Here, $\langle ... \rangle_{L^2}$ denotes the average over all the $L^2$ sites at a given time $t$, and $\langle ... \rangle_{t,L^2}$ denotes averaged over time and $L^2$ sites for a fixed $\xi$ (meaning a fixed ``nonequilibrium temperature''). Notably, a nonzero $\bar{\theta}_{L^2}$ in the steady state indicates global rotation of the spins, consistent with the identification of the model as an {\em active oscillator model}.

We numerically calculate $\sigma(N_0)$ by 
\begin{equation}
\sigma (N_0)=\sqrt{\langle(N-N_0)^2\rangle}
\label{N_0}
\end{equation}
in the steady state, where $N$ is the  total number of {particles} in an open square box of increasing linear length $l_s << L$, and $N_0\equiv c_0 \ l_s^2$ is the average number of particles within that box; $\langle ... \rangle_{t}$ denotes averaged over time.

All the averages are performed in the steady states.

\subsection{Steady states in the agent-based model}

Our simulations clearly reveal the existence of distinct ordered and disordered states. The most spectacular results from our simulations are in fact manifested in the movies MOVIE1, MOVIE2, MOVIE3 and MOVIE4 in the Supplemental Material~\cite{supp2}. Movies MOVIE1 and MOVIE2 respectively represent the temporal evolution of the spin configurations in the ordered steady states and states without any ordering, whereas the corresponding movies showing the time evolution of the density distributions are MOVIE3 (ordered state) and MOVIE4 (no order).

To motivate the reader further and to give him/her an bird's eye view of the ensuing results, we also refer to the snapshots in Fig.~\ref{snapshot-visual}(a), (b), (d), and (e) (ordered state), and (c) and (f) (disordered state). In the spin configuration plots (Fig.~\ref{snapshot-visual}(a)-(c)), an arrow represents the local coarse-grained spin orientation at a time after the steady state has reached, whereas the colorbars in Fig.~\ref{snapshot-visual}(d)-(f) represent the local density fluctuation. Fig.~\ref{snapshot-visual}(a)-(b) show a nearly ordered spin configurations for the given parameter values which is validated by the corresponding density colormaps shown in Fig.~\ref{snapshot-visual}(d)-(e) where the uniform color gradient signifies a homogeneous ordered phase. In Fig.~\ref{snapshot-visual}(a) and (d), we have shown the plots for $L=64$ wheres  Fig.~\ref{snapshot-visual}(b) and (e) are for $L=256$. These figures clearly demonstrate that ordering persists even with a 4 fold increase in $L$, as expected in ordered states. In contrast, Fig.~\ref{snapshot-visual}(c) and (f) signifies the disordered states (for the parameter values mentioned) as evident from the disorderly orientations of the spin vectors in (c) and from the density colormap in (f) which displays wide fluctuations in the local density, including some empty or near empty regions. Unsurprisingly, the movies mentioned above directly correspond to the model's configuration snapshots. For instance, movies MOVIE1 and MOVIE3 correspond  to the snapshots in Fig.~\ref{snapshot-visual}(a) and (d), which are clearly phase-ordered states, together with small density fluctuations around the mean density. In contrast, movies MOVIE2 and MOVIE4 correspond to the snapshots in Fig.~\ref{snapshot-visual}(c) and (f), which clearly lack any phase order, and have large density fluctuations.

\begin{figure*}[!hbt]
\includegraphics[width=\linewidth]{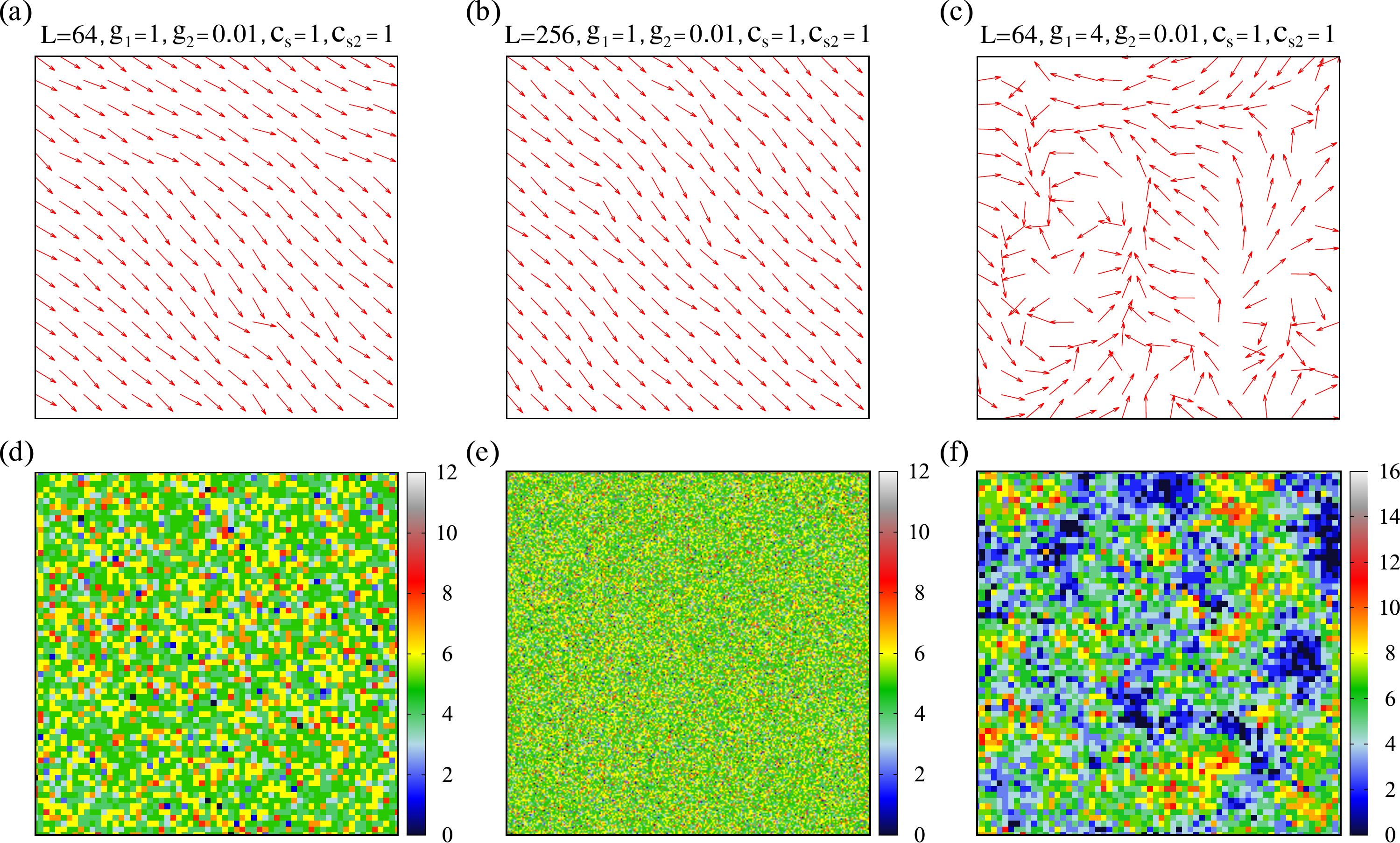}
 \caption{Representative snapshots showing ordered [(a) and (b)], and disordered (c) spin configurations, and ordered [(d) and (e)], and disordered (f) density distributions. The model parameters are mentioned on top of each figure. For better visual clarity, we have shown the phase configurations taken from a small part of each system (a $16^2$ section is shown); the density distributions are shown over full system sizes. Subplots (a) and (b) show ordered configurations for $L=64,\,256$, showing the persistence of order even after 4 fold increase in $L$; subplot (c) shows clear lack of any order. Similarly, subplots (d) and (e) show small density fluctuations about the mean density $c_0=5$ in the snapshots for $L=64,\,256$, whereas in subplot (f) much bigger density fluctuations can be seen, signifying a disordered state.  Other parameters: $c_0=5$, $\xi=0.1$, and $p_\theta=p_c=0.5$.
}
 \label{snapshot-visual}
\end{figure*}

The ordered and disordered states can be further distinguished by considering the probability distribution $P(\theta)$ of the orientation $\theta \in [0,2\pi]$ and the probability distribution $P(c)$ of the density $c$. The corresponding plots are shown in Fig.~\ref{histo1}. In a perfectly ordered uniform state, $P(\theta)\propto \delta(\theta-\theta_0)$ and $P(c)\propto \delta (c-c_0)$, where $\theta_0$ is an arbitrary reference state $\in [0,2\pi]$. Nonetheless, it is reasonable to expect that in the ordered states, both $P(\theta)$ and $P(c)$ will continue to be sufficiently narrow and will have peaks around the respective mean values of $\theta$ and $c$, as shown in Fig.~\ref{histo1}(a) and (b), respectively. In Fig.~\ref{histo1}, we observe that $P(\theta)$ and $P(c)$ are sharply peaked around angle $\theta_0\approx 330^\circ$ and at the mean density $c_0=5$. In general, the mean density, as obtained from and related to the peak of $P(c)$, should depend on the simulation parameter for average density. While the specific value of $\theta_0$ (calculated in the rotating frame) is of no physical significance, the narrow distribution does imply an orientationally ordered state, accompanied by a nearly smooth distribution of the particle number, corresponding to $P(c)$ being sharply peaked around $c=c_0$. Quite obviously, the distributions, $P(\theta)$ and $P(c)$, would be sufficiently broad with larger variance in disordered states as also demonstrated in Fig.~\ref{histo1} where we observe that the phase distribution is flat, indicating all values of the phase is equally probable, along with a wide distribution of the density implying large fluctuations of the density. Note that we have also observed another form of disordered phase in which the density distribution has two peaks, one near zero and another smaller peak at a density far greater than the mean density $c_0$ (see Fig.~\ref{roughAndaggregate} { below}). Two kinds of disordered phases are discussed in detail later.

\begin{figure}[!hbt]
\includegraphics[width=\columnwidth]{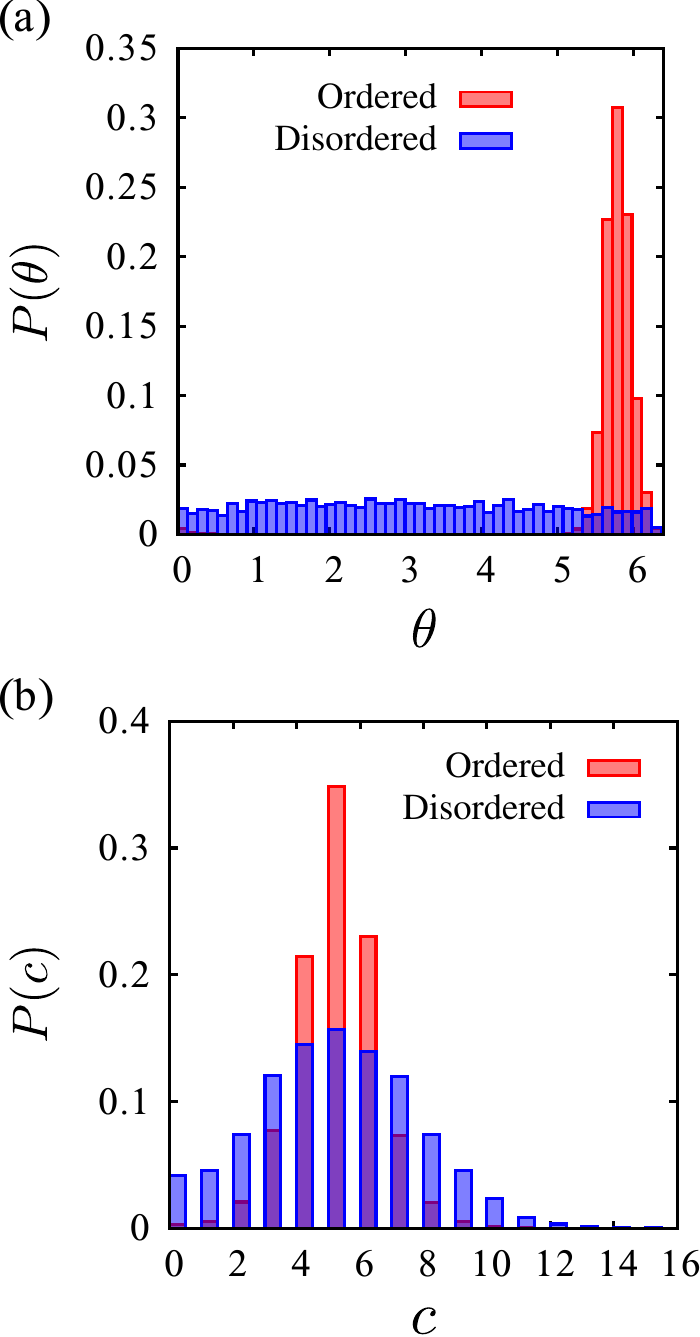}
\caption{Plots of (a) $P(\theta)$ versus $\theta$ and (b) $P(c)$ versus $c$ in the ordered (red) and disordered (blue) states. As argued, in the ordered state [corresponding to the snapshots in Fig.~\ref{snapshot-visual}(a) and (d)], both $P(\theta)$ and $P(c)$ are narrow, peaked around their mean values, whereas in the disordered state [corresponding to the snapshots in Fig.~\ref{snapshot-visual}(c) and (f)], $P(\theta)$ is nearly flat, and $P(c)$ is much broader than for the ordered state (see text). }\label{histo1}
\end{figure}


\subsubsection{Ordered states}

  Are all the ordered states in the agent-based model,  parametrized by the simulation parameters $g_1,\,g_2,\,c_s,\,c_{s2}$ statistically equivalent? That is to ask: Are all of them characterized by the same scaling exponents?  We now set out to address this generic question by quantitatively studying these ordered states below by using the agent-based model.

The natures of order in the NESS for the different parameters are investigated in Fig.~\ref{fig:DiffXYOrdering_zeta0_1}. Fig.~\ref{fig:DiffXYOrdering_zeta0_1}(a) and Fig.~\ref{fig:DiffXYOrdering_zeta0_1}(b) demonstrate instantaneous spin configurations, and Fig.~\ref{fig:DiffXYOrdering_zeta0_1}(c) and Fig.~\ref{fig:DiffXYOrdering_zeta0_1}(d) show the corresponding densities on a $32^2$ segment of a $256^2$ lattice for better visualization. The system exhibits SQLRO/MNF for the active coefficients (a \& c) $g_1=1.0$, $g_2=0.02$ and WQLRO/GNF for (b \& d)  $g_1=2.0$, $g_2=0.03$ respectively, the quantitative validation of which is provided below.

 We have numerically calculated $\langle\theta^2\rangle$ from (\ref{thetasq-def-num}) and extracted its dependence on $\ln L$ and plotted $\langle \theta^2\rangle/\ln L$ versus $\ln L$ in Fig.~\ref{num-plot1-nofit}(a). Since in equilibrium $\langle\theta^2\rangle$ scales as $\ln L$ meaning QLRO, we investigate if the nonlinear effects alter the $\ln L$-dependence of $\langle\theta^2\rangle$  in the active problem. In fact, we detect two distinctly different kinds of dependence, {\em both} in the phase-ordered steady states: Noting that for QLRO or in the equilibrium limit, $\langle\theta^2\rangle/\ln L$ is a constant independent of $ L$, we find $\langle\theta^2\rangle$ necessarily increases with $\ln L$. While our numerical results emphatically rule out SRO, it is also not LRO, since  $\langle\theta^2\rangle$ grows with $\ln L$ (hence with $L$). This behavior is then reminiscent of QLRO, being a ``middle ground'' between SRO and LRO. However, a careful observation of Fig.~\ref{num-plot1-nofit}(a) reveals that depending upon the parameters, there are actually two {\em qualitatively distinctly different} scaling behaviors - in one case $\langle\theta^2\rangle$ grows {\em slower} than $\ln L$, and in the other case {\em faster} than $\ln L$. We name them {\em stronger} than QLRO or {\em SQLRO} and {\em weaker} than QLRO or WQLRO, respectively. 
  However, the system eventually becomes disordered for large enough values of $\xi$. For a detailed picture of the nature of ordering as a function of the control parameters, we refer to Table~\ref{tab4Simulation1}. 
 

In parallel to calculating $\langle\theta^2\rangle$, we further investigate the number fluctuations $\sigma (N_0)$ as defined in (\ref{N_0}), and extract its dependence on $N_0$. In equilibrium systems away from any critical point $\sigma(N_0)\sim \sqrt{N_0}$. To see if active effects can change this result, we plot $\sigma(N_0)/\sqrt{N_0}$ versus $\sqrt{N_0}$ in Fig.~\ref{num-plot1-nofit}(b). This, just like $\langle\theta^2\rangle$ as a function of $\ln L$, displays two qualitatively different kind of scaling with $\sqrt{N_0}$. In one case, $\sigma(N_0)$ grows {\em slower} than $\sqrt{N_0}$ for model parameters for which $\langle\theta^2\rangle$ shows SQLRO. This case clearly implies {\em miniscule} (with respect to its equilibrium counterpart) number fluctuations or {\em MNF}. For other choices, $\sigma(N_0)$ grows faster than $\sqrt{N_0}$ for model parameters for which $\langle\theta^2\rangle$ shows WQLRO. This case gives {\em giant} (again with respect to its equilibrium counterpart) number fluctuations or {\em GNF}.
Our investigation confirms that MNF and GNF in Fig.~\ref{num-plot1-nofit}(b) respectively correspond to SQLRO and WQLRO of Fig.~\ref{num-plot1-nofit}(a), and therefore establishes the SQLRO $\Leftrightarrow$ MNF and WQLRO $\Leftrightarrow$ GNF correspondence in the agent-based model.

\begin{figure*}[!hbt]
\includegraphics[width=\linewidth]{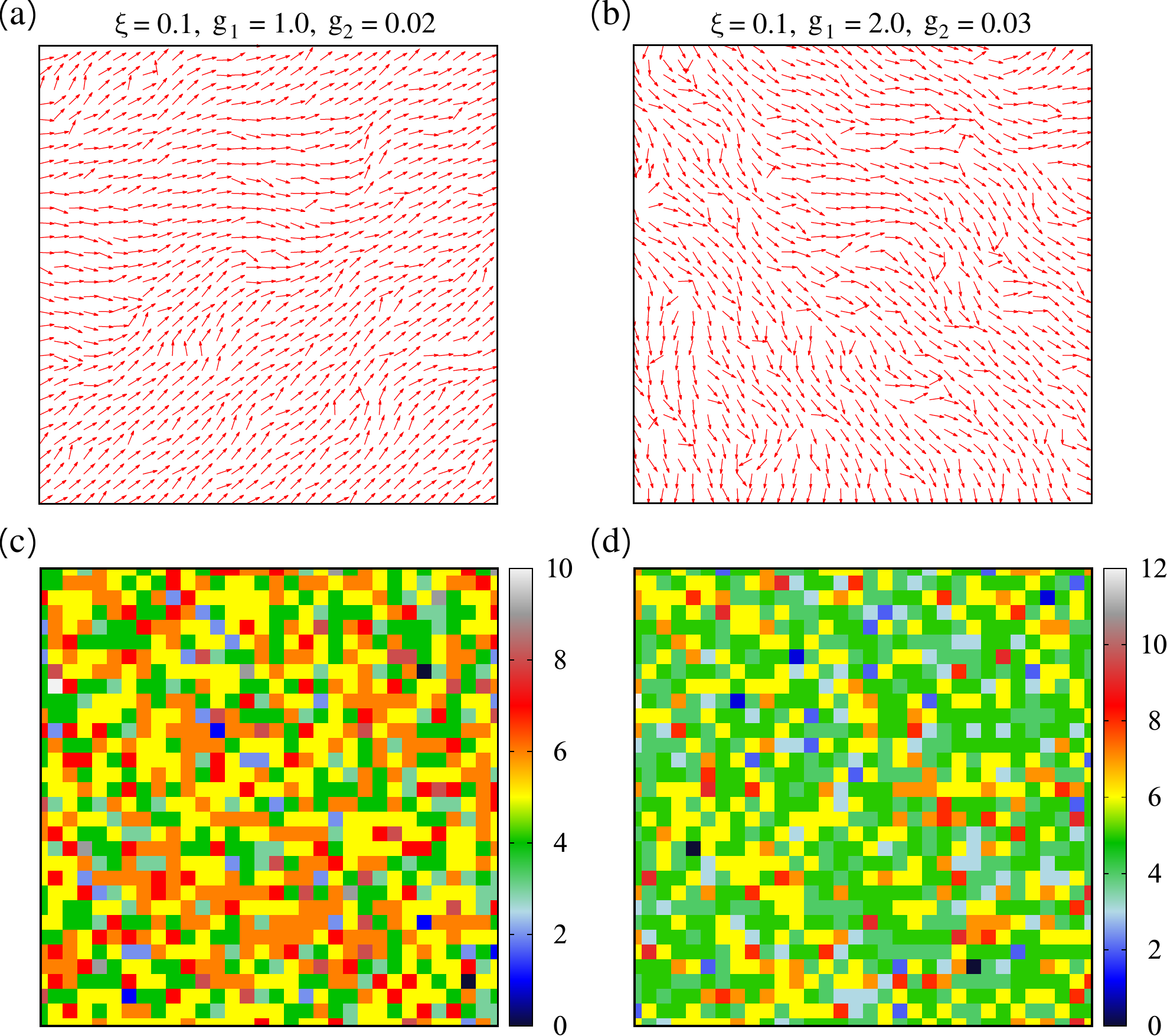}\vskip0.2cm
\caption{NESS snapshots showing (a-b) the ordered spin configurations and (c-d) the corresponding instantaneous density profiles in a $32^2$ segment of a system with $L=256$, shown for two different combinations of active coefficients $g_1$ and $g_2$: (a \& c) $g_1=1.0$, $g_2=0.02$  and (b \& d) $g_1=2.0$, $g_2=0.03$ respectively. Parameters: $c_0=5$, $\xi=0.1$, $p_\theta=p_c=0.5$, $c_s=1$, and $c_{s2}=1$.
}
\label{fig:DiffXYOrdering_zeta0_1}
\end{figure*}

 \begin{figure*}[hbt!]
\includegraphics[width=\linewidth]{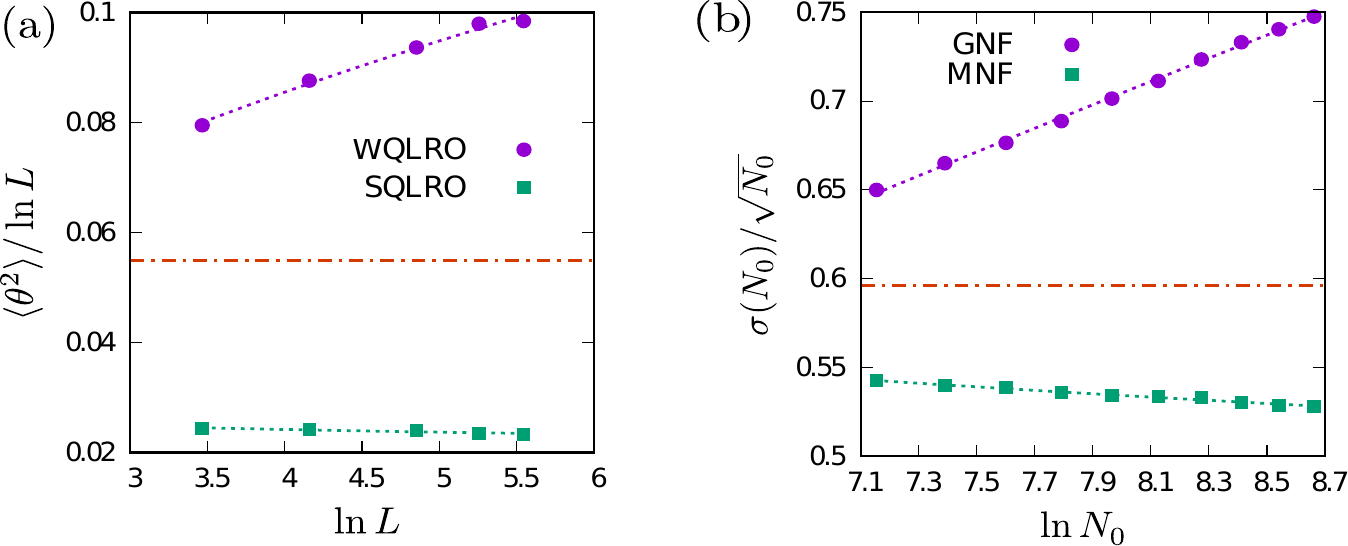}
\caption{ Plots showing scaling in the ordered phase obtained from the agent-based simulations: (a)  $\langle\theta^2\rangle/\ln L$ versus  $\ln L$ showing SQLRO and WQLRO, and  (b) $\sigma(N_0)/\sqrt{N_0}$ versus $\ln N_0$ showing MNF and GNF ($L=128$) with parameters : $c_0=5$, $\xi=0.1$,  $p_\theta=p_c=0.5$, $c_s=1$, $c_{s2}=1$, (SQLRO/MNF): $g_1=1.0$, $g_2=0.02$  and (WQLRO/GNF): $g_1=2.0$, $g_2=0.03$  respectively. The red broken horizontal lines denote the behavior of plots in the equilibrium limit, or the (a) QLRO and (b) NNF correspondence associated with the plots. The SQLRO-MNF and WQLRO-GNF correspondences are clearly established in the agent-based model (see text).}
\label{num-plot1-nofit}
\end{figure*}



\begin{figure}[hbt!]
\includegraphics[width=8 cm]{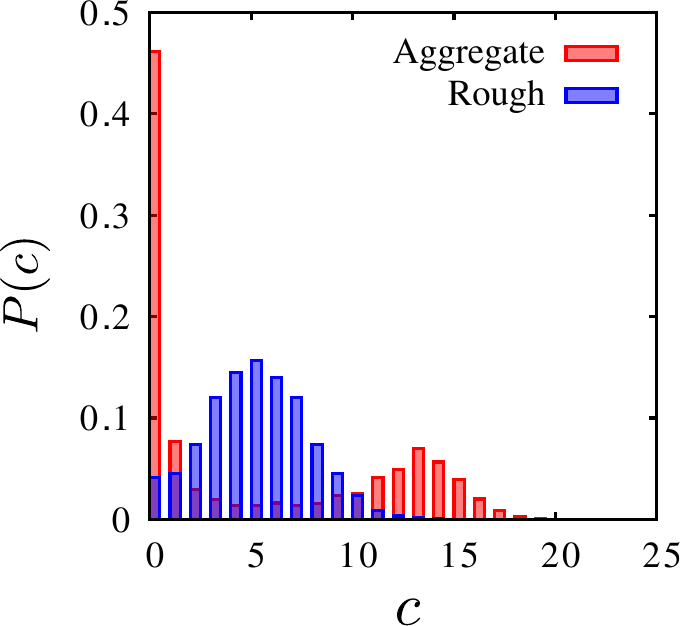}
\caption{Plot of $P (c)$ versus $c$ showing two distinct types of disordered phase obtained in the agent-based model, referred as rough and aggregate phases. Parameters: $L=64$, $c_0=5,\,\xi=0.1$, $p_{\theta} = p_{c} = 0.5$, (`Rough'): $g_1 = 4.0$, $g_2 = 0.01$, $c_s = 1$, and $c_{s2} = 1$, and (`Aggregate'): $g_1 = - 1.0$, $g_2 = 0.01$, $c_s = -1$, and $c_{s2} = -1$, respectively.}
\label{roughAndaggregate}
\end{figure}

\begin{table}[h!]
\begin{center}
\begin{tabular}{|p{1.0cm}|p{1.0cm}|p{2.2cm}|}
 \hline
  \multicolumn{3}{|c|}{Types of Order} \\
 \hline
 \hline
  $g_1$ & $g_2$   & Order\\
 \hline
 \vspace{0.05mm} & \vspace{0.05mm}  & \vspace{0.05mm} \\

  $1.0$   &   & SQLRO\vspace{0.5mm} \\

  $1.5$  & $0.01$ & SQLRO\vspace{0.5mm} \\

  $2.0$  &  & SQLRO\vspace{3.5mm} \\
\cline{1-3}
\vspace{0.01mm} & \vspace{0.01mm}  & \vspace{0.01mm} \\
   $1.0$ &   &  SQLRO \vspace{0.5mm}\\

   $1.5$  &$0.02$  & SQLRO \vspace{0.5mm}\\

   $2.0$  & & QLRO$^*$ \vspace{3.5mm}\\
\cline{1-3}
\vspace{0.01mm} & \vspace{0.01mm} & \vspace{0.01mm}  \\
  $1.0$ &    & SQLRO \vspace{0.5mm}\\

 $1.5$   &  $0.03$  & QLRO$^*$ \vspace{0.5mm}\\

  $2.0$ &    & WQLRO \vspace{0.5mm}\\
  
  \vspace{0.01mm} & \vspace{0.01mm} & \vspace{0.01mm}  \\
\hline
\vspace{0.01mm} & \vspace{0.01mm}  & \vspace{0.01mm} \\
   & $0.01$ &   \vspace{0.5mm} \\
$4.0$ & $0.02$ &   Rough phase \vspace{0.5mm} \\
& $0.03$ &   \vspace{0.5mm} \\
\vspace{0.01mm} & \vspace{0.01mm}  & \vspace{0.01mm} \\
\hline
\end{tabular}
\caption{The natures of ordering obtained for different values of the parameters  $g_1, \; \text{and}\; g_2$ respectively. The values of the coefficients $c_s$ and $c_{s2}$ are kept $1$. Other parameters:  $c_0=5$, $\xi=0.1$, and $p_\theta=p_c=0.5$. Increasing the magnitude of $g_1$ can result in a transition from ordered to disordered rough phase in our simulations (see text). }
 \label{tab4Simulation1}
\end{center}
\end{table}

\subsubsection{Disordered states}
We now use our agent-based model to study the nature of the disordered states and characterize them. Surprisingly, our simulation studies reveal the existence of {\em two} distinct types of disordered states, characterized by their distinctive density distributions (see Fig.~\ref{roughAndaggregate}). In one case, the probability distribution of the number density is still peaked about $c_0$, but is much wider (and also asymmetric about $c_0$). Clearly, a broad density distribution peaked at $c_0$ is found in our simulations implies lack of stable cluster formations of any preferred sizes.  We name this {\em rough phase}. In the second case, that we call as the {\em aggregate} phase, it is not peaked about $c_0$; instead, a large peak close to zero, and a smaller peak at a density much larger than $c_0$ are observed deep inside the aggregate phases. 

Visual inspection of movies of the disordered states (see Supplemental Material~\cite{supp2}) reveals that the life times of the density clusters deep inside the aggregate phases clearly exceed the simulations times. In the rough phase, however, no persistent clusters of preferred sizes are visible in NESS. See movies MOVIE2 and MOVIE4  for the time-evolution of the phase and density configurations in the rough phase, and the corresponding movies MOVIE5 and MOVIE6 for the aggregate phase in Ref.~\cite{supp2}. Macroscopic distinctions between the rough and aggregate phase steady states are clearly visible from these movies. 


A more detailed analysis of the different kind of disordered states (see Fig.~\ref{fig:LinUns}, Fig.~\ref{rough-aggregate} and Fig.~\ref{aggregate3}), changes in the nature of the disordered states depending upon $c_s$ (see Fig.~\ref{crossover} and Fig.~\ref{crossover1}), and the existence of the  rough phases for $g_2=0$  as obtained from our simulations (see Fig.~\ref{fig:KPZ} and Fig.~\ref{fig:KPZ1}), are discussed in details in Appendix~\ref{add-numericalresults}. 

We find that the magnitudes and signs of the coefficients in the agent-based model determine the nature of the disordered states. Depending upon the sign of these model parameters ($g_2$, $c_s$, and $c_{s2}$) the nearly phase-ordered system can get disordered corresponding to the aggregate phases or the rough phases in our simulations, which are manifested by disordered spin morphology and the formation of clusters of particles of typical sizes in NESS (see Fig.~\ref{fig:LinUns}; Appendix~\ref{disd-st}). 
 Further, we also note that it is thus possible to switch between the ordered and rough phases by tuning the values of the active coefficients. For instance, it is found that increasing only the magnitude of $g_1$ can results in a transition from ordered to rough phase as noted in Table~\ref{tab4Simulation1} (also see Fig.~\ref{snapshot-visual}). See Table~\ref{tab5Simulation} for a general classification of the steady states, parametrized by the signs of the model parameters.

\begin{table*}[!hbt]
\begin{center}
\begin{tabular}{|p{1.5cm}|p{1.5cm}|p{1.5cm}|p{1.5cm}|p{4.0 cm}|}
 \hline
 \multicolumn{5}{|c|}{Nature of ordering as parametrized by the model parameters} \\
 \hline
 \hline
 \hfil$g_1$ & \hfil$g_2$ &  \hfil$c_s$ & \hfil$c_{s2}$ &\hfil Observed phase \\
 \hline
\hfil$+ve$ & \hfil$+ve$ &  \hfil$+ve$ & \hfil$+ve$ &\hfil \textbf{Ordered phase}   \vspace{1.0mm}\\

 \hfil $+ve$ &\hfil $+ve$ &\hfil $+ve$ &\hfil $-ve$ &\hfil Aggregate phase  \vspace{1.0mm}\\

 \hfil$+ve$ &\hfil $+ve$ &\hfil  $-ve$ &\hfil $+ve$ &\hfil \textbf{Ordered phase}  \vspace{1.0mm} \\

 \hfil$+ve$ &\hfil $+ve$ &\hfil  $-ve$ &\hfil $-ve$ &\hfil Aggregate phase  \vspace{1.0mm}\\

\hfil $-ve$ &\hfil $+ve$ &\hfil  $+ve$ &\hfil $+ve$ &\hfil \textbf{Ordered phase}  \vspace{1.0mm}\\

 \hfil$-ve$ &\hfil $+ve$ &\hfil  $+ve$ &\hfil $-ve$ &\hfil Aggregate phase  \vspace{1.0mm} \\

\hfil$-ve$ &\hfil $+ve$ &\hfil  $-ve$ &\hfil $+ve$ &\hfil \textbf{Ordered phase}  \vspace{1.0mm} \\

\hfil$-ve$ &\hfil $+ve$  &\hfil  $-ve$ &\hfil $-ve$ &\hfil Aggregate phase  \vspace{1.0mm}\\

\hfil $+ve$ &\hfil $-ve$ &\hfil  $+ve$ &\hfil $+ve$ &\hfil Aggregate phase  \vspace{1.0mm}\\

 \hfil$+ve$ &\hfil $-ve$ &\hfil  $+ve$ &\hfil $-ve$ &\hfil \textbf{Ordered phase}   \vspace{1.0mm}\\

\hfil$+ve$ &\hfil $-ve$ &\hfil  $-ve$ &\hfil $+ve$ &\hfil Aggregate phase  \vspace{1.0mm}\\

\hfil $+ve$ &\hfil  $-ve$ &\hfil $-ve$ &\hfil $-ve$ &\hfil \textbf{Ordered phase}  \vspace{1.0mm}\\

\hfil $-ve$ &\hfil $-ve$ &\hfil $+ve$ &\hfil $+ve$ &\hfil Aggregate phase \vspace{1.0mm}\\

\hfil$-ve$ &\hfil $-ve$  & \hfil $+ve$ &\hfil $-ve$ &\hfil \textbf{Ordered phase}  \vspace{1.0mm}\\

\hfil $-ve$ &\hfil $-ve$ &\hfil  $-ve$ &\hfil $+ve$ &\hfil Aggregate phase \vspace{1.0mm}\\

\hfil$-ve$ &\hfil $-ve$ &\hfil  $-ve$ &\hfil $-ve$ &\hfil \textbf{Ordered phase}  \vspace{1.0mm}\\
\hline
\end{tabular}
\caption{{Nature of order (or lack thereof) in the NESS in  our agent-based model, depending upon the possible signs of the active coefficients. For the phases written in bold, the absolute values of the model parameters correspond to the ordered phases of TABLE~\ref{tab4Simulation1}. The existence of the aggregate phase does not depend on the magnitudes of the parameter values, rather it depends on the sign of the active coefficients
.}}
 \label{tab5Simulation}
\end{center}
\end{table*}

In the equilibrium limit of the model, we find the phase variance to display the $\langle \theta^2 \rangle \sim \ln L$ behavior at sufficiently low noises, which is the hallmark of the two-dimensional equilibrium XY model (see Fig.~\ref{fig:PureXY}; Appendix~\ref{equilibriumXYmodel}).

\subsection{Summary of the numerical results}

We now summarize our results obtained from the Monte-Carlo simulations of the agent-based models.

\noindent
(i) The steady states in the agent-based model can be (a) phase-ordered, or (b) disordered.\\

\noindent
(ii) In the phase-ordered states, the variance of the phase fluctuations grows with the system size $L$. Depending upon the model parameters that determine the various active dynamical processes, $\langle \theta^2\rangle$ can grow faster or slower than $\ln L$. 
Similarly, the standard deviation of the number fluctuations $\sigma(N_0)$ can grow with $N_0$ faster or slower than $\sqrt{N_0}$. 
Our numerical results reveal a surprising and hitherto unknown correspondence between the phase and density fluctuations - the model exhibits the simultaneous occurrence of SQLRO (WQLRO) with MNF (GNF).


\noindent
(iii) The disordered states are found to be of two distinct types: (a) aggregate phase characterized by stable density clusters of distinctive size, and a probability distribution of the density that is {\em not} peaked at $c_0$, the mean density, but is bimodal with one peak close to zero, and another at a density much larger than $c_0$, (b) rough phase with {\em no} stable density clusters, corresponding to a density probability distribution that {\em is} peaked around $c_0$, but much broader than those for the phase-ordered states.

}

\section{Derivation of the Hydrodynamic theory}\label{hydro}

{ We have demonstrated the existence of stable phase-ordered states mediated by a mobile conserved species density in our agent-based model.  It should be remembered that this model is a particular model in the general class of models with many parameters arising from, e.g., more complex forms for the phase different-dependent and density dependent local rotations, and phase difference-dependent particle hopping in the update rules. One may additionally consider replacing the simple spin alignment rules by other alignment rules involving for instance slightly longer ranged (longer than nearest neighbor) attractions. If we include all of these, we can make a microscopic model with practically as many parameters as we may want. However, so long as all these new variants of the original update rules are all ``short ranged'', i.e., fall off sufficiently fast as a function of the separation, and retain all the basic invariances of the original update rules, they should ultimately show the same long wavelength properties as the original model. A hydrodynamic theory is a particularly powerful tool to establish that, which ultimately classify the universality class(es). Such a theory necessarily has a finite number of relevant parameters for a given class of microscopic models having arbitrary (and practically unlimited) number of parameters. A hydrodynamic theory for the agent-based model should also provides firmer footing to the plethora of remarkable results obtained from the agent-based simulations, by validating them analytically. To that end, we set up the hydrodynamic theory for the stable phase-ordered  states, and calculate the scaling properties  of fluctuations in those states. We also study the stability conditions of the phase-ordered states, which are expected to give us clues about the nature of the disordered states found in the agent-based model.}


\subsection{Identifying the hydrodynamic variables}

As in our agent-based model, we consider an active system consisting of a collection of mobile particles of fixed total number on a substrate, with each lattice site containing one XY spin.  Thus each  site is assigned an ``XY spin vector'' of fixed length. In order to formulate hydrodynamics, we must coarse-grain this particle-based description into a position-dependent two-component spin vector field, or equivalently a complex scalar field $Z({\bf r},t)$, characterized by an amplitude and a phase. In principle, the coarse-grain procedure should lead to fluctuations in both the amplitude and phase of $Z$. However, since we are considering a nearly phase-ordered collection of XY spins, and {\em not} at any continuous transition between the orientationally ordered and disordered phases, the spin amplitude fluctuations are {\em massive}, or have relaxation times that remain finite in the long wavelength limit. This is due to the fact that these amplitude fluctuations, unlike the fluctuations in the phases, are {\em not} Goldstone modes associated with spontaneous breaking of continuous rotation invariance of the spin space. Thus, in the absence of any symmetry arguments for the amplitude dynamics to be slow, we expect it to be generically fast. Hence, in the spirit of hydrodynamics, we ignore the amplitude fluctuations, and set $Z^2=1$ without any loss of generality.  In contrast, invariance under a global rotation of the { XY spins} (equivalently, a rotation of the chosen reference state) implies that the broken symmetry phase fluctuations are {\em slow} variables with { relaxation} time-scales diverging in the long wavelength limit. Furthermore, since we do not consider birth or death of the mobile species, their number density must be a conserved quantity, corresponding to it being a slow variable as well. Furthermore, due to the friction from the substrate, there is no momentum conservation. 
%
Therefore, in the long wavelength limit, we neglect the amplitude fluctuations and retain
 phase $\theta ({\bf x},t)$, and  number density $c({\bf x},t)$ as the hydrodynamic variables.  
 
In another different physical realization of the system, we could imagine the system to have mobile particles of fixed numbers, each carrying an XY spins on a $2d$ substrate, which has the {\em same} hydrodynamic equations for the phase-ordered states, owing to the identical number of broken symmetry mode and conserved variable (one each), and have the same symmetries. 
 
 \subsection{Hydrodynamic equations of motion}
 
 We now systematically formulate the hydrodynamic equations of motion of the phase $\theta ({\bf x},t)$, and  number density $c({\bf x},t)$.  
 For a driven system as ours, in the absence of any free energy functional to derive the equation of motions, we must write down the equations of motion by appealing to the general symmetries and conservation laws. Here, symmetries not only include the symmetries of the underlying microscopic dynamics, they should also include the symmetries of the states as well. This in particular means it depends on the specific symmetries broken in the ordered states. { The relevant symmetries are as follows:
 
 (i) For a collection of phase-ordered  { XY spins}, the relevant broken symmetry is the rotational invariance in the order parameter space}~\footnote{This is analogous to the breaking of the rotational invariance in the ``spin'' space of the classical spin models, e.g., the XY or the Heisenberg model, in the low temperature ferromagnetic phase. This is not to be confused with the rotational invariance of the physical space, which is assumed to be isotropic (i.e., unbroken) throughout this article}. 
 
 { (ii) Further, the system should be invariant under a translation and rotation in the physical space and a constant shift of the phase (equivalently, under rotation of all the spins, or of the reference state by an arbitrary amount). }

 These symmetry considerations dictate that the equation of motion for $\theta$ should have the general form
\begin{equation}
 \partial_t \theta = \kappa \nabla^2 \theta + \frac{\lambda}{2} ({\boldsymbol\nabla} \theta)^2 + \Omega(c) + f_\theta,\label{theta-full}
\end{equation}
where, $\kappa$ is the stiffness, $\Omega (c)$ is a general function of the density, and we have retained the lowest order nonlinear terms in $\theta$ and spatial gradients, in the spirit of hydrodynamics assuming small phase fluctuations in a nearly phase-ordered state. The $\kappa \nabla^2 \theta$ represents the standard spin relaxation term of equilibrium origin. The remaining two terms on the rhs of (\ref{theta-full}) represent {\em active rotation} in response to the local phase difference and number density. None of these terms can be obtained from a free energy functional. If we set $c=c_0=const.$, i.e., consider uniform density, then (\ref{theta-full}), the equation of motion of $\theta$, reduces to the well-known Kardar-Parisi-Zhang equation~\cite{kpz}, after absorbing a constant term $\Omega(c_0)$ by going to a rotating frame.

The equation for the number density $c$ should follow a conservation law and have the form
\begin{equation}
 \partial_t c = D \nabla^2 c +\lambda_0 {\boldsymbol\nabla}\cdot(\tilde \Omega(c){\boldsymbol\nabla}\theta) + f_c.\label{c-full}
\end{equation}
Equation~(\ref{c-full}) corresponds to a particle current
\begin{equation}
 {\bf J}_c=-D{\boldsymbol\nabla}c - \lambda_0\tilde \Omega(c){\boldsymbol\nabla}\theta.\label{c-curr}
\end{equation}
Here, $D$ is the diffusivity of the { mobile species}, and $\lambda_0$ is a coupling constant that can be positive or negative. Again we have truncated up to the lowest order in $\theta$ and spatial gradients; $\tilde \Omega(c)$ is another general functions of $c$; in general $\Omega(c)$ and $\tilde\Omega(c)$ are two different independent nonlinear functions of $c$. The $\lambda_0\tilde \Omega(c){\boldsymbol\nabla}\theta$-term describes particle current in response to a local phase difference; it may enhance or suppress the diffusive current $D{\boldsymbol\nabla}c$. The two signs of $\lambda_0$ for a given $\tilde\Omega(c)$ are reminiscent of contractile and extensile active matters~\cite{sriram-RMP}. Indeed, we shall see below that the two possible signs of $\tilde\lambda_0$ corresponds to {\em two different physical} states of the system.
 Gaussian noises $f_\theta$ and $f_c$ are white and conserved noises, respectively, which are added phenomenologically to model the inherent underlying stochastic dynamics. These are the nonequilibrium analogs of thermal noises in equilibrium models. 
 Their variances are independent of the damping coefficients in a driven system as ours in the absence of any Fluctuations-Dissipation Theorem (FDT)~\cite{chaikin}. We assume the variances to have the forms as given in (\ref{f-theta}) and (\ref{noise-c}) above.

Equations~(\ref{theta-full}) and (\ref{c-full}) can be obtained from a complex Ginzburg-Landau equation (CGLE) coupled with a density field $c$; see Appendix~\ref{app-cgle}. {The } advection by a flow field is irrelevant in the RG sense is discussed in Appendix~\ref{irre-adv}.

Now write $c({\bf x},t)$ as a sum of the mean concentration $c_0$ and local fluctuations around it, i.e., $c({\bf x},t)=c_0+\delta c({\bf x},t)$.  We expand the functions $\Omega(c)$ and $\tilde \Omega (c)$ about $c_0$. With this,  
the hydrodynamic equations retaining up to the lowest order nonlinearities take the form
\begin{subequations}
 \begin{align}
 & \frac{\partial\theta}{\partial t}=\kappa\nabla^2\theta + \Omega_1\delta c+\frac{\lambda}{2} ({\boldsymbol\nabla}\theta)^2+ \Omega_2 (\delta c)^2 + f_\theta,\label{theta-full-eq}\\
 &\frac{\partial\delta c}{\partial t}=\lambda_0\tilde\Omega_0 \nabla^2\theta + D\nabla^2\delta c + \lambda_0 \tilde \Omega_1 {\boldsymbol\nabla}\cdot(\delta c{\boldsymbol\nabla}\theta) + f_c.\label{c-full-eq}
 \end{align}
\end{subequations}
Here, $\Omega_1\equiv \partial \Omega/\partial c|_{c=c_0},\,\Omega_2 \equiv \partial^2 \Omega/\partial^2 c|_{c=c_0},\,\tilde \Omega_0\equiv  \tilde \Omega|_{c=c_0},\,\tilde\Omega_1 \equiv \partial \tilde\Omega/\partial c|_{c=c_0}$. Without any loss of generality, we set $\Omega_1 >0$; $\tilde \Omega_0$ can be positive or negative.  Using the fact that the phase $\theta$, an angle, is dimensionless, and $c$ has the dimension of surface (areal) density, we can fix the dimensions of the various model parameters in (\ref{theta-full-eq}) and (\ref{c-full-eq}). Parameters $\kappa$ and $D$ have the dimensions of diffusivity, $\lambda_0\tilde\Omega_0$ has the dimensions of frequency, $\lambda,\,\Omega_1$ and $\lambda_0\tilde\Omega_1$ have the dimensions of diffusivity, and $\Omega_2$ has the dimension of hyperdiffusivity.  
 Lastly, a term $\Omega (c_0)$ in the hydrodynamic equation for $\theta$ has been removed by going to a rotating frame: We shift phase $\theta({\bf r},t)$ by $\Omega(c_0)$, $\theta({\bf r},t)\rightarrow \theta({\bf r},t) + \Omega (c_0)t$, which removes the term $\Omega(c_0)$ from (\ref{theta-full}), that gets generated upon expanding $\Omega(c)$ about $c=c_0$.  { While we have derived the hydrodynamic equations (\ref{theta-full-eq}) and (\ref{c-full-eq}) from symmetry principles and conservation laws, these equations could also be derived from a general Smoluchowski-like Equation for a basic kinetic model corresponding to this system, and then by calculating the relevant moments of the distributions. We should however keep in mind that the orientational order in the present problem is in the order parameter space, distinct from the physical space; see Ref.~\cite{saint2,saint3} for detailed discussions.   }



{We note that the noises $f_\theta$ and $f_c$ in Eqs.~(\ref{theta-full}) and (\ref{c-full}), or (\ref{theta-full-eq}) and (\ref{c-full-eq}) have their origins in the stochasticity of the agent-based model. For example, noise $f_\theta$ is generated upon coarse-graining the microscopic phase update rules, which are the stochastic updates of the phase $\theta$, characterized by $p_\theta$ and the randomness in the sites to be updated in a given Monte-Carlo step, and the ``Vicsek''-like noise $\xi$}~ { \footnote{ {We make a technical point here for interested readers. It is expected that such a coarse-graining procedure should also generated multiplicative noises which will multiply the different terms in the hydrodynamic equations (as well as the subleading terms neglected in the hydrodynamic limit). However, since the microscopic random processes are all {\em short range}, $f_\theta$ too is short range. This then gives that all the multiplicative noise terms are {\em irrelevant} in the scaling sense. This consideration leaves $f_\theta$ as the only relevant noise in (\ref{theta-full-eq}). Similarly, the noise $f_c$ is generated upon coarse-graining the microscopic density update rules characterized by $p_c$, which include random hopping to a neighboring site with a lower density and the randomness in selecting the target site in the phase difference-dependent current. Again, coarse-graining should generate multiplicative noises, all of which are irrelevant in the scaling sense for reasons similar to the above. This leaves $f_c$ as the only relevant noise in (\ref{c-full-eq}).}}}
{ We have set the noise crosscorrelations to zero, keeping only the auto-correlations (\ref{noise-theta}) and (\ref{noise-c}) non-zero. The model at hand being a driven model, the noise variances and the damping coefficients are mutually independent. This allows us to choose the noise variances independently. Whether or not noise crosscorrelations exist depends upon whether the microscopic stochastic processes affecting the phase and density updates are crosscorrelated or not. For simplicity, we have chosen them not to be crosscorrelated.}

The hydrodynamic theory for the active XY model in contact with a conserved species on a substrate developed here is the hydrodynamic theory for active superfluids on a substrate as well. Hydrodynamic theory for an equilibrium superfluid has four hydrodynamic variables, {\em viz.}, phase, density, momentum density  and energy density (equivalently, entropy density). In its driven, nonequilibrium analogue on a substrate the last two are no longer hydrodynamic variables, leaving only the phase and density as the two hydrodynamic variables in the problem. This justifies the applicability of Eqs.~(\ref{theta-full-eq}) and (\ref{c-full-eq}) for an active superfluid on a substrate. 

 For a growing surface or a moving biomembrane, $\theta({\bf x},t)$ is the local height measured in the Monge gauge with respect to an arbitrary base plane. Due to the arbitrariness of the location of the base plane, the dynamics of the surface fluctuations is unaffected by the position of the base plane. This in turn implies that the dynamics must be invariant under a constant shift $\theta ({\bf x},t)\rightarrow \theta({\bf x},t) + const.$, which is the same symmetry as the phase. As in a given dimension all systems having the same symmetry are described by the same hydrodynamic theory, Eqs.~(\ref{theta-full-eq}) and (\ref{c-full-eq}) are the hydrodynamic equations for a fluctuating surface with a conserved species on it. Indeed, due to these symmetry reasons these equations also serve as the hydrodynamic theory of an active fluid membrane with a finite tension but without momentum conservation that contains an active species in it~\cite{act-mem}, as mentioned earlier. Nonetheless, there is a crucial difference between the active XY model problem and the surface dynamics problem. In the former case, $\theta({\bf x},t)$ is a {\em compact} variable, being confined between 0 and $2\pi$~\cite{chaikin}. In contrast, for a surface, $\theta({\bf x},t)$ is {\em non-compact} or unbounded. This has the consequence that there can be topological defects in the spin model, which have no analog in the height version of the problem. However, since we are considering smoothly varying systems, this distinction between the spin and the height versions of the problem does not arise.

Notice that the density fluctuations cease to become hydrodynamic variables in two distinct limits: (i) In the immobile limit with conserved but strictly uniform density, which is achieved when the  spins are rigidly grafted on the lattice points with no other conserved species present, and (ii) particle non-conservation via birth-death events~\cite{john-superfluid}. In each of these cases, the phase $\theta$ satisfies the pure KPZ equation~\cite{john-superfluid}.

 Although we motivate this study by principally considering active XY spins grafted on the lattice rigidly together with a diffusing conserved density trapped near the surface, the hydrodynamic equations (\ref{theta-full-eq}) and (\ref{c-full-eq}) apply to any system having the same symmetries, i.e., invariance under a translation and rotation in the physical space and a constant shift of the phase (equivalently, under rotation of all the spins, or of the reference state by an arbitrary amount). We may consider, for instance, a collection of diffusively mobile XY spins. This can happen when the mobile particles themselves carry a XY spin. In the phase-ordered states of this version of the model that has the same symmetry as above, we again have one broken symmetry mode (the phase), and one conserved density (of the mobile spins). The hydrodynamic equations for this model are same as (\ref{theta-full-eq}) and (\ref{c-full-eq}). Consider further an active system consisting of {\em immobile} XY spins grafted on lattice points and an incompressible asymmetric binary fluid on a substrate. An asymmetric binary fluid is described by a scalar order parameter $\phi$ that {\em does not} have the Ising symmetry, having a non-zero mean. Due to the conservation of the molecules of the two components that make up the binary fluid, $\phi$ follows a conserved dynamics. Further, overall incompressiblility implies that the local mean density is a constant. Lastly, the presence of a substrate ensures non-conservation of momentum, and the latter drops out of the hydrodynamics theory. This leaves us with the phase $\theta$ and the order parameter $\phi$ as the only two hydrodynamic variables. The rotational symmetry of the phase, absence of the Ising symmetry of $\phi$ and  the conservation of the latter ensures that the coupled dynamics of $\theta$ and $\phi$ are given by Eqs.~(\ref{theta-full-eq}) and (\ref{c-full-eq}); see Refs.~\cite{safran,cates}.  Interestingly, one-dimensional versions of Eqs.~(\ref{theta-full-eq}) and (\ref{c-full-eq}) map onto the hydrodynamic equations for a sedimenting one-dimensional crystal~\cite{rangan1} due to symmetry reasons.  Equations  (\ref{theta-full-eq}) and (\ref{c-full-eq}) also apply to a chiral active hexatic on a substrate again due to symmetry reasons~\cite{maitra}.

\section{Scaling and correlations in the linear theory}\label{lin-theo}

\begin{figure}[htb]
\includegraphics[width=0.9\columnwidth]{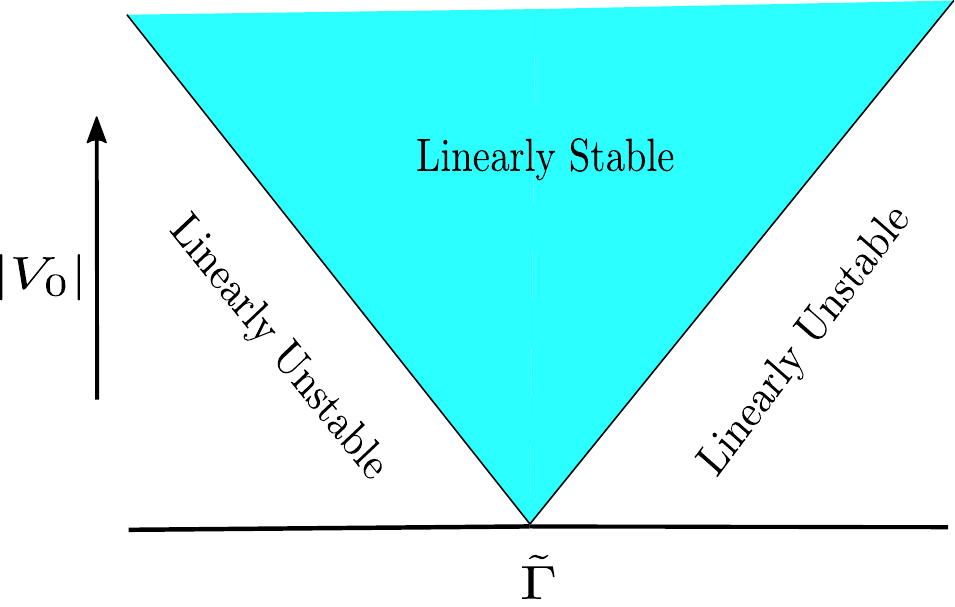}
 \caption{Schematic phase diagram in the $\tilde\Gamma-|V_0|$ plane for a given wavevector $k$. Linearly stable and unstable regions are marked. The magnitude of the slopes of the lines demarcating the two is $k/2$. We focus on the linearly stable region (see text).}\label{v-gammaT}
\end{figure}

\begin{figure}[htb]
\includegraphics[width=0.7\columnwidth]{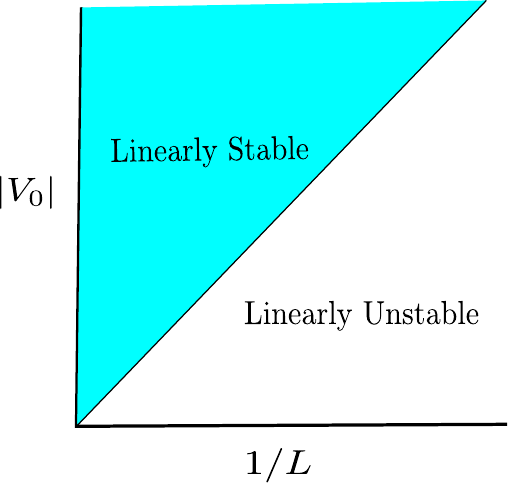}
 \caption{Phase diagram in the $1/L-|V_0|$ plane. Linearly stable and unstable regions are marked. Slope of the boundary line between the linearly stable and unstable regions is $\pi\tilde{\Gamma}$. We focus on the linearly stable region in the hydrodynamic theory (see text).}\label{v-L}
\end{figure}

It is instructive to first focus on the linearized hydrodynamics  [set $\lambda=\Omega_2=\tilde\Omega_1=0$ in Eqs.~(\ref{theta-full-eq}) and (\ref{c-full-eq})], and study their properties.  It is convenient to scale $\delta c$ by $\alpha$, which brings the linearized equations to the forms
\begin{subequations}
 \begin{align}
  &\frac{\partial\theta}{\partial t}=\kappa\nabla^2\theta + \frac{\Omega_1}{\alpha}\delta c + f_\theta,\label{lin1}\\
 &\frac{\partial\delta c}{\partial t}=\alpha\lambda_0\tilde\Omega_0 \nabla^2\theta + D\nabla^2\delta c + f_c,\label{lin2}
 \end{align}
\end{subequations}
where we have absorbed a factor of $\alpha$ in $f_c$.
We now fix $\alpha$ by setting 
{
\begin{equation}
\Omega_1/\alpha = \alpha |\lambda_0 \tilde \Omega_0| = V_0 \label{V0-defn}
\end{equation}
}
without any loss of generality. Let us now analyze the mode structure of the coupled linear equations (\ref{lin1}) and (\ref{lin2}). { By Fourier transforming (\ref{lin1}) and (\ref{lin2}) in space and time  we get 
the dispersion relation}
\begin{equation}
 \omega = \frac{i}{2}\left[-(\kappa +D)k^2 \pm \sqrt{(\kappa - D)^2 k^4 - 4 \lambda_0 \tilde \Omega_0\Omega_1 k^2}\right].\label{disp1}
\end{equation}
There are two distinct cases, delineated by the sign of the product $\lambda_0 \tilde \Omega_0 \Omega_1$. We first consider $\lambda_0 \tilde \Omega_0 \Omega_1>0$. We are interested in the long wavelength limit. We then get the dispersion relation for Fourier frequency $\omega$ as
\begin{equation}
 \omega = \pm V_0 k - i\Gamma k^2,\label{lin-disp}
\end{equation}
for sufficiently small wavevector $k$, a wavevector regime for which $V_0^2 \gg \tilde\Gamma^2 k^2/4,\;\Gamma = (\kappa+D)/2$ and further $\tilde\Gamma = \kappa-D$. This
corresponds to a linearly stable system with underdamped propagating waves with speed $V_0$; see Fig.~\ref{v-gammaT}. 
In other words, the underdamped propagating waves will be visible provided the linear system size $L$ is large enough to ensure $V_0\gg  \pi\tilde\Gamma/L$ for a fixed $\tilde\Gamma$; $k_{min}=2\pi/L$ is the smallest wavevector. See Fig.~\ref{v-L}.

 We have found that the sign of the product of the coefficients of the  linear cross coupling   terms in (\ref{lin1}) and (\ref{lin2}) controls the linear stability. This can be understood as follows. Assume $\Omega_1/\alpha>0$ without any loss of generality. Now consider a local region with excess particles, i.e., $\delta c>0$, surrounded by regions of deficit particles ($\delta c<0$).  This will ``speed up'' the local rotation of the spins, $\partial\theta/\partial t>0$ in the local patch where $\delta c>0$, surrounded by regions with $\partial\theta/\partial t<0$, implying slowing down there. This results into the orientation of the spins in the excess particle region to be ``locally ahead'' of the neighboring spins, giving a negative Laplacian of the phase locally. This will {\em enhance} $\delta c$ locally (i.e., {\em more} excess particles locally) if $\lambda_0 \tilde \Omega_0 \Omega_1<0$, which in turn increases $\partial\theta/\partial t$ further, eventually creating a run away process leading to the linear instability discussed above. On the other hand, if $\lambda_0 \tilde \Omega_0 \Omega_1>0$, an initial local positive $\delta c$ will {\em reduce} it via the linear $\nabla^2\theta$-term in (\ref{lin2}), ruling out any linear instability. See Ref.~\cite{rangan1,rangan2} for a similar mechanism in a one-dimensional sedimenting lattice.  In the surface growth version of the model, this linear instability implies that local excess or deficit of particles (i.e., $\delta c\neq 0$) enhances the local curvature; see Ref.~\cite{tirtha-active} for a similar mechanism of instability in an active membrane. 

 The correlation functions of $\theta({\bf k},\omega)$ and $\delta c({\bf k},\omega)$ in the linearly stable case in the linear theory can be calculated from (\ref{lin1}) and (\ref{lin2}) in a straightforward manner:
\begin{eqnarray}
 &&\langle |\theta({\bf k},\omega)|^2 \rangle = \frac{2D_\theta(\omega^2+D^2k^4)+2V_0^2D_ck^2}{(V_0^2k^2-\omega^2)^2+\omega^2k^4(\kappa +D)^2},\label{theta-corr-lin}\\
 &&\langle |\delta c({\bf k},\omega)|^2 \rangle = \frac{2D_\theta V_0^2k^4+2D_ck^2(\omega^2+\kappa^2k^4)}{(V_0^2k^2-\omega^2)^2+\omega^2k^4(\kappa +D)^2},\label{c-corr-lin}\\
 &&\langle \theta(-{\bf k},-\omega) \delta c({\bf k},\omega) \rangle \equiv \langle \delta c(-{\bf k},-\omega) \theta({\bf k},\omega) \rangle^*=\nonumber\\&&~~~~\frac{2V_0k^2\left[-D_\theta(-i\omega+Dk^2)+D_c(i\omega+\kappa k^2)\right]}{(V_0^2k^2-\omega^2)^2+\omega^2k^4(\kappa +D)^2}.\label{cross-corr-lin}
\end{eqnarray}
Unsurprisingly, both $\langle |\theta({\bf k},\omega)|^2 \rangle$ and $\langle |\delta c({\bf k},\omega)|^2 \rangle$ are real and positive definite, but the cross-correlation function $\langle \theta(-{\bf k},-\omega) \delta c({\bf k},\omega) \rangle$ is in general complex. All the correlators $\langle |\theta({\bf k},\omega)|^2 \rangle,\,\langle |\delta c({\bf k},\omega)|^2 \rangle$ and $\langle \theta(-{\bf k},-\omega) \delta c({\bf k},\omega) \rangle$ are peaked at $\omega = \pm V_0 k$ in the long wavevlength limit, with the width of each peak scaling with $k$ as $k^2$ (again in the long wavelength limit); the latter implies a dynamic exponent $z=2$ in the linear theory. This further implies that in the limit $k\rightarrow 0$, the ratio of the distance between the two peaks ($=2V_0k$) and their widths ($=2\Gamma k^2$) diverges: define { \em Quality factor} $\varphi=V_0/(\Gamma k)$ as the ratio of the two, which diverges for $k\rightarrow 0$. Although Eqs.~(\ref{theta-full-eq}) and (\ref{c-full-eq}) have their individual damping coefficients, respectively, $\kappa$ and $D$, in the linear coupled system of (\ref{lin1}) and (\ref{lin2}) it is their sum $\kappa+D\equiv 2\Gamma$ that acts as the effective damping. Thus for stability reasons, we must have $\Gamma >0$; individual positivity of $\kappa$ and $D$ are formally not necessary for stability, although since $\kappa$ and $D$ are physically identified with the microscopic spin stiffness and particle diffusivity, respectively, they are expected to be individually positive.  

We now calculate the equal-time correlation functions of $\theta({\bf x},t),\,\delta c({\bf x},t)$. We start from the definition
\begin{equation}
 C^0_{\theta\theta}(k)\equiv \langle |\theta({\bf k},t)|^2\rangle = \int_{-\infty}^\infty\frac{d\omega}{2\pi}\langle |\theta({\bf k},\omega)|^2\rangle,
\end{equation}
where a superscript ``0'' refers to this being a linear theory result.
We now use (\ref{theta-corr-lin}) above. By applying the residue  theorem, we get for the equal-time phase autocorrelation function in the Fourier space in the linear theory
\begin{equation}
  C^0_{\theta\theta}(k) \approx \frac{D_\theta + D_c}{2 \Gamma k^2},\label{theta-k-corr}
\end{equation}
see Appendix \ref{same-time-corr}.
Correlator (\ref{theta-k-corr}) implies 
\begin{equation}
 \Delta_\theta^0\equiv \langle\theta^2 ({\bf x},t)\rangle = \frac{D_\theta + D_c}{4\pi \Gamma} \ln \left(\frac{L}{a_0}\right),\label{qlro-l}
\end{equation}
corresponding to QLRO.  As mentioned earlier, (\ref{qlro-l}) holds for the variance of the surface fluctuations in the EW equation~\cite{stanley} as well. 

It is useful to compare the linear theory result in (\ref{theta-k-corr}) with the corresponding result for the equilibrium XY model 
in its QLRO phase at temperature $T$ as given in (\ref{qlro}) above: We note that the $k$-dependence is unchanged from its equilibrium counterpart. In fact 
(\ref{qlro}) and (\ref{qlro-l}) have the same $\ln L$-dependence.
 \, Direct comparison with (\ref{qlro-l}) allows us to define an effective temperature $T_{eff}^\theta$, an effective spring constant $\kappa_{eff}$ and a dimensionless or reduced temperature ${\cal T}_\theta$ for the phase fluctuations by
\begin{subequations}
 \begin{align}
   T_{eff}^\theta&\equiv D_\theta + D_c,\\
 \kappa_{eff}&\equiv 2\Gamma,\\
 {\cal T}_\theta &\equiv \frac{T^\theta_{eff}}{\kappa_{eff}}=\frac{D_\theta + D_c}{2\Gamma}.\label{red-temp}
 \end{align}
\end{subequations}

Equation~(\ref{qlro-l}) allows us to calculate the {\em 
Debye-Waller factor} (DWF) $W$, which for an equilibrium system tells us about the depression of the order parameter due to thermal fluctuations in the ordered phase; here, it would reveal the corresponding depression of the order parameter due to the noises, which generalize the role of temperature in a driven nonequilibrium system. This is defined as~\cite{chaikin} 
\begin{equation}
 {\cal R}e\langle\exp(i\theta)\rangle = \exp (-\langle\theta^2({\bf x},t)\rangle/2)\equiv \exp (-W)
\end{equation}
Thus we find
\begin{equation}
 W\equiv \frac{1}{2}\langle \theta^2 ({\bf x},t)\rangle = \frac{D_\theta + D_c}{8\pi \Gamma} \ln \left(\frac{L}{a_0}\right),
\end{equation}
such that the order parameter 
\begin{equation}
{\cal R}e\langle \exp (i\theta)\rangle = \exp[ -W]=\left(\frac{L}{a_0}\right)^{-\eta_0},\label{dw-lin}
\end{equation}
where 
\begin{equation}
\eta_0=\frac{D_\theta + D_c}{8\pi \Gamma}\label{dw-lin1}
\end{equation}
is model-dependent.
Thus, the order parameter decays {\em algebraically} with $L$ as $L^{-\eta_0}$, eventually vanishing in thermodynamic limit $L\rightarrow \infty$. Defining the   spin variable as a complex function $Z({\bf r},t)\equiv Z_0 \exp\left[i\theta({\bf x},t)\right]$ and noting that deep in the phase-ordered state, amplitude $Z_0\approx const.$, we can calculate the spin-spin equal-time correlation function in the real space in the linear theory:
\begin{eqnarray}
 C_{ZZ}^0({ r})&\equiv& \langle \cos[\theta({\bf x},t) - \theta({\bf x'},t)]\rangle \nonumber \\&=&{\cal R}e\langle \exp[i\{\theta({\bf x},t)-\theta({\bf x'},t)\}]\rangle\nonumber \\&=&\exp [-\langle\{\theta({\bf x},t)-\theta({\bf x'},t)\}^2\rangle/2],
\end{eqnarray}
where $r=|\bf x-x'|$. As shown in Appendix~\ref{real-corr},
\begin{equation}
 C_{\theta\theta}^0(r)\equiv \langle\left[\theta({\bf x},t)-\theta({\bf x'},t)\right]^2\rangle\approx \frac{D_\theta+D_c}{2\pi\Gamma}\ln (r/a_0)
\end{equation}
for large $r$. This gives
\begin{equation}
 C_{ZZ}^0(r)\approx\exp\left[-\frac{D_\theta+D_c}{4\pi\Gamma}\ln (r/a_0)\right]\equiv \left(r/a_0\right)^{-\tilde\gamma_0}
\end{equation}
for large $r$, where $\tilde\gamma_0=(D_\theta+D_c)/(4\pi\Gamma)$.
Thus the equal-time spin correlation function decays {\em algebraically} with distance, eventually vanishing at infinite distance. The above forms of $W$, $C^0_{\theta\theta}(r)$ and $C_{ZZ}^0(r)$ are the hallmarks of QLRO, and are identical in form with their counterparts for the $2d$ XY model in its low temperature QLRO phase.

We now focus on the number density fluctuations and calculate the equal-time density correlator defined as:
\begin{equation}
  \langle |\delta c({\bf k},t)|^2\rangle = \int_{-\infty}^\infty\frac{d\omega}{2\pi}\langle |\delta c({\bf k},\omega)|^2\rangle.
\end{equation} 
We now use (\ref{c-corr-lin}) above and integrate over $\omega$, see Appendix \ref{same-time-corr}. This gives in the linear theory
\begin{equation}
  C_{cc}^0(k)\equiv\langle |\delta c({\bf k},t)|^2\rangle \approx \frac{D_\theta + D_c}{2 \Gamma }\label{c-k-corr}
\end{equation}
in the long wavelength limit, that is independent of $k$. It is useful to compare (\ref{c-k-corr})  with the equilibrium density auto correlation function of a non-critical system, which at temperature $T$ has the form
\begin{equation}
 C_{cc}(k)_{eq}\equiv \langle  |\delta c({\bf k},t)|^2\rangle|_{eq}=\frac{T}{\chi},\label{c-k-equi}
\end{equation}
which is independent of $k$. Here, $\chi$ is the effective equilibrium susceptibility that is finite in the long wavelength limit away from any critical point. Comparing (\ref{c-k-corr}) with (\ref{c-k-equi}), we find
\begin{subequations}
 \begin{align}
  & T_{eff}^c \equiv D_\theta+D_c,\\
 & \chi \equiv 2\Gamma,
 \end{align}
\end{subequations}
for the effective temperature $T_{eff}^c$ and effective susceptibility $\chi$, respectively, in the long wavelength limit.
%
The equal-time real space density auto correlation function
\begin{equation}
 C_{cc}^0(r)\equiv\langle \delta c({\bf x},t)\delta c({\bf x'},t)\rangle
\end{equation}
is the inverse Fourier transform of $C_{cc}^0(k)$. { For the convenience of evaluating this inverse Fourier transform, we now include a hyperdiffusion term $-\zeta\nabla^4 \delta c$ in (\ref{c-full-eq}) to get
\begin{equation}
 \frac{\partial\delta c}{\partial t}=\lambda_0\tilde\Omega_0 \nabla^2\theta + D\nabla^2\delta c -\zeta\nabla^4 \delta c+ \lambda_0 \tilde \Omega_1 {\boldsymbol\nabla}\cdot(\delta c{\boldsymbol\nabla}\theta) + f_c.\label{c-full-eq1}
\end{equation}
}
This gives
\begin{eqnarray}
&&\langle \delta c({\bf x},t)\delta c({\bf x'},t)\rangle\approx \frac{D_c+D_\theta}{2\Gamma}\int_{k=0}^\infty \frac{d^2k}{(2\pi)^2}\frac{\exp(i{\bf k\cdot r})}{1+\zeta^2 k^2}\nonumber \\
&&\approx \frac{D_c+D_\theta}{4\Gamma}\exp(- r/\zeta)\frac{1}{\sqrt r},
\label{den-corr-real}
\end{eqnarray}
for $r\gg \zeta$;
  a convergence factor $1/(1+\zeta^2k^2)$ has been introduced to ensure convergence of the integral, where we have used the asymptotic form of the Bessel function for large $r$.
\, Here, $\zeta$ is a length scale that sets the scale of the short range interaction; for $r\gg \zeta$, $\langle \delta c({\bf x},t)\delta c({\bf x'},t)\rangle$ decreases rapidly. 

Lastly, we consider the cross-correlation function $\langle \delta c({\bf -k},-\omega)\theta({\bf k},\omega)\rangle$. We start from (\ref{cross-corr-lin}) above and use the definition
\begin{equation}
 \langle \delta c({\bf -k},t)\theta({\bf k},t)\rangle = \int_{-\infty}^\infty\frac{d\omega}{2\pi}\langle \delta c({\bf -k},-\omega)\theta({\bf k},\omega)\rangle.
\end{equation}
Staying within the linear theory and
again integrate over $\omega$, we find
\begin{equation}
 C_\times^0(k)\equiv\langle \delta c({\bf -k},t)\theta({\bf k},t)\rangle=\frac{D_c\kappa - D_\theta D}{2\Gamma V_0};
\end{equation}
see Appendix \ref{same-time-corr} for a detailed derivation.
This immediately gives
\begin{equation}
 C_\times^0(r)\equiv\langle \delta c({\bf x},t)\theta({\bf x'},t)\rangle \approx \frac{D_c\kappa - D_\theta D}{4\pi\Gamma V_0}\exp(- r/\zeta)\frac{1}{\sqrt r}
\end{equation}
for $r\gg \zeta$. We define a dimensionless ratio $\varpi$ that will be useful in our subsequent discussions. It is defined in the linear theory as
\begin{equation}
 \varpi = \frac{C_{cc}^0(k)}{C_\times^0 (k)}\label{dimless-rat}
\end{equation}
that is independent of $k$ and is an ${\cal O}(1)$ number when all the model parameters are themselves ${\cal O}(1)$ numbers.

At this stage, it is interesting to note that we can extract the same effective temperature from both $\langle\theta({\bf x},t)^2\rangle$ and $\langle (\delta c({\bf x},t))^2\rangle$: $T_{eff}^\theta=T_{eff}^c\equiv T_{eff}$, apparently implying an effective equilibrium-like behavior for the linearly coupled system of $\theta$ and $\delta c$ at a temperature $T_{eff}$. Nonetheless, a non-zero cross-correlation function $ \langle \delta c({\bf x},t)\theta({\bf x'},t)\rangle$ breaks the condition of thermal equilibrium, or the Fluctuation-Dissipation Theorem (FDT)~\cite{chaikin}; see Appendix~\ref{fdt1}.

Let us further examine the consequence of (\ref{den-corr-real}). Let us consider a finite area $\tilde L\times \tilde L$ and study the density fluctuations in it. If we consider two such small areas of equal size with a linear size $\tilde L$, one centered at ${\bf x}$ and the other at ${\bf x}'$, we can calculate the variance of the number fluctuations in each of the areas and how that variance scale with the area. Now writing ${\bf r}=(x-x',\,y-y')$, we get for the variance 
of the number fluctuations $\delta N$ about the mean $N_0$ in an area $\tilde L\times \tilde L$
\begin{eqnarray}
 &&\langle (\delta N)^2\rangle\nonumber =\int_0^{\tilde L} dx\int_0^{\tilde L}dy \int_0^{\tilde L}dx' \int_0^{\tilde L} dy' \langle \delta c({\bf x},t)\delta c({\bf x'},t)\rangle\nonumber \\
 &\approx& \int_0^{\tilde L} dx\int_0^{\tilde L}dy \int_0^{\tilde L}dx' \int_0^{\tilde L} dy' \frac{D_c+D_\theta}{4\Gamma} \exp(- r/\zeta)\frac{1}{\sqrt r}\,d\phi\nonumber \\
 &\approx& \tilde L^2 \frac{D_c+D_\theta}{4\Gamma},     \label{numb-fluc-lin}
\end{eqnarray}
assuming $\tilde L \gg\zeta$, where $\phi$ is the polar angle.
Thus $\langle (\delta N)^2\rangle$ scales with $\tilde L$ as $\tilde L^2$. Since the mean number $N_0$ in an area $\tilde L\times \tilde L$ scales with $\tilde L^2$, we find that $\langle(\delta N)^2\rangle$ scales as $N_0$, giving the standard deviation of the number fluctuations $\sigma(N_0)\propto \sqrt {N_0}$, corresponding to  {\em normal number fluctuation} (NNF), as expected in an equilibrium system with short range interactions away from any critical points.

On the other hand, for $\lambda_0 \tilde \Omega_0\Omega_1<0$, the dispersion relation by using (\ref{disp1}) becomes
\begin{equation}
 \omega = \pm i V_0 k,
\end{equation}
to the lowest order in $k$. Thus one of the modes grows with a growth rate that scales with $k$
corresponding to {\em linear instability}. { These instabilities are {\em static}.}

{We thus conclude that in the linearized theory, a nearly phase-ordered state, characterized by $\langle \theta({\bf x},t)^2\rangle$ and $\langle (\delta c({\bf x}.t)^2\rangle$, is either statistically indistinguishable from the $2d$ XY model in its QLRO phase and ordinary diffusing particles in a lattice-gas system, respectively, with both being at temperature $T_{eff}$, in so far as the phase fluctuations and the density fluctuations are concerned, or the system is linearly unstable implying absence of any ordering}.  In the next Section, we focus on the linearly stable phase-ordered states and ask whether these are also nonlinearly stable or not. If these phase-ordered states are indeed nonlinearly stable, then we ask: what are their universal scaling properties? We address this issue systematically below.

\section{Nonlinear effects}\label{nonlin}

To know whether nonlinear effects preserve or destabilize the linearly stable uniform states in $2d$, we start from the hydrodynamic equations that contain the lowest order nonlinear terms in fields and gradients:
\begin{subequations}
 \begin{align}
  &\partial_t \theta = \kappa \nabla^2 \theta + \frac{\lambda}{2} ({\boldsymbol \nabla} \theta)^2 + V_0 \delta c+ \Omega_2 (\delta c)^2 + f_\theta,\label{theta}\\
 &\partial_t \delta c = D \nabla^2 \delta C + V_0 \nabla^2 \theta +\lambda_1 {\boldsymbol\nabla}\cdot (\delta c{\boldsymbol\nabla}\theta) + f_c,\label{C}
 \end{align}
\end{subequations}
where, we have retained only the terms lowest order in fields and gradients; $\lambda_1\equiv \lambda_0\tilde\Omega_1$. Noises $f_\theta$ and $f_c$ are still assumed to be zero-mean Gaussian-distributed with variances given by (\ref{noise-theta}) and (\ref{noise-c}); we ignore any cross-correlations between $f_\theta$ and $f_c$ for simplicity. We are interested in the scaling properties in the long wavelength limit, by which we specifically mean wavevector $k$ satisfying  $k\ll \frac{V_0}{\kappa-D}$, such that there are underdamped propagating waves in the wavevector regime of interest.  Identifying ${\boldsymbol\nabla}\theta$ with a ``velocity'' ${\bf v}_s\equiv {\boldsymbol\nabla}\theta$, the resulting equations have structural similarities with the coupled Burgers (BMHD) equations; see Refs.~\cite{ab-jkb-sr,abfrey,abfrey1} and Appendix~\ref{bmhd}.


Equations~(\ref{theta}) and (\ref{C}) are invariant under ${\bf x}\rightarrow -{\bf x},\,\theta({\bf x},t)\rightarrow \theta(-{\bf x},t),\,\delta c({\bf x},t)\rightarrow \delta c(-{\bf x},t)$. Further, 
if $\lambda=\lambda_1$, the model equations (\ref{theta}) and (\ref{C}) admit invariance under a pseudo-Galilean transformation~\cite{fns,stanley,abfrey}:
\begin{equation}
 {\bf x}\rightarrow {\bf x} + {\bf U}_0t,\,{\boldsymbol\nabla}\theta \rightarrow {\boldsymbol\nabla}\theta + {\bf U}_0,\,\frac{\partial}{\partial t} \rightarrow \frac{\partial}{\partial t} + \lambda {\bf U}_0\cdot {\boldsymbol\nabla}, \label{gal-trans}
\end{equation}
where $U_0$ is a constant.
In general however there is no physical requirement of imposing the Galilean invariance on the model equations (\ref{theta}) and (\ref{C}). Hence, in this model there are no restrictions on the magnitudes and signs of $\lambda,\,\lambda_1$; in fact, they can be of the same or opposite signs.  In the limit of immobile {particles}, i.e.,
if we set $V_0=\Omega_2=0$, Eq.~(\ref{theta}) reduces to the well-known Kardar-Parisi-Zhang (KPZ) equation~\cite{kpz,stanley}. It is known that at $2d$ the KPZ equation only has a {\em rough} phase, which for the phase variable $\theta({\bf x},t)$ means destruction of the phase-ordered state. We will now study whether the additional nonlinear terms and the propagating modes can alter this picture,  and make  stable order possible.

\subsection{Renormalization group analysis}\label{rg}

The presence of the nonlinear terms in (\ref{theta}) and (\ref{C}) preclude any exact enumeration of the correlation functions. Instead, perturbative approaches are needed. Na\"ive perturbation theory, however, produces diverging corrections to the model parameters, similar to the $2d$ KPZ equation. These divergences are systematically handled within the dynamic RG framework~\cite{fns,halpin,janssen,stanley}, which is conveniently implemented  by using a path integral description, equivalent to and constructed from Eqs.~(\ref{theta}) and (\ref{C}), in terms of the fields $\theta({\bf x},t)$ and $\delta c({\bf x},t)$, and their dynamic conjugate fields, respectively, $\hat\theta({\bf x},t)$ and $\hat c({\bf x},t)$~\cite{janssen}, after averaging over the noise distributions with variances given by (\ref{noise-theta}) and (\ref{noise-c}).

The momentum shell RG procedure consists of integrating over the short wavelength 
Fourier modes
of $\theta({\bf r},t)$, $\delta c({\bf r},t)$, $\hat \theta({\bf r},t)$ and $\hat c({\bf r},t)$, followed by  rescaling of lengths and times~\cite{fns,halpin,stanley}.
In particular, we  follow the usual approach of initially restricting the wavevectors  
to be within a bounded circular Brillouin zone: $|{\bf k}|<\Lambda$. However, the precise value of the upper cutoff $\Lambda$ has no effect on our final  results. The fields $\theta({\bf r},t)$, $\delta c({\bf r},t)$, $\hat \theta({\bf r},t)$ and $\hat c^({\bf r},t)$
are separated into the high and low wave vector parts
$\theta({\bf r},t)=\theta^<({\bf r},t)+\theta^>({\bf r},t)$, $\delta c({\bf r},t)=\delta c^<({\bf r},t)+\delta c^>({\bf r},t)$, $\hat\theta({\bf r},t)=\hat\theta^<({\bf r},t)+\hat\theta^>({\bf r},t)$, and $\hat c({\bf r},t)=\hat c^<({\bf r},t)+\hat c^>({\bf r},t)$
where $\theta^>({\bf r},t)$, $\delta c^>({\bf r},t)$, $\hat\theta^>({\bf r},t)$ and $\hat c^>({\bf r},t)$ have support in the large wave vector  (short wavelength) 
range $\Lambda
e^{-l}<|{\bf k}|<\Lambda$, while $\theta^<({\bf r},t)$, $\delta c^<({\bf r},t)$, $\hat\theta^<({\bf r},t)$ and $\hat c^<({\bf r},t)$ have support in the small 
wave vector (long wavelength) range $|{\bf k}|<e^{-l}\Lambda$; $b\equiv e^{l}>1$.
We then integrate out $\theta^>({\bf r},t)$, $\delta c^>({\bf r},t)$, $\hat\theta^>({\bf r},t)$ and $\hat c^>({\bf r},t)$ perturbatively 
in 
 the anhamornic coupling $\lambda,\,\Omega_2$ and $\lambda_1$; as usual, this resulting perturbation theory of $\theta^<({\bf r},t)$,  $\delta c^<({\bf r},t)$,  $\hat\theta^<({\bf r},t)$ and $\hat c^<({\bf r},t)$
can be represented by Feynman graphs, with the order of perturbation theory 
reflected by the number of loops in the graphs we consider.  
 { The Feynman vertices representing the anharmonic couplings  are illustrated in 
Fig.~\ref{vertex} in Appendix \ref{action-function}}. After this 
perturbative step, we rescale lengths, 
with ${\bf r}\rightarrow{\bf r }' e^{l}$,  which restores the UV cutoff back to 
$\Lambda$, together with rescaling of time $t\rightarrow t'e^{z l}$  (equivalently in the Fourier space, the momentum and frequency are rescaled as ${\bf k}\rightarrow {\bf k}'/b$ and $\omega \rightarrow \omega'/b^z$, respectively), where $z$ is dynamic exponent. This is then followed
by rescaling the long wave length part of the fields that we define in the Fourier space for calculational convenience. We fix the scale factors by demanding that certain 
harmonic part of the action functional (\ref{action}) in Appendix~\ref{action-function}  {\em do not scale}. 
In $d$-space dimensions, we have
\begin{eqnarray}
 &&\theta({\bf k},\omega)=b^{\chi_\theta}\theta'({\bf k'},\omega'),\, \delta c({\bf k},\omega)=b^{\chi_c}\delta c'({\bf k'},\omega'),\nonumber \\
 &&\hat\theta({\bf k},\omega)=b^{\hat\chi_\theta}\hat\theta'({\bf k'},\omega'),\, \hat c({\bf k},\omega)=b^{\hat\chi_c}\hat c'({\bf k'},\omega').\label{basic-scale}
\end{eqnarray}
This gives the exponents  $\chi_\theta=4,\chi_c=3,\hat\chi_\theta=2,\hat \chi_c=3$ at $2d$; see also Appendix \ref{rescale}.

We restrict ourselves here to a one-loop perturbative RG calculation. See Appendix~\ref{perturbative} for the detailed calculations.

Before we proceed further, let us quickly revisit the dynamic RG calculation for the $2d$ KPZ equation~\cite{natterman,stanley,uwe,kpz}. Our model reduces to the KPZ equation if $\delta c=0=V_0$. One defines an effective dimensionless coupling constant $\tilde g = \frac{\lambda^2 D_\theta}{4\kappa^3}\frac{k_d}{(2\pi)^d}$ in $d$ dimensions; $k_d$ is the area of a unit hypersphere in $d$ dimension. The RG flow equation for the coupling $\tilde g$ at $2d$ (which is the lower critical dimension of the KPZ equation) reads
\begin{equation}
 \frac{d\tilde g}{dl}=\tilde g^2.\label{kpz-flow}
\end{equation}
This can be easily integrated to give a scale-dependent $\tilde g(l)$
\begin{equation}
 \tilde g(l)=\frac{\tilde g(0)}{1-\tilde g_0l},
\end{equation}
where $\tilde g(l=0)$ is a constant of integration, and is the value of $\tilde g$ at the smallest scale or its unrenormalized value;  $l$ is the logarithm of a length-scale, and is the ``RG time''. Clearly at the RG fixed point, $d\tilde g(l)/dl=0$, giving $\tilde g=0$ is the {\em only} fixed point, which is an unstable fixed point. In fact, as $l$ approaches a finite {\em nonuniversal} value determined by the microscopic or unrenormalized parameters, $\tilde g(l)$ diverges at a finite $l$, signaling breakdown of the perturbation theory. In terms of a height field that satisfies the $2d$ KPZ equation, this is interpreted as the existence of a rough phase, inaccessible to perturbative calculations, or for the phase variable $\theta$, the non-existence of any phase ordered state at $2d$ in the thermodynamic limit.    Notice that the flow of the coupling $\tilde g$ to infinity (as a result of $d\tilde g/dl>0$) is crucial to this breakdown of perturbation theory and the consequent lack of any phase-ordered states. Flow equation (\ref{kpz-flow}) shows that at $2d$, coupling $\tilde g$ is marginally relevant (consistent with the lack of a smooth phase in the $2d$ KPZ equation, or the lack of an ordered state at $2d$ here); $\tilde g=0$, which is the only fixed point of (\ref{kpz-flow}), is {\em marginally unstable}.

We calculate the relevant one-loop corrections to all the model parameters of our model equations (\ref{theta}) and (\ref{C}); see Appendix~\ref{one-loop} for details. It turns out that only $\kappa,D,D_\theta,D_c$ have relevant fluctuation-corrections at the one-loop order due to the nonlinear terms. These corrections are $\log$-divergent at $2d$. However, there are {\em no} diverging fluctuation corrections to the wave speed $V_0$ or the coupling constants $\lambda,\,\lambda_1$ and $\Omega_2$ at the one-loop order.

Rescaling space, time and the fields in the standard way as discussed above (see also Appendix~\ref{one-loop}), we get the RG recursion relations for the various model parameters at the one-loop order.
{
We define two dimensionless parameters 
\begin{equation}
\mu_1=\frac{\Omega_2}{\lambda},\;\, \mu_2=\frac{\lambda_1}{\lambda}.\label{mu-defn} 
 \end{equation}
 
  We also define 
 \begin{align}
  \overline{D}=D_\theta+D_c,\, \tilde{D}=D_\theta-D_c.
 \end{align}

We find
\begin{subequations}
 \begin{align}
 &\frac{d\tilde{D}}{dl}=A(\mu_1,\mu_2)g\tilde{D},\label{tilde-D}\\
 &\frac{d\overline{D}}{dl}=G(\mu_1,\mu_2)g\overline{D},\label{overline-D}
 \end{align}
\end{subequations}
}
where 
\begin{subequations}
 \begin{align}
 &A(\mu_1,\mu_2)=\mu_1^2-\mu_2^2/2+\mu_1+1/4,\\ 
 &G(\mu_1,\mu_2)=\mu_1^2+\mu_2^2/2+\mu_1+1/4.
 \end{align}
\end{subequations}

Both $A$ and $G$ are dimensionless ratios; $A$ can be of any sign, but $G$ is non-negative by construction. We further define the dimensionless coupling constant
\begin{equation}
g=\frac{\lambda^2\overline{D}}{8\Gamma^3}\frac{k_d}{(2\pi)^d} \label{dim-coup-here}
\end{equation}
in $d$-dimensions, that is the analogue of the dimensionless coupling constant $\tilde g$ in the KPZ problem. 

By using the rescaling factors defined in (\ref{basic-scale}), we find the scaling of $g$ as 
\begin{equation}
 g'=b^{2-d}g,\label{scale-g}
\end{equation}
showing that at $2d$, $g$ does not scale at all under na\"ive rescaling of space, time and the long wavelength parts of the fields. This means $2d$ is the critical dimension, just as $2d$ is the critical dimension of the KPZ equation. We further define {
\begin{equation}
{\cal N}\equiv \frac{\tilde D}{\overline D}=\frac{D_\theta-D_c}{D_\theta+D_c},\label{N-defn}
\end{equation}
}
which is a dimensionless ratio. We find
\begin{equation}
 \frac{d {\cal N}}{dl}=g\left[ A(\mu_1,\mu_2)- G(\mu_1,\mu_2){\cal N} \right].
\end{equation}
Thus, the flow of $\cal N$ is independent of all other parameters (apart from $\mu_1$ and $\mu_2$).
At the RG fixed point
\begin{equation}
{\cal N}^*=\frac{A(\mu_1,\mu_2)}{G(\mu_1,\mu_2)},\label{fixed-N}
\end{equation}
which  is linearly stable. Here and below a ``superscript $^*$'' represents a RG fixed point value.  The RG recursion equations for $\Gamma,\tilde\Gamma$ read (we set ${\cal N}={\cal N}^*$ in the long RG time limit)

\begin{widetext}
 
\begin{subequations}
 \begin{align}
   &\frac{d\Gamma}{dl}=\Gamma\left[g\left\{B(\mu_1,\mu_2)-\frac{C(\mu_1,\mu_2){\cal N}^*}{4}+\left(\frac{\mu_2^2-1}{8}+\frac{C(\mu_1,\mu_2
  ){\cal N}^*}{4}\right)\frac{\tilde\Gamma}{\Gamma} \right\}
  + \tilde{g_1}\frac{C(\mu_1,\mu_2)}{2}\frac{\tilde\Gamma}{\Gamma} \right],\label{gam1}\\
  &\frac{d\tilde\Gamma}{dl}=\Gamma\left[g\left\{F(\mu_1,\mu_2)-\frac{H(\mu_1,\mu_2){\cal N}^*}{4}+\left(\frac{4\mu_1\mu_2-\mu_2^2-1}{8}+\frac{H(\mu_1,\mu_2){\cal N}^*}{4}\right)\frac{\tilde\Gamma}{\Gamma}
  \right \} + \tilde{g_1}\frac{H(\mu_1,\mu_2)}{2}\frac{\tilde\Gamma}{\Gamma}\right],\label{gam2}
 \end{align}

\end{subequations}

\end{widetext}

where
\begin{eqnarray}
 && \tilde{g_1}=
 \frac{G(\mu_1,\mu_2)-A(\mu_1,\mu_2)}{2G(\mu_1,\mu_2)}g.
\end{eqnarray}

\begin{subequations}
 \begin{align}
  & B(\mu_1,\mu_2)=\frac{7\mu_1\mu_2}{2}+\frac{9\mu_2^2}{8}+\mu_1+\mu_2+\frac{1}{8},\\
 & C(\mu_1,\mu_2)=\mu_1\mu_2-\mu_2^2/2+\mu_2/2-\mu_1,\\
 & H(\mu_1,\mu_2)=\mu_2^2/2-\mu_1\mu_2-\mu_1+\mu_2/2,\\
 & F(\mu_1,\mu_2)=-\mu_1\mu_2-\frac{9\mu_2^2}{8}+\mu_1+\mu_2+\frac{1}{8}.
 \end{align}

\end{subequations}
Parameters $A, G, B, C, H$ and $F$ depend only on $\mu_1(l=0)$ and $\mu_2(l=0)$, and hence are all marginal as well as dimensionless.

Since the nonlinear coupling constants $\lambda,\,\Omega_2$ and $\lambda_1$ do not receive any relevant fluctuation corrections themselves at the one-loop order (see Appendix \ref{one-loop}), the dimensionless parameters $\mu_1$ and $\mu_2$ too do not receive any diverging fluctuation corrections at the one-loop order, i.e., { $\mu_1$ and $\mu_2$ are constants along the RG flow.}  This implies
 \begin{equation}
  \frac{d\mu_1}{dl}=0=\frac{d\mu_2}{dl}.
 \end{equation}
 Thus, if the other renormalized model parameters and the scaling exponents depend upon $\mu_1$ and $\mu_2$, they will actually depend upon $\mu_1(l=0)$ and $\mu_2(l=0)$, i.e., the initial or unrenormalized values of $\mu_1$ and $\mu_2$, which are the tuning parameters in this theory as we see below. 

Furthermore, we obtain

\begin{figure}[htb]
\includegraphics[width=0.8\columnwidth]{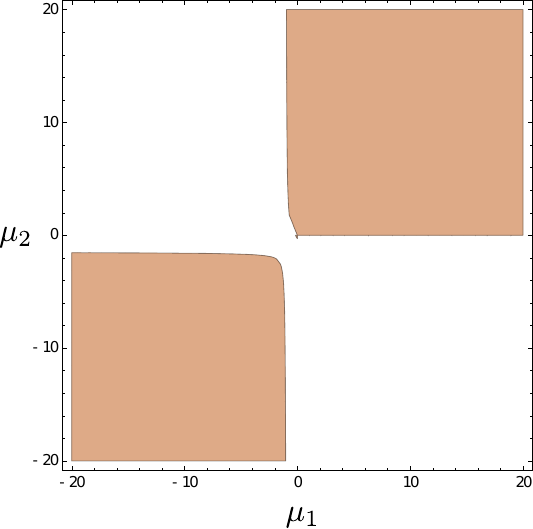}
\caption{Regions with $-1\leq\psi\leq1$ in the $\mu_1-\mu_2$ plane: Light colored regions have bounded values of $\psi$. In the remaining region, $\psi$   has unphysical solutions.} \label{psi-plot}
\end{figure}

\begin{widetext} 
 
\begin{eqnarray}
 \frac{d}{dl}\left(\frac{\tilde \Gamma}{\Gamma}\right) &=& g\left[F-\frac{AH}{4G} +\frac{\tilde\Gamma}{\Gamma}\left\{-B+\frac{AC}{4G}+\frac{H}{4}+\frac{4\mu_1\mu_2-\mu_2^2-1}{8}\right\}-\left(\frac{\tilde\Gamma}{\Gamma}\right)^2\left(\frac{\mu_2^2-1}{8} +\frac{C}{4}\right)\right]\nonumber\\&=& g\left[ a +m\left(\frac{\tilde\Gamma}{\Gamma}\right) + \overline{C}\left(\frac{\tilde\Gamma}{\Gamma}\right)^2 \right].\label{gamma-ratio}
 \end{eqnarray}
 
 \end{widetext}
 
 where 
\begin{subequations}
 \begin{align}
  &a=F-AH/(4G),\\
 &\overline C=-C/4 - (\mu_2^2-1)/8,\\
&m=-B+AC/(4G)+H/4 \nonumber \\&+ (4\mu_1\mu_2 - \mu_2^2 -1)/8
 \end{align}
\end{subequations}
are yet another set of dimensionless constants.
The ratio $\tilde \Gamma/\Gamma$ is yet another dimensionless number. In the spirit of hydrodynamics and scaling, we assume scaling solutions 
\begin{equation}
\Gamma(l)=\Gamma (0) l^{\eta_2},\;\tilde \Gamma(l)=\tilde\Gamma (0)l^{\eta_2},
\end{equation}
 in the long wavelength limit in a putative stable ordered phase, which satisfy the flow equations (\ref{gam1}) and (\ref{gam2}) for some yet unknown scaling exponent $\eta_2$.
Thence,  $\psi\equiv \tilde\Gamma(l)/\Gamma(l)=\tilde\Gamma_0/\Gamma_0$ is a constant at the RG fixed point. By using (\ref{gamma-ratio}), we find that $\psi$ follows the equation
\begin{equation}
 \psi=\frac{1}{2\overline C}\left[-m\pm \sqrt{m^2 -4a\overline C}\right],\label{psieq}
\end{equation}
 For a physically meaningful solution, {$\psi$ must be real and also be in the range between -1 and 1, since $|\tilde \Gamma|\leq \Gamma$ by definition} ~\footnote{ We have assumed the same scaling exponents $\eta_2$ for both $\Gamma(l)$ and $\tilde\Gamma(l)$. If we instead assume different scaling exponents given by $\Gamma(l)=\Gamma_0l^{\eta_a}$ and $\tilde \Gamma(l)=\tilde\Gamma_0 l^{\eta_b}$, then if $\eta_a>\eta_b$, then $\psi=0$, else if $\eta_a<\eta_b$, then $\psi$ diverges, which is unphysical}. {We note that  the condition for $\psi$ to be real  is $m^2 >4a\overline C$, together with a choice for $\psi$ in eq (\ref{psieq}) to have stable values is $\frac{1}{2\overline C}\left[-m- \sqrt{m^2 -4a\overline C}\right]$ }; see Fig.~\ref{psi-plot} for regions in the $\mu_1-\mu_2$ plane where ${\cal I}m\,\psi=0$.

\begin{figure}[htb]
\includegraphics[width=0.8\columnwidth]{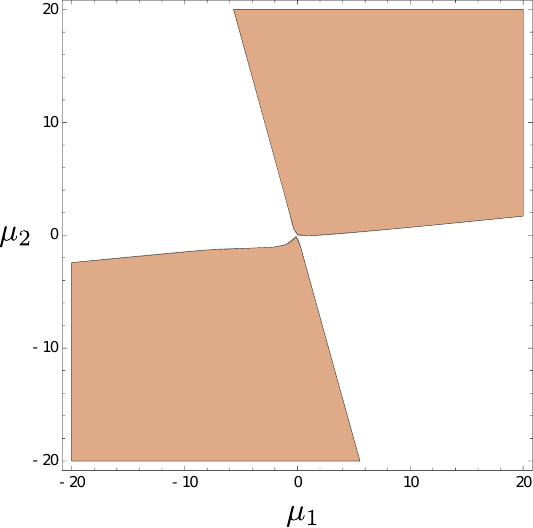}
 \caption{Regions with $\Delta_1>0$ in the $\mu_1-\mu_2$ plane: Light colored regions have $\Delta_1>0$; in the remaining region $\Delta_1<0$.}\label{del-plot}
\end{figure}

This is to be supplemented by the RG flow equation for $g$ itself, which reads:
\begin{equation}
 \frac{dg}{dl}=g^2\left[G-3\left(B-\frac{AC}{4G}\right)+3\psi \overline C\right]=-\Delta_1g^2,\label{gflow}
\end{equation}
where 
\begin{equation}
\Delta_1\equiv -\left[G-3\left(B-\frac{AC}{4G}\right)+3\psi \overline C\right].\label{del1}
\end{equation}
If $\Delta_1>0$, $dg/dl<0$, $g(l)$ decreases as the ``RG time'' $l$ increases, paving the way for possible phase ordered states. Assuming positive $\Delta_1$, solving (\ref{gflow}) we get
\begin{equation}
 g(l)=\frac{g_0}{g_0\Delta_1l +1}\approx \frac{1}{\Delta_1l} \label{g-sol}
\end{equation}
for large RG time $l$. Here, $g_0\equiv g(l=0)$ is the ``initial'' or unrenormalized value of $g$. Here and below, we use a suffix ``0'' to represent quantities evaluated at the ``initial RG time'' ($l=0$), or the corresponding unrenormalized, microscopic values. In contrast, if $\Delta_1<0$, then the solution of (\ref{gflow}) gives
\begin{equation}
 g(l)=\frac{g_0}{-g_0|\Delta_1|l +1}.\label{g-sol-un}
\end{equation}
Evidently, as $l\rightarrow 1/\Delta_1g_0$, a non-universal value controlled by the unrenormalized model parameters, $g(l)\rightarrow \infty$, signaling breakdown of the perturbation theory, which is reminiscent of the flow of $\tilde g(l)$ in the $2d$ KPZ problem. This in turn implies loss of the phase-ordered state. We do not discuss this further here, and below we focus on $\Delta_1>0$. Therefore, with $\Delta_1>0$ in (\ref{gflow}), coupling $g$ is {\em marginally irrelevant}, and the fixed point $g^*=0$ is {\em marginally stable}, whereas for $\Delta_1<0$ coupling $g$ is {\em marginally relevant} and $g^*=0$ is a {\em marginally unstable} fixed point; see Fig.~\ref{del-plot} for regions in the $\mu_1-\mu_2$ plane with $\Delta_1>0$. However, $\Delta_1>0$ is {\em not} the necessary and sufficient condition for the existence of phase-ordered states, as we discuss below. { Notice that in principle, one could define several other dimensionless coupling constants by using $\Omega_2$ and $\lambda_1$ in place of $\lambda$ in the definition of $g$ in (\ref{dim-coup-here}), all of which will be proportional to $g$, the proportionality constants being combinations of $\mu_1,\,\mu_2$. Thus, their RG flow equations can be obtained straightforwardly from the RG flow equation (\ref{gflow}) of $g$, using the fact that $\mu_1$ and $\mu_2$ are {\em marginal}. We however choose to work in terms of $g,\,\mu_1,\,\mu_2$ for reasons of convenience and simplicity of explanations; see also Ref.~\cite{astik-qckpz} for similar technical issues.} 

We thus conclude that
we must have $\Delta_1>0$ together with a real $-1\leq \psi\leq 1$ for the existence of stable uniform phase-ordered steady states, and for which the above scaling solutions hold good~\footnote{ What is the nature of the steady state when $\psi$ is  complex? From (\ref{gam1}) and (\ref{gam2}), both $\Gamma(l)$ and $\tilde \Gamma(l)$ are complex when $\psi$ is complex. This means (\ref{lin-disp}) will now have a higher order in $k$ imaginary term of the form $i\tilde\Gamma^2k^3/V_0$, which triggers linear instability at intermediate wavevectors. Of course, these instabilities would be cutoff by even higher order linear damping terms of equilibrium origin, neglected in the hydrodynamic approach. These can potentially lead to patterned steady states.}. 


Now  substitute (\ref{g-sol}) in (\ref{overline-D}) to get
\begin{equation}
 \frac{d\overline D}{dl}= \frac{G(\mu_1,\mu_2)}{\Delta_1(\mu_1,\mu_2)l}\overline D,
\end{equation}
solving which we find
\begin{equation}
 \overline D(l)=\overline D (0) l^{\eta_1},
\end{equation}
with 
\begin{equation}
\eta_1\equiv \frac{G(\mu_1,\mu_2)}{\Delta_1(\mu_1,\mu_2)}>0,\label{eta1}
\end{equation}
since $G(\mu_1,\mu_2)>0$ by construction and $\Delta_1(\mu_1,\mu_2)>0$ for the stability.
Thus $\overline D(l)/\overline D(0)$ necessarily diverges in the entire stable region of the phase space. Similarly, substituting for ${\cal N}^*$ and $\psi=\tilde \Gamma(l)/\Gamma(l)$ in (\ref{gam1}) we obtain


\begin{eqnarray}
  \frac{d\Gamma}{dl}&=&\Gamma\frac{1}{\Delta_1 l}\left[B - \frac{AC}{4G} -\overline{C}\psi  \right].\label{gam3}
\end{eqnarray}
  Solving Eq.~(\ref{gam3}) we get
  \begin{equation}
   \Gamma(l) = \Gamma(0)l^{\eta_2},
  \end{equation}
with


\begin{eqnarray}
 \eta_2&&\equiv \frac{1}{\Delta_1 }\left[B - \frac{AC}{4G} -\overline{C}\psi  \right] =\frac{1}{3}\left[1+
 \frac{G}{\Delta_1} \right]\nonumber\\
 &&=\frac{1}{3}(1+\eta_1).\label{eta2}
  \end{eqnarray}
  

Equation~(\ref{eta2}) shows that $\eta_2>0$. This implies that $\Gamma(l)\gg \Gamma(0)$ in the renormalized theory in the long wavelength limit. This further means that the dynamics in the long wavelength limit of the renormalized theory is {\em faster} (albeit logarithmically) than that in the linear or noninteracting theory;  see also Eq.~(\ref{time-dyn}) below in this context.  As mentioned above, $ \overline D(0)$ and $\overline \Gamma (0)$ are to be identified with the respective unrenormalized or bare parameters $\overline D$ and $\Gamma$.
 
Clearly, the scaling exponents $\eta_1,\eta_2$ that characterize the stable, ordered phase are parametrized by the real values of ${\cal N}^*$ and $\psi$, together with the constraint $\Delta_1>0$, which in turn are functions of $\mu_1,\,\mu_2$.



As we have argued above, in the spirit of scaling solutions
\begin{subequations}
 \begin{align}
 & \tilde D (l)= \tilde D(0)l^{ \eta_1(\mu_1,\mu_2)},\\
&\tilde\Gamma(l)=\tilde\Gamma (0) l^{\eta_2 (\mu_1,\mu_2)};
 \end{align}
\end{subequations}
here, $\tilde D(0)\equiv (D_\theta(l=0)-D_c(l=0))$. 
The above results for scale-dependent $\overline D(l)$ and $\Gamma(l)$ imply wavevector-dependent renormalized parameters
\begin{subequations}
 \begin{align}
   &\overline D(k)\equiv D_\theta (k) + D_c(k)=\overline D (0) \left[\ln \left(\frac{\Lambda}{k}\right)\right]^{\eta_1},\\
 &\Gamma(k)\equiv \left[\kappa(k)+D(k)\right]/2=\Gamma (0)\left[\ln \left(\frac{\Lambda}{k}\right)\right]^{\eta_2}.\label{renorm-scal1}
 \end{align}

\end{subequations}

Assuming finite ${\cal N}^*$ and $\psi$, we further find that
\begin{equation}
 \tilde D(k)\approx\tilde D_0 \left[\ln \left(\frac{\Lambda}{k}\right)\right]^{\eta_1},\,\tilde\Gamma(k)\approx\tilde \Gamma_0\left[\ln \left(\frac{\Lambda}{k}\right)\right]^{\eta_2},
\end{equation}
in the limit of small wavevector.

As an explicit example, let us work out the case with ${\cal N}^*=0$ and $\psi =0$ (these conditions imply $D_\theta^*=D_c^*$ and $\kappa^*=D^*$ at the RG fixed point). These conditions are satisfied at isolated points in the $\mu_1-\mu_2$ plane. These are calculated from the simultaneous solutions of ${\cal N}^*=0$ and $\psi=0$. { The condition ${\cal N}^*=0$ is satisfied by setting $A(\mu_1,\mu_2)=0$. On the other hand in eq. (\ref{gamma-ratio}), if 
\begin{equation}
F(\mu_1,\mu_2)=0,\; {\rm and}\; -B+H/4+(4\mu_1\mu_2 - \mu_2^2 -1)/8=m<0,
\end{equation}
then the solution $\psi=0$ gives a stable RG fixed point.
These conditions are used to find
\begin{equation}
\mu_1=\frac{\sqrt{2}-1}{2},\mu_2=1.
\end{equation}
We now use the RG flow equations (\ref{overline-D}) and (\ref{gam1}) to obtain
\begin{subequations}
 \begin{align}
 & \frac{d\overline D}{dl}= \frac{2\sqrt{2}}{27-2\sqrt{2}}\frac{\overline D}{l},\\
 & \frac{d\Gamma}{dl}= \frac{9}{27-2\sqrt{2}} \frac{\Gamma}{l}.
 \end{align}
\end{subequations}

Solving, we obtain 
\begin{equation}
\eta_1\approx 0.117,\eta_2 \approx 0.372
\end{equation}
in this special case.} We next find the scaling exponents given any general ${\cal N}^*\neq 0$ and  any real $|\psi|\leq 1$, maintaining $\Delta_1>0$. We present the results in a tabular form; see Table~\ref{tab1}.

{

{ The existence of continuously varying scaling exponents found above depends crucially on the nonrenormalization of 
the coupling constants, and hence $\mu_1,\,\mu_2$ at the one-loop order. In the absence of any symmetries, there is no guarantee that $\mu_1,\mu_2$ should remain unrenormalized at higher loop orders. We now argue that these actually do not matter. This is so because they will involve higher powers of $g$, and will, hence, vanish like $1/l$  to a power greater than 1. Therefore, their integrals over $l$ from zero to infinity will be finite, so they will not change the anomalous behavior of  $D$ and $\kappa$. Similarly, they cannot make any divergent contribution to  $\mu_1$ and $\mu_2$.
Therefore, our results, even though they were only derived to one loop order, are, in fact, asymptotically exact. This automatically implies that the continuous variation of the scaling exponents, making them nonuniversal, is also asymptotically exact~\cite{john-pvt}.}

Wavevector-dependent $\overline D(k)$ and $\Gamma(k)$ may be used to calculate the renormalized correlation functions in the stable regimes of the theory, which we discuss in the next Section~\ref{scal}. 

\subsection{Scaling of the correlation functions and order}\label{scal}

Since in the renormalized theory, coupling constant $g(k)\rightarrow 0$ as $k\rightarrow 0$, the renormalized theory is effectively linear, albeit with renormalized model parameter. This consideration gives the renormalized correlation functions straight forwardly.

 We start by constructing the renormalized correlation functions in the Fourier space as a function of $\bf k$ and $\omega$. We get
\begin{align}
 &\langle |\theta({\bf k},\omega)|^2 \rangle_R = \frac{2D_\theta(k)[\omega^2+D(k)^2k^4]+2V_0^2D_c(k)k^2}{(V_0^2k^2-\omega^2)^2+4\omega^2k^4\Gamma(k)^2},\label{theta-corr-R}\\
 &\langle |\delta c({\bf k},\omega)|^2 \rangle_R = \frac{2D_\theta(k) V_0^2k^4+2D_c(k)k^2[w^2+\kappa(k)^2k^4]}{(V_0^2k^2-\omega^2)^2+4\omega^2k^4\Gamma(k)^2},\label{c-corr-R}\\
 &\langle \theta(-{\bf k},-\omega) \delta c({\bf k},\omega) \rangle_R  \equiv\langle \delta c(-{\bf k},-\omega) \theta({\bf k},\omega) \rangle_R^*= \nonumber\\
 &\frac{2V_0k^2\left[-D_\theta(k)(-i\omega+D(k)k^2)+D_c(k)(i\omega+\kappa(k) k^2)\right]}{(V_0^2k^2-\omega^2)^2+4\omega^2k^4\Gamma(k)^2};\label{cross-corr-R}
\end{align}
 ``$R$'' refers to the quantity being evaluated in the renormalized theory.
It is straightforward to see that the renormalized correlation functions in the Fourier space qualitatively retain their corresponding forms in the linear theory. All of them in the renormalized theory in the long wavelength limit have two well-separated peaks which get narrower relative to their distances in the linear theory: we find  the {\em renormalized Quality factor} $\varphi_R=[V_0/\Gamma(k) k]$ still diverges for $k\rightarrow 0$, albeit {\em logarithmically slower} for $\eta_2>0$, which can also be seen from the ratio $\varphi_R/\varphi$, that {\em vanishes} logarithmically as $k\rightarrow 0$:
\begin{equation}
 \frac{\varphi_R}{\varphi}= \left[\ln\left(\frac{\Lambda}{k}\right)\right]^{-\eta_2}.
\end{equation}
The corresponding time-dependent renormalized correlation functions may be calculated by inverse Fourier transform.

From the renormalized, scale-dependent $\Gamma(k)$, we may extract a time-scale $\tau(k)$ of relaxation in the long wavelength limit:
\begin{equation}
\tau(k)^{-1}\equiv 2\Gamma(k)k^2=2\Gamma(0) \left[\ln\left(\frac{\Lambda}{k}\right)\right]^{\eta_2}k^2.\label{time-scale}
\end{equation}
This implies breakdown of conventional dynamic scaling and instead suggests an ``extended scaling relation'' between time $t$ and length-scales $r$:
\begin{equation}
 t\sim r^2 \left[\ln \left(r/a_0\right)\right]^{-\eta_2}. \label{time-dyn}
\end{equation}
Therefore, the dynamic scaling relation $t\sim r^2$ in the linear theory is now modulated by a logarithmic correction, that {\em speeds up} (since $\eta_2>0$) relaxation of the fluctuations, as already mentioned above.


\subsubsection{Phase correlation function}

The above results on the renormalized model parameters are used to obtain the following result for the phase fluctuations in the renormalized theory. We obtain for the renormalized equal-time phase correlator $C^R_{\theta\theta}(k)$ in the Fourier space 
\begin{equation}
 C^R_{\theta\theta}(k)\equiv\langle |\theta({\bf k})|^2\rangle_R = \frac{\overline D(0)}{4\pi\Gamma(0)\,k^2\left[\ln \left(\frac{\Lambda}{k}\right)\right]^\eta }\label{theta-k-ren-corr}
\end{equation}
where 
\begin{eqnarray}
&&\eta\equiv \eta_2-\eta_1=\frac{1}{3}\left(1-\frac{2G}{\Delta_1}\right)\label{eta};\\
&& \eta_1 = \frac{1-3\eta}{2},\,\eta_2=\frac{1-\eta}{2}.
\end{eqnarray} 
 Since $\Delta_1>0$ for stability, and $G$ is non-negative by definition, we must have $\eta<1/3$. 
We note that, if $\eta_1=\eta_2$, i.e., $\eta=0$ when $\Delta_1=2G$, $\langle |\theta({\bf k})|^2\rangle_R$ scales in the same way as its equilibrium analog, whereas if $\eta>0$, i.e., $\eta_1<\eta_2$ with $\Delta_1>2G$, renormalized correlator
\begin{equation}
C^R_{\theta\theta}(k)\ll C_{\theta\theta}(k)_{eq},
\end{equation}
its equilibrium analog and if $\eta<0$, i.e., $\eta_1>\eta_2$ with $\Delta_1<2G$, 
\begin{equation}
C^R_{\theta\theta}(k)\gg C_{\theta\theta}(k)_{eq}
\end{equation}
for small enough $k$, i.e., in the long wavelength limit $k\rightarrow 0$. In general,  $\eta(\mu_1,\mu_2)$ gives the scaling of $C^R_{\theta\theta}(k)$ as a function of $\mu_1,\mu_2$ {, subject to the condition $\Delta_1(\mu_1,\mu_2)>0$ and $\psi(\mu_1,\mu_2)$ is real with value between -1 and +1}. Clearly, $\eta(\mu_1,\mu_2)$ varies as $\mu_1$ and $\mu_2$ change.

Having worked out the correlation functions in the Fourier space, we now calculate the variances and correlations of the fluctuations in the real space.

The renormalized variance $\Delta^R_\theta$ of the phase fluctuations is given by
\begin{eqnarray}
\Delta^R_\theta&\equiv& \langle \theta^2({\bf x},t)\rangle_R\nonumber\\
&=& \int \frac{d^2k}{(2\pi)^2} \frac{\overline D(k)} {2\Gamma(k)}\frac{1}{k^2}\approx \frac{\overline D(0)}{4\pi\Gamma(0)}(\ln L/a_0)^{1-\eta}.
\label{DeltaR}
\end{eqnarray}
in the renormalized theory.

We find $\eta$ by varying $\mu_1,\,\mu_2$ in the stable region of the phase diagram, defined by $\Delta_1>0$ and $-1\leq \psi \leq 1$. We in fact find two distinct scaling behavior, depending upon whether $\eta>0$, or $\eta<0$. In the renormalized theory, in the phase space region with $\eta>0$,  $\langle \theta ({\bf x},t)^2\rangle_R$ grows with $L$ {\em logarithmically slower} than QLRO:
\begin{eqnarray}
 \Delta_\theta^R&\equiv& \langle \theta ({\bf x},t)^2\rangle_R 
 \nonumber \\
 &\ll& \langle \theta ({\bf x},t)^2\rangle_{eq}=\frac{T}{2\pi\kappa_0}\ln \left(\frac{L}{a_0}\right)
\end{eqnarray}
for large $L$ (i.e., system size). This represents
a type of order {\em logarithmically stronger} than the usual QLRO - hence we name it {\em stronger than QLRO} or SQLRO.

In contrast, for $\eta<0$,  $\langle \theta ({\bf x},t)^2\rangle$ grows with $L$ {\em logarithmically faster} than QLRO:
\begin{equation}
 \Delta_\theta^R \gg \langle \theta ({\bf x},t)^2\rangle_{eq}
\end{equation}
for large $L$. This clearly 
represents an order {\em weaker} than the usual QLRO - hence we name it {\em logarithmically weaker than QLRO} or WQLRO.

 In the surface version of the model, the above results for $\Delta_\theta^R$ implies surfaces that are logarithmically smoother or logarithmically rougher than the EW surface, respectively, for $\eta>0$ and $\eta<0$.

We further calculate the Debye-Waller factor $W_R$ in the renormalized theory, giving 
\begin{equation}
W_R=\langle \theta^2({\bf x},t)\rangle_R/2 =\frac{\overline D(0)}{8\pi\Gamma(0)}[\ln (L/a_0)]^{(1-\eta)}.
\end{equation}
This gives for the order parameter
\begin{eqnarray}
 &&\langle\cos\theta({\bf x},t)\rangle =\exp[-W_R]\nonumber \\&\approx& \exp\left[-\frac{\overline{D}(0)}{8\pi\Gamma(0)}[\ln (L/a_0)]^{(1-\eta)}\right].
\end{eqnarray}
Thus, the order parameter decays with $L$, but remains finite for finite $L$; it vanishes only as $L\rightarrow\infty$. With $\eta<0$ this decay is {\em faster} for WQLRO and with $\eta>0$ it is {\em slower} for SQLRO.

In order to calculate the equal-time phase correlation function, we first calculate $C_{\theta\theta}^R(r)$ as defined above [see Eq.~(\ref{theta-corr-intro1})]. We get 
\begin{eqnarray}
 C^R_{\theta\theta}(r)&=&\frac{\overline{D}(0)}{\Gamma(0)}\int \frac{d^2k}{(2\pi)^2} \frac{1-\exp[i{\bf k}\cdot {\bf r}]}{k^2\left[\ln \left(\frac{\Lambda}{k}\right)\right]^{\eta}}\nonumber \\&\approx& 
 \frac{\overline{D}(0)}{2\pi\Gamma(0)}[\ln (r/a_0)]^{1-\eta}\label{theta-corr-ren1}
 \end{eqnarray}
 in the limit of large $r$; see Appendix~\ref{real-corr}. This in turn gives for the renormalized equal-time spin correlation function in the real space
\begin{align}
 & C_{ZZ}^R(r)\equiv \langle \cos [\theta({\bf x},t)-\theta({\bf x'},t)]\rangle_R \nonumber \\
 & \approx \exp\left[-\frac{\overline{D}(0)}{2\pi\Gamma(0)}[\ln (r/a_0)]^{1-\eta}\right]\equiv (r/a_0)^{-\tilde\gamma(r)}
 \end{align}
  for large $r$ that decays as the separation $r$ approaches infinity.
 Here, 
 \begin{equation}
  \tilde\gamma(r)\equiv \frac{\overline D(0)}{4\pi\Gamma(0)}[\ln(r/a_0)]^{-\eta}.\label{gamma-tilde}
 \end{equation}
  This decay is {\em faster} for $\eta<0$ corresponding to WQLRO, or {\em slower} for $\eta>0$ corresponding to SQLRO, than the usual QLRO for the $2d$ XY model, which corresponds to the intermediate case of $\eta=0$. 
  For $\eta=0$, $\tilde\gamma(r)=\tilde\gamma_0$, and $C_{ZZ}^R(r)$ takes a form identical with that for $2d$ equilibrium XY model in its QLRO phase. { Furthermore, $\tilde\gamma(r)$ clearly varies continuously with $\mu_1,\,\mu_2$, forming a rare example of continuously varying, parameter-dependent scaling in a theory with {\em relevant} interactions. In contrast, the phase fluctuations in the $2d$ XY model at equilibrium in its QLRO phase, which also shows parameter-dependent scaling, is described by a Gaussian, and hence, a noninteracting theory. See Ref.~\cite{chate11} for an active matter example of quasi-long-range polar order with
continuously-varying scaling exponents. }  
  
   Once again, in the context of the surface version of the model, Eq.~(\ref{theta-corr-ren1}) means a surface with fluctuations having correlations that decays {\em slower} and {\em faster} than that for an EW surface, respectively, for $\eta>0$ and $\eta<0$.

  As we shall see below, SQLRO and WQLRO for the phase fluctuations are intimately connected to the nature of the number density fluctuations: In this model, SQLRO is associated with miniscule number fluctuations (also known as hyperuniformity), whereas WQLRO is associated with giant number fluctuations. We derive these results next.
 
 \subsubsection{Density correlation function}
 
 Next, we calculate the renormalized equal-time density correlation function $C_{c c}^R( k)\equiv \langle |\delta c({\bf k},t)|^2\rangle_R$. We find
\begin{equation}
 C_{cc}^R(k)= \frac{\overline D(k)}{2\Gamma(k)}=\frac{\overline D(0)}{2\Gamma(0)} \left[\ln \left(\frac{\Lambda}{k}\right)\right]^{-\eta}.\label{c-renorm-corr}
\end{equation}
Clearly, if $\eta=0$, $C_{c c}^R(k)$ has no $k$-dependence in the renormalized theory, similar to the corresponding linear theory results, resembling {\em normal number fluctuations} (NNF) observed in an off-critical equilibrium gas with short-ranged interactions. On the other hand, if $\eta>0$, 
\begin{equation}
 C_{cc}^R(k)\ll C_{cc}(k)_{eq}
\end{equation}
in the renormalized theory in the limit $k\rightarrow 0$, indicating significant suppression of the density fluctuations {\em vis-a-vis} the equilibrium results in the long wavelength limit. This is an example of {\em miniscule number fluctations (MNF)} or {\em hyperuniformity}, originally proposed in Ref.~\cite{hyper}. In contrast, for $\eta<0$, 
\begin{equation}
 C_{cc}^R(k)\gg C_{cc}(k)_{eq}
\end{equation}
 in the limit $k\rightarrow 0$, implying divergent density fluctuations relative to its equilibrium analog in the long wavelength limit. This is the analog of {\em giant number fluctuations} (GNF) in polar-ordered active matter systems~\cite{2d-polar,sriram-RMP}. 
 
 More insight may be obtained from the equal-time correlation function in the real space. This is given by the inverse Fourier transform of (\ref{c-renorm-corr})
\begin{eqnarray}
 C_{c c}^R(r)&=& \frac{\overline D(0)}{2\Gamma(0)}\int \frac{d^2k}{(2\pi)^2} \frac{\exp[i{\bf k}\cdot {\bf r}]}{\left[\ln \left(\frac{\Lambda}{k}\right)\right]^{\eta}},
  \end{eqnarray}
  where the upper limit of $k$ is  understood to be restricted to a finite value. 
  
  {
  
  We now calculate the form of $C_{cc}^R(r)$ in the asymptotic limit of large $r$. 
 Clearly, if $\eta>0$, $C_{cc}^R(r)$ is significantly reduced vis-a-vis for $\eta<0$ in the limit of large $r$. In the former case, fluctuations are suppressed and in the latter case they are enhanced in the limit of a large $r$. 
  

To proceed further we now make the change of variables ${\bf k}={\bf Q}/r$. We get

\begin{equation}
 \langle \delta c(r) \delta c(0)\rangle  =  \frac{\overline D(0)}{4\pi\Gamma(0)}\frac{1}{r^2} \int d^2Q \frac{\exp (i {\bf Q}\cdot \hat {\bf r})}{[\ln(\Lambda r)-\ln Q]^\eta}.
 \label{int2a}
\end{equation}
Because of the $\exp (i {\bf Q}\cdot\hat {\bf r})$-term in the above integral, it  is dominated by $Q\sim {\cal O}(1)$. In that range, for large $r$ (specifically, $r\gg\Lambda^{-1}$), $\ln(Q)\ll \ln(\Lambda r)$. Hence we can expand:
\begin{equation}
 \frac{1}{[\ln(\Lambda r)-\ln Q]^\eta}\approx \ln(\Lambda r)]^{-\eta}+\eta[\ln(\Lambda r)]^{-1-\eta} \ln Q.
\end{equation}
 
Using this in the integral in (\ref{int2a}) gives
\begin{eqnarray}
 &&\int d^2Q \exp (i {\bf Q}\cdot \hat {\bf r})(1/[\ln(\Lambda r)-\ln Q]^\eta)  \nonumber \\&\approx& \frac{1}{[\ln(\Lambda r)]^{\eta}} \left(\int d^2Q \exp [i {\bf Q}\cdot \hat {\bf r}]\right)   \nonumber\\ &+&\frac{\eta}{[\ln(\Lambda r)]^{1+\eta}} \left(\int d^2Q \exp (i {\bf Q}\cdot \hat{\bf r}) \ln Q \right).
\end{eqnarray}

{ Now the first integral vanishes for large $r$, since $\int d^2Q \exp (i {\bf Q} \cdot \hat {\bf r})$ is a short-ranged function of $r$, which rapidly vanishes for large $r$ [in fact, this contribution gives the equilibrium result (\ref{den-corr-real}) above, which vanishes for large $r$]. 
The second integral, $\int d^2Q \exp (i {\bf Q}\cdot \hat {\bf r}) \ln Q$, is however a non-zero constant of ${\cal O}(1)$} (it is in fact the Fourier transform of a $2d$ Coulomb potential). Thus, we ultimately get for the renormalized number density real space correlation function $C_{cc}^R(r)$:
\begin{equation}
C_{cc}^R(r)\approx \frac{-\overline D(0)}{4\pi\Gamma(0)}\frac{1}{r^2} \eta[\ln(\Lambda r)]^{-1-\eta}.\label{ren-c-corr}
\end{equation}
The alert reader will identify (\ref{ren-c-corr}) with what we have mentioned in Eq.~(\ref{c-corr-intro}) in the beginning of this paper. 

At this point it is worth noting that for large $r\Lambda \gg 1$, the sign of $C_{cc}^R(r)$ is determined by and opposite to the sign of $\eta$: It is positive for negative $\eta$, and negative for positive $\eta$. In both cases, $C_{cc}^R(r)$ shows a power law decay in $r$ as $1/r^2$, modulated by an $\eta$-dependent logarithmic factor.  Furthermore, $C_{cc}^R(r)|_{\eta >0} \ll C_{cc}^R(r)|_{\eta <0}$ for large $r$, meaning in that limit fluctuations are strongly suppressed with positive $\eta$, relative to negative $\eta$. Since $\eta >(<)0$ corresponds to MNF (GNF), unsurprisingly  $C_{cc}^R(r)$ with MNF is vanishingly small relative to its value with GNF in the limit of large $r$. 

Clearly, $C_{cc}^R(r)$, as given in (\ref{ren-c-corr}) vanishes when $\eta=0$. This is not surprising, since (\ref{ren-c-corr}) holds for $r\Lambda\gg 1$, whereas, for $\eta=0$, $C_{cc}^R$ is identical to its linear theory result which is a short range function that indeed vanishes as $r$ exceeds a finite convergence length $\xi$; see Eq.~(\ref{den-corr-real}) above.



  We now calculate the variance $\langle (\delta N)^2\rangle$ of the number fluctuations in an open finite square area of size $\tilde L\times \tilde L$:
  \begin{equation}
   \langle (\delta N)^2\rangle=\int_0^{\tilde L} d^2 r\int_0^{\tilde L} d^2 r'C_{cc}(|{\bf r-r'}|).
  \end{equation}
  With $\delta N\equiv N- \langle N\rangle$, we find that
    \begin{equation}
 \langle(\delta N)^2\rangle=  \int d^2 r d^2 r' \frac{-\eta \overline D(0)}{4\pi\Gamma (0)|{\bf r-\bf r'|}^2}([\ln(\Lambda |{\bf r}-{\bf r'}|)]^{-1-\eta}.
    \end{equation}
Changing variables of integration from $\bf r'$ to ${\bf R}\equiv {\bf r'-r}$, we get
\begin{equation}
 \langle (\delta N)^2\rangle= \int d^2 r \int d^2 R \frac{-\eta\overline D(0)}{4\pi\Gamma(0) R^2}[\ln(\Lambda |R|)]^{-1-\eta}.\label{del-N}
 \end{equation}
 The integral over $R$ is actually an integral over an oddly shaped region: whatever the shape of the counting area is, it is not centered at $R=0$, but at $R=r$, since the $r'$ integral is centered around 0. Nonetheless, because, as we shall see, everything goes logarithmically, this integral is going to be quite insensitive to the precise limits, and so can be approximated, for most $r$, by an integral over a circle of radius $L$; that is, 
\begin{equation}
 \int  \frac{d^2 R}{|R|^2 [\ln(\Lambda |R|)]^{1+\eta}}\approx 2 \pi \int_{ 1/\Lambda}^{\tilde L}   \frac{R dR}{R^2[\ln(\Lambda R)]^{1+\eta}}.
\end{equation}
 

This integral is elementary, and gives
\begin{equation}
 \int d^2 R \frac{1}{|R|^2}[\ln(\Lambda |R|)]^{-1-\eta}\approx \frac{-2 \pi}{\eta} [\ln(\Lambda \tilde L)]^{-\eta}.
\end{equation}

Using this in (\ref{del-N}), we get
\begin{equation}
 \langle (\delta N)^2\rangle\propto \int d^2 r [\ln(\Lambda \tilde L)]^{-\eta} \propto \tilde L^2 [\ln (\Lambda \tilde L)]^{-\eta}.
 \end{equation}

Now using $N_0\equiv\langle N\rangle =C_0 \tilde L^2$, we get
\begin{equation}
\sigma(N_0)\equiv \sqrt{\langle (\delta N)^2\rangle} \propto \sqrt{N_0} [\ln N_0]^{-\eta/2}.
\label{numb-fluc}
\end{equation}
 Equation~(\ref{numb-fluc}) shows the standard deviation of the number fluctuations in an open area $\tilde L\times \tilde L$. It shows that
\begin{equation}
  \sigma (N_0)|_{\eta >0} \ll \sigma (N_0)|_{\eta =0} \ll \sigma (N_0)|_{\eta <0}. 
 \label{numb-fluc1}
\end{equation}
Thus, if $\eta>0$ (i.e., SQLRO), $\sigma(N_0)$ is much smaller (i.e., MNF) than its equilibrium counterpart ($\eta=0$, or NNF) that scales as $\sqrt{N_0}$, whereas for $\eta<0$ (i.e., WQLRO), it is much bigger (i.e., GNF), for large enough  $N_0$.
 }

Equations~(\ref{theta-k-ren-corr}) and (\ref{c-renorm-corr}) imply that the two different scaling behaviors of SQLRO and WQLRO, associated respectively, with MNF and GNF, delineated respectively by $\eta>0$ and $\eta<0$, can also be interpreted  in terms of ``scale-dependent'' renormalized reduced temperature ${\cal T}_\theta^R$, in analogy with  the reduced temperature ${\cal T}_\theta$ defined by (\ref{red-temp}).  Analogously, from (\ref{theta-k-ren-corr}), we can extract an {\em effective} or {\em renormalized} reduced temperature 
\begin{equation}
{\cal T}_{\theta}^R(k)\equiv {\cal T_{\theta}}\left[\ln \left(\frac{\Lambda}{k}\right)\right]^{-\eta}. \label{red-tempR}
\end{equation}
Thus, if $\eta>0$, ${\cal T}^R_{\theta}(k)$ decreases as $k\rightarrow 0$, whereas if $\eta<0$, ${\cal T}^R_\theta(k)$ increases. In other words, if $\eta >(<) 0$, the system effectively gets cooled (heated) as $k\rightarrow 0$.

\subsubsection{Cross-correlation function}

Lastly, we calculate the renormalized equal-time cross-correlation function $C_\times^R (k)\equiv \langle \delta c  ({\bf k},t)\theta(-{\bf k},t)\rangle_R$,  which, in contrast to its linear theory expression, now picks up a weak $k$-dependence in the renormalized theory. We find
\begin{equation}
 C_\times^R (k)\approx\frac{D_c(k)\kappa(k)-D_\theta(k)D(k)}{2\Gamma(k)V_0}.
\end{equation}
Thus, $C_\times^R (k)$ no longer depends purely on $\overline D(k)$ and $\Gamma(k)$, but depends rather on all of $D_\theta(k),D_c(k),\kappa(k)$ and $D(k)$ separately. Using the scaling above we find
\begin{equation}
 C_\times^R (k)\approx\frac{D_c(0)\kappa(0)-D_\theta(0) D(0)}{2\Gamma(0) V_0}\left[\ln \left(\frac{\Lambda}{k}\right)\right]^{\eta_1}\label{cross-ren}
 \end{equation}
 in the hydrodynamic limit. Interestingly, thus, renormalized $C_\times^R(k)$ carries only the scale-dependence of the renormalized noise strengths, but is independent of the scale-dependence of the renormalized damping coefficients. Since $\eta_1>0$, in the long wavelength limit, $C_\times^R (k)\gg C_\times(k)$, the equal-time cross-correlation function in the linear theory. In the special case where ${\cal N}^*=0$ and $\psi=0$, we have $C_\times^R(k)=0$. We further recall the definition of the dimensionless number $\varpi$ and consider its renormalized analogue $\varpi_R$
\begin{equation}
 \varpi_R = \frac{C_{c c }^R(k)}{C_{\times }^R (k)}=\left[\ln \Lambda/k\right]^{-\eta_2}\times {\cal O}(1),
\end{equation}
using $\eta=\eta_2-\eta_1$. Since $\eta_2>0$ necessarily, 
 $\varpi_R$ vanishes in the long wavelength limit $k\rightarrow 0$. If we take $\varpi^{-1}$ as a measure of the extent of the breakdown of FDT, such that $\varpi^{-1}=0$, e.g., $C_\times^R(k)=0$ implies restoration of FDT in the renormalized theory, we note that in general the degree of violation of FDT is much greater in the renormalized theory, since $\varpi_R\ll \varpi$ vanishes, albeit logarithmically, in the long wavelength limit.

 Unsurprisingly, (\ref{cross-ren}) implies a renormalized cross-correlation function $C_\times^R(r)$ in the real space for large $r,\,r\Lambda \gg 1$ is
\begin{align}
 C^R_\times (r) &\approx  \frac{D_c(0)\kappa(0)-D_\theta(0) D(0)}{4\pi\Gamma(0) V_0}\nonumber \\ &\times\frac{\eta_1}{r^2} [\ln(\Lambda r)]^{-1+\eta_1}.\label{cross-corr-r}
\end{align}
  Thus, $C^R_\times (r)$ has the same functional form as the equal-time density correlation function $C_{cc}^R(r)$ for $r\Lambda \gg 1$.

   Equations~(\ref{time-dyn}), (\ref{theta-k-ren-corr}), (\ref{DeltaR}), (\ref{theta-corr-ren1})-(\ref{c-renorm-corr}), (\ref{ren-c-corr}), (\ref{numb-fluc}) and (\ref{cross-corr-r}) collectively delineate the new universality class in which the stable ordered phases of this model belong. Furthermore, this new universality class is parametrized by $\mu_1,\,\mu_2$ that can vary continuously, leading to {\em continuously varying scaling exponents} in the ordered states.


\subsubsection{Rotation frequency}

We now define rotation frequency $\omega_0 \equiv \langle \frac{\partial\theta}{\partial t}\rangle$ and { calculate it in the renormalized theory} for (\ref{theta}) and (\ref{C}). Unsurprisingly, $\omega_0=0$ in the linear theory as can be easily seen from (\ref{lin1}). However, $\omega_0\neq 0$ in general in the nonlinear theory. We get 
\begin{equation}\label{ang_freq}
 \omega_0=\frac{1}{2}\langle ({\boldsymbol\nabla}\theta)^2\rangle +\mu_1 \langle (\delta c)^2\rangle, 
\end{equation}
where, we have set $\lambda=1$ by rescaling the unit of time. Using our results obtained above, we find
\begin{equation}\label{ang_freq_2}
 \omega_0= \left(\frac{1}{2} +\mu_1\right) \frac{\overline D(0)}{4\pi\Gamma (0)}\int dk\frac{k}{\left[\ln\left( \frac{\Lambda}{k}\right)\right]^\eta}.
\end{equation}
Thus, $\omega_0$ is finite and nonuniversal even in the renormalized theory (due to its explicit dependence on $\overline D(0)$ and $\Gamma(0)$), and has a magnitude which clearly depends upon the sign of $\eta$, also on $\mu_1$ explicitly.  It should be noted that $\omega_0$ is {\em not} the rotation frequency with respect to the lab frame; rather it is the {\em nonlinear correction} to the rotation frequency $\tilde\omega\equiv \Omega (c_0)$ in the uniform state.
Thus, this nonlinearity-controlled correction to the rotation frequency can speed up or slow down the { XY spins}, and is tunable by the active effects, a feature measurable in relevant experiments, and can be used to infer information about the coupling constants. This is reminiscent of bacterial dynamics and synchronization in response to changing environments; see, e.g., Refs.~\cite{quorum1,quorum2}. In the agent-based model, calculation of the { rotation frequency} also leads to a periodic global rotation of the spins in the ordered phase (See Fig.~\ref{fig:rotation} in Appendix~\ref{noise-effects-active-rot} for details).


Let us now calculate the scaling of the renormalized correlation functions  for ${\cal N}^*=0$ and $\psi=0$ as explicit examples. Using the corresponding values of $\eta_1$ and $\eta_2$ as obtained above, we get 
\begin{eqnarray}
&&\eta_1=0.117,\,\eta_2=0.372,\,\,\eta=0.255,\\
 C_{\theta\theta}^R(k)&=&\frac{\overline D_0}{2\Gamma(0)\,k^2\left[\ln \left(\frac{\Lambda}{k}\right)\right]^{0.255} },\\
 C_{c c}^R(k)&=&\frac{\overline D(k)}{2\Gamma(k)}=\frac{\overline D(0)}{2\Gamma(0)} \left[\ln \left(\frac{\Lambda}{k}\right)\right]^{-0.255},\\
 C_\times^R (k)&=&\frac{D_c(0)\kappa(0)-D_\theta(0) D(0)}{2\Gamma(0) V_0}\left[\ln \left(\frac{\Lambda}{k}\right)\right]^{0.117},\\
 {\cal T}_{\theta}(k)&=&{\cal T_{\theta}}\left[\ln \left(\frac{\Lambda}{k}\right)\right]^{-0.255},\\
 \omega_0&=&\left(\frac{1}{2} +\mu_1\right) \frac{\overline D(0)}{4\pi\Gamma_0}\int dk\frac{k}{\left[\ln\left( \frac{\Lambda}{k}\right)\right]^{0.255}}.
\end{eqnarray}

 While $\omega_0$ is the rotation frequency of the XY spins, in the height-interpretations of $\theta$ for the surface version of the model, $\omega_0$ is the nonlinear contribution to the mean growth rate of the surface, which of course is nonuniversal.

\subsubsection{Different scaling regimes and representative scaling exponents}


 Values of the scaling exponents $\eta_1,\,\eta_2$ and $\eta$ as functions of $\mu_1,\,\mu_2$ and the corresponding natures of phase orders and density fluctuations are listed in Table~\ref{tab1}. We further study the possibility of order when $|\Omega_2|,\,|\lambda_1|\gg |\lambda|$. To study this we consider the asymptotic limit $\mu_1,\mu_2\rightarrow\pm \infty$. To proceed, we further define another dimensionless number $\tilde\mu\equiv \mu_1/\mu_2$; $\tilde\mu$ itself can take any real value, positive or negative: $-\infty \leq\tilde\mu\leq \infty$. We present our results in Table~\ref{tab2}. Lastly, we summarize our general results on different types of orders and density fluctuations in Table~\ref{tab3},  which reveals continuously varying scaling exponents in the ordered states.
\linebreak


\begin{table*}[!hbt]
 \begin{center}
  \begin{tabular}{|p{1.1cm}|p{1.1cm}|p{1.1cm}|p{1.1cm}|p{1.4cm}|p{1.2cm}|p{1.4cm}|p{1.9cm}|p{2.6cm}|}
   \hline
   \multicolumn{9}{|c|}{Exponents}\\
   \hline
   $\mu_1$ & $\mu_2$ & ${\cal N}$ & $\psi$ & $\eta_1$ & $\eta_2$ & $\eta$ & Phase order & Density fluctuations\\
   \hline
1.1 & 1.5  & 0.389 & -0.127 & 0.125 & 0.375 & 0.25 & SQLRO & MNF\\
1.0 & 5.0 &-0.694 & -0.525 & 0.109 & 0.369 & 0.26 & SQLRO & MNF\\
1.55 & 2.95 & -0.117 & -0.334 & 0.105 & 0.368 & 0.263 & SQLRO & MNF\\
2.2 & 1.25 & 0.806 & -0.014 & 0.221 & 0.407 & 0.186 & SQLRO & MNF\\
3.6 & 0.93  & 0.949 & 0.141 & 0.494 & 0.498 & 0.001 & QLRO & NNF\\
4.0 & 0.6 & 0.998 & 0.237 & 0.976 & 0.659 & -0.317 & WQLRO & GNF\\
8.0 & 1.47 &0.97 & -0.004 & 0.876 & 0.625 & -0.251 & WQLRO & GNF\\
-3.0 & -5.0  & -0.333 & -0.771 & 0.098 & 0.366 & 0.268 & SQLRO & MNF\\
-4.0 & -3.0  & 0.462 & -0.748 & 0.155 & 0.385 & 0.23 & SQLRO & MNF\\
-4.0 & -4.98 &-0.006 & -0.71  & 0.107 & 0.369 & 0.262 & SQLRO & MNF\\
-7.7 & -2.5 & 0.886 & -0.704 & 0.5 & 0.5 & 0.0 & QLRO & NNF\\
-9.0 & -2.5  & 0.917 & -0.687 & 0.662 & 0.554 & -0.108 & WQLRO & GNF\\
$\frac{\sqrt{2}-1}{2}$ & 1 & 0& 0 & 0.117 & 0.372 & 0.255 & SQLRO & MNF\\
\hline
  \end{tabular}
  \caption{Values of $\cal N$, $\psi$, $\eta_1$, $\eta_2$ and $\eta$ as functions of $\mu_1,\,\mu_2$, and the corresponding nature of order in the stable region. This clearly suggests continuously varying scaling exponents in the ordered states (see text).}\label{tab1}\vskip0.5cm
  \begin{tabular}{ |p{1.2cm}|p{1.4cm}|p{1.4cm}|p{1.7cm}|p{1.4cm}|p{2.2cm}| }
 \hline
 \multicolumn{6}{|c|}{Scaling exponents when $\mu_1,\mu_2\rightarrow$ $+\infty$ or $-\infty$, $\tilde\mu=\mu_1/\mu_2$ } \\
 \hline
$\tilde\mu$ &  $\eta_1$ & $\eta_2$  & $\eta$ & Phase Order & Density fluctuations\\
 \hline
$\gg$ 1 &  undefined & undefined  & undefined & disorder & - \\
$\ll$ 1 &  0.2 & 0.4  & 0.2 & SQLRO & MNF\\
1 & 0.126 & 0.375 & 0.249 & SQLRO & MNF\\
-1 & undefined & undefined & undefined & disorder & - \\
\hline
\end{tabular}
\caption{Scaling exponents in the limits $\mu_1,\,\mu_2\rightarrow$ $+\infty$ or $-\infty$ with $\tilde\mu\equiv \mu_1/\mu_2$ can take any real value.}
 \label{tab2}

 \end{center}

\end{table*}





\begin{table*}[!hbt]
\begin{center}
\begin{tabular}{ |p{1.5cm}|p{1.5cm}|p{3.5cm}|p{3.5cm}|p{2.5cm }|}
 \hline
  \multicolumn{5}{|c|}{Types of orders} \\
 \hline
 \hline
$\eta_1$,\ $\eta_2$ & $\eta$ & Phase fluctuations  & Number fluctuations & $\tau_\theta^R(k\rightarrow 0)$ \\
 \hline
 $\eta_1>\eta_2$  & $\eta<0$ & $\Delta_\theta/\Delta_{eq}\gg 1$  for large $L\implies$WQLRO & $\sigma(\tilde L)\gg \sigma(\tilde L)_{eq}\implies $ GNF & $\tau_\theta^R(k\rightarrow 0)\gg \tau_\theta$ \\
\hline
 $\eta_1=\eta_2$  & $\eta=0$ & $\Delta_\theta/\Delta_{eq}\sim {\cal O}(1)\implies $QLRO & $\sigma(\tilde L)\sim\sigma(\tilde L)_{eq}\implies $ NNF & $\tau_\theta^R(k\rightarrow 0)= \tau_\theta$\\
\hline
 $\eta_1<\eta_2$  & $\eta>0$ & $\Delta_\theta/\Delta_{eq}\ll 1$ for large $L\implies$SQLRO &$\sigma(\tilde L)\ll \sigma(\tilde L)_{eq}\implies $  MNF & $\tau_\theta^R(k\rightarrow 0)\ll \tau_\theta$\\
\hline
\end{tabular}
\caption{Summary of results: different types of orders and number fluctuations}
 \label{tab3}
\end{center}
\end{table*}


{
We now revisit the scaling exponents in the ordered phase obtained from the MCS studies in the light of  the hydrodynamic theory predictions. Our hydrodynamic theory predictions have prompted us to phenomenologically fit our numerical results on the phase variance in the ordered phase to   $\langle \theta^2 \rangle \sim (\ln L)^{\gamma_1}$. We find good fitting with $\gamma_1=0.900 \pm 0.026$ ($g_1=1.0$, $g_2=0.02$) and $1.453 \pm 0.033$ ($g_1=2.0$, $g_2=0.03$) respectively. We get $\gamma_1 \simeq (1-\eta)$ where $\gamma_1=0.900 \pm 0.026$ (with $g_1 = 1.0$, $g_2 = 0.02$ in Fig.~\ref{num-plot1-nofit}(a) signifies $\eta>0$ or an ordering {\em stronger} than QLRO (SQLRO) whereas on the other hand $\gamma_1=1.453 \pm 0.033$ (with $g_1 = 2.0$,  $g_2 = 0.03$ in Fig.~\ref{num-plot1-nofit}(a) signifies $\eta<0$ or an ordering {\em weaker} than QLRO (WQLRO). In general, since $\gamma_1$ can be more or less than 1, $\langle\theta^2\rangle$ can grow faster (WQLRO), or slower (SQLRO) than $\ln L$; $\gamma_1=1$ corresponds to QLRO that delineates SQLRO and WQLRO. In the same way, being inspired by our hydrodynamic theory results, to quantitatively analyze the density fluctuations, we have plotted $\sigma(N_0)/\sqrt{N_0}$ versus $\ln N_0$ in Fig.~\ref{num-plot1-nofit}(b), and fitted with a phenomenological form $\sigma(N_0)\sim \sqrt{N_0/ (\ln N_0)^{\gamma_2}}$, which allows us to extract $\gamma_2$. Noting that with $\gamma_2=1$, i.e., in the equilibrium limit, $\sigma(N_0)$ scales as $\sqrt{N_0}$, $\sigma(N_0)/\sqrt{N_0}$ {\em decreases} as $\ln N_0$ increases for MNF, whereas it rises with $\ln N_0$ for GNF, as shown in Fig.~\ref{num-plot1-nofit}(b)} \footnote{ The slopes in Fig.~\ref{num-plot1-nofit}(b) directly yield the values of the scaling exponent $\eta$. According to the predictions of the RG calculations on the hydrodynamic theory the values of $\eta$ extracted from Fig.~\ref{num-plot1-nofit}(a) (SQLRO and WQLRO, respectively) should be same as those extracted from Fig.~\ref{num-plot1-nofit}(b) (MNF and GNF, respectively). We however find quantitative discrepancies between the exponents obtained from these two plots, which are possibly due to the relatively small system sizes $L$ used in our simulations due to computational limitations. They could also be due to the low order perturbation theory used to obtain the analytical result. Nonetheless, the agent-based model clearly corroborates the SQLRO-MNF and WQLRO-GNF correspondences, as predicted by the hydrodynamic theory, thereby putting our RG calculations on a firmer footing.}.



\begin{table}[h!]
\begin{center}
\begin{tabular}{|p{1.0cm}|p{1.0cm}|p{2.2cm}|p{2.0 cm}|}
 \hline
  \multicolumn{4}{|c|}{Types of Order and the Exponents ($\gamma_1$)} \\
 \hline
 \hline
  $g_1$ & $g_2$  & $\gamma_1$ & Order\\
 \hline
 \vspace{0.05mm} & \vspace{0.05mm} & \vspace{0.05mm} & \vspace{0.05mm} \\

  $1.0$   &  & $0.961 \pm 0.019$ & SQLRO\vspace{0.5mm} \\

  $1.5$  & $0.01$ &  $0.957 \pm 0.025$ & SQLRO\vspace{0.5mm} \\

  $2.0$  &  & $0.946 \pm 0.024$ & SQLRO\vspace{3.5mm} \\
\cline{1-4}
\vspace{0.01mm} & \vspace{0.01mm} & \vspace{0.01mm} & \vspace{0.01mm} \\
   $1.0$ &   & $0.900 \pm 0.026$ & SQLRO \vspace{0.5mm}\\

   $1.5$  &$0.02$ &  $0.954 \pm 0.034$ & SQLRO \vspace{0.5mm}\\

   $2.0$  & & $1.028 \pm 0.021$ & QLRO$^*$ \vspace{3.5mm}\\
\cline{1-4}
\vspace{0.01mm} & \vspace{0.01mm} & \vspace{0.01mm} & \vspace{0.01mm} \\
  $1.0$ &   & $0.892 \pm 0.035$ & SQLRO \vspace{0.5mm}\\

 $1.5$   &  $0.03$ & $1.040 \pm 0.016$ & QLRO$^*$ \vspace{0.5mm}\\

  $2.0$ &    & $1.453 \pm 0.033$ & WQLRO \vspace{0.5mm}\\
  \vspace{0.01mm} & \vspace{0.01mm} & \vspace{0.01mm} & \vspace{0.01mm} \\
\hline
\vspace{0.01mm} & \vspace{0.01mm} & \vspace{0.01mm} & \vspace{0.01mm} \\
   & $0.01$ & \hspace{8mm}----- &  \vspace{0.5mm} \\
$4.0$ & $0.02$ & \hspace{8mm}----- &  Rough phase \vspace{0.5mm} \\
& $0.03$ & \hspace{8mm}----- &  \vspace{0.5mm} \\
\vspace{0.01mm} & \vspace{0.01mm} & \vspace{0.01mm} & \vspace{0.01mm} \\
\hline
\end{tabular}
\caption{Quantification of the exponent $\gamma_1 (\Leftrightarrow 1-\eta)$ and the nature of ordering obtained for different values of the active coefficients  $g_1, \; \text{and}\; g_2$ respectively. The values of the coefficients $c_s$ and $c_{s2}$ are kept $1$. Other parameters:  $c_0=5$, $\xi=0.1$, and $p_\theta=p_c=0.5$. Increasing the magnitude of $g_1$ can result in a transition from ordered to disordered rough phase in our simulations (see text). ($^*$ Although $\gamma_1$ within the error is slightly larger than unity, but we are still inclined to call the ordering QLRO, keeping in mind the statistical fluctuations).}
 \label{tab4Simulation}
\end{center}
\end{table}


\subsection{Destabilization of order}\label{dest-order}

So far in the above, we have discussed the statistical properties of the ordered states within the hydrodynamic theory. These ordered states are however not the only states of the system. The system can also be in the disordered states that are characterized by SRO phase-order, as revealed by our MCS studies of the agent-based model. While the hydrodynamic theory cannot be used to study the disordered states, they still provide important clue to the ``paths to disorder'', i.e., destabilization of the ordered states, which in turn can be used to understand the MCS results on the disordered states. In fact, there are definite indications in the hydrodynamic theory of the existence of macroscopically two distinct kinds of disordered states of the agent-based model, which we discuss below.

According to the hydrodynamic equations (\ref{theta-full-eq}) and (\ref{c-full-eq}), the system can get disordered in {\em two distinct} ways, which should result into two types of distinct disordered states. First of all, when the condition of linear stability is not met, i.e., when $\lambda_0\tilde \Omega_0\tilde \Omega_1 <0$, there is a growing mode with a growth rate proportional to wavevector $k$. Of course, there should be stabilizing higher order damping terms at ${\cal O}(k^2)$,  such that there would be a preferred wavevector $k_c$ at which the growth rate is maximum. Emergence of patterns or structures of length scales $\sim 2\pi/k_c$ is then expected. Needless to say, the growth at this preferred wavevector $k_c$ will eventually saturate in the long time limit due to the nonlinear effects, e.g., the nonlinear terms included in (\ref{theta-full-eq}) and (\ref{c-full-eq}), or other subleading nonlinear terms not included in the hydrodynamic theory. All in all, then the disordered steady states should consist of clusters of typical size $2\pi/k_c$, which is an ${\cal O}(1)$ number, precluding any conventional scaling behavior. This is nothing but the ``aggregate phase'' found in the MCS studies of the agent-based model.  On the other hand, the system can become disordered even in the linearly stable case when $\Delta_1$, as given in (\ref{del1}), becomes negative. In that case, the scale-dependent coupling constant $g(l)$ flows to infinity in a manner  reminiscent of the $2d$ KPZ equation. The $2d$ KPZ equation shows only a rough phase that has no intrinsic scale; it  cannot however be studied in perturbation theories~\cite{stanley}. Nonetheless, the rough phase is still believed to display standard universal scaling~\cite{stanley}. Drawing analogy with the instability of the smooth phase of the $2d$ KPZ equation, we expect similar scaling behavior here as well for $\Delta_1<0$. This means the absence of a preferred wavevector. This further implies that there should be no clusters of any ``preferred macroscopic size'', and we name them ``rough phase'' as found in the agent-based model, in direct analogy with the rough phase of the $2d$ KPZ equation. Indeed, the spin configurations in the rough phase appear very similar to that obtained from the $2d$ KPZ equation, valid in the immobile limit or without number conservation; see Fig.~\ref{fig:KPZ} (a) and (c) in Appendix~\ref{KPZ} for simulation snapshots of the phase in the $2d$ KPZ limit of the model. Fluctuations in the rough phase are much larger than in the ordered phases, as can be seen from the much broader $P(c)$ found in the MCS studies of the rough phase; see Fig.~\ref{histo1}. Evidently, these rough phases appear very different than the aggregate phase formed in the linearly unstable case. Of course, in both the cases, there will be complete loss of any phase order. We study these disordered states numerically by using our agent-based model, which clearly confirm the hydrodynamic theory speculations on the disordered states. These two classes of disordered states visibly appear different in their respective snapshots of the densities (compare the snapshot in Fig.~\ref{snapshot-visual}(f) with those in Fig.~\ref{fig:LinUns}; Appendix~\ref{disd-st}) and can be characterized by their respective density distribution $P(c)$ (see Fig.~\ref{roughAndaggregate}). Like $P(c)$, a similar observations on the disordered states can also be made by finding the average number of clusters and the corresponding average cluster sizes specific to particular  $c$'s in both rough and aggregate phases; see Fig.~\ref{rough-aggregate} and Fig.~\ref{aggregate3} in Appendix~\ref{disd-st} for details.  


See Table~\ref{tab4Simulation1} or \ref{tab4Simulation}, and Table~\ref{tab5Simulation} for a general classification of the steady states (ordered or disordered) based on the magnitudes and signs of the simulation parameters.


We note that in the hydrodynamic theory the  nonlinear stability or the robustness against noises of the  linearly stable states is not destroyed by the reversal of the sign of the coupling constant $\lambda$ in Eq.~(\ref{theta-full-eq}). There is however no symmetry of Eq.~(\ref{theta-full-eq}) under $\lambda\rightarrow -\lambda$. Nonetheless, the phase diagram in Fig.~\ref{phase-diag} reveals that within  restricted ranges of $\mu_1$ and $\mu_2$, a steady state with a given pair of $(\mu_1,\,\mu_2)$ with one particular sign of $\lambda$ implies the existence of another steady state with $(-\mu_1,\,-\mu_2)$ obtained with the opposite sign of $\lambda$. This however does not imply that  these two pair of steady states are necessarily statistically identical. Further studies are required to establish generic relations, if any, between these pair of steady states. 

{ Due to the independent nature of the origins of the agent-based model and the hydrodynamic theory, at this level it is not possible to identify any one-to-one correspondence between the terms in the hydrodynamic equations and the microscopic update rules. Nonetheless, we have noticed interesting parallels. For instance, we find that phase-ordered states can exist even in certain extreme limits of the model parameters. For instance in the agent-based model, with $g_1=0$ our simulation results exhibit an ordered NESS, with certain values of the other active coefficients (see Fig.~\ref{infinite}; Appendix~\ref{limitingcase-ordered}). Similarly, in our hydrodynamic theory in the limit of $\mu_1,\mu_2\rightarrow \infty$, phase-ordered states are obtained {(see Table~\ref{tab2})}. What these parallels suggest concerning the inter-relations between the agent-based model and the hydrodynamic theory is an interesting questions beyond the scope of the present study.}

\subsection{Phase diagram}\label{phase-sec}

We now present the phase diagram of the linearly stable sector of the model (i.e., $\lambda_0\tilde \Omega_0\tilde \Omega_1 >0$) in the $\mu_1-\mu_2$ plane over a limited range of $\mu_1,\,\mu_2$  in Fig.~\ref{phase-diag} below. The different regions in the $\mu_1-\mu_2$ plane corresponding to ordered states are essentially the common regions satisfying the constraints $\Delta_1>0$ and $-1\leq\psi\leq1$, i.e., the regions common to the shaded regions in Fig.~\ref{del-plot} and Fig.~\ref{psi-plot}. Different regions displaying SQLRO with MNF, and WQLRO with GNF are marked. The boundary lines between the SQLRO and WQLRO regions correspond to QLRO with NNF or normal density fluctuations. 
In the remaining regions of the phase space in the $\mu_1-\mu_2$ plane no stable scaling behavior pertaining to uniform ordered phases is found, signaling destruction of the uniform phase-ordered states.
\begin{figure}[htb]
 \includegraphics[width=0.9\columnwidth]{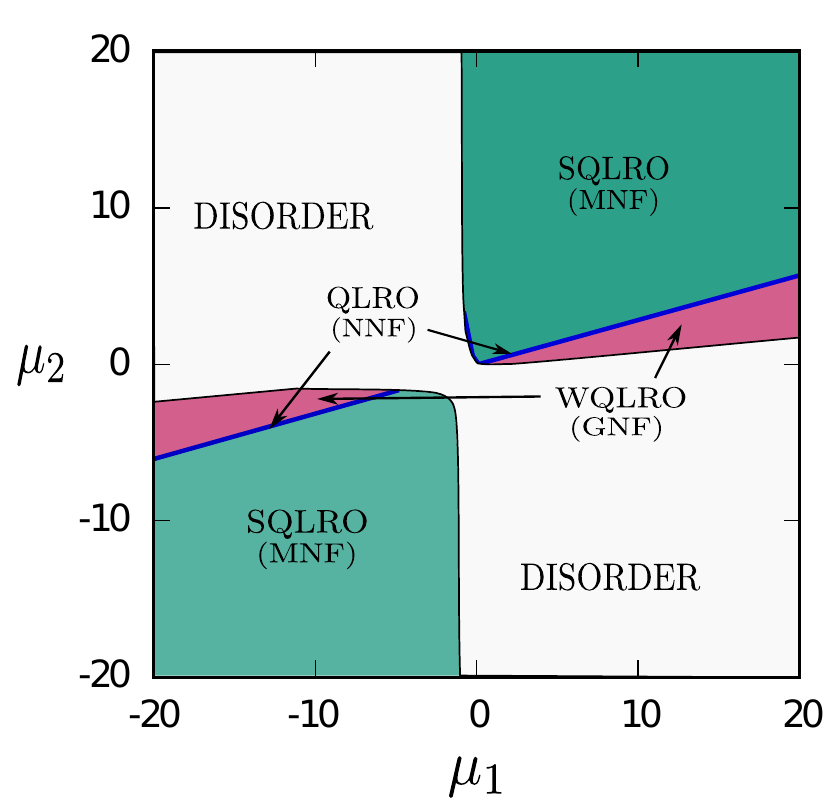}
 \caption{Phase diagram in the $\mu_1-\mu_2$ plane for the linearly stable sector of the model ($\lambda_0\tilde \Omega_0\tilde \Omega_1 >0$). SQLRO with MNF (WQLRO with GNF) are marked in magenta (green). The blue colored lines between the SQLRO and WQLRO regions are QLRO phases with NNF. In the remaining regions of this plane, the no uniform order is found (see text).} \label{phase-diag}
\end{figure}

In principle, an equivalent phase diagram of the agent-based model could be obtained by varying the different model parameters and identifying the stable phase-ordered states in the Monte-Carlo simulations. We did not do this here {because of the high computing costs involved, and even if done, it may not provide any additional significant insights to the physics of our model.}

In physical realizations of the model, the 
magnitudes of the model parameters, which parametrize scaling, should depend on $c_0$ (or equivalently $N$, the total particle number for a given system size), and $a_0$ (a rough measure of the particle size). For instance, $\kappa,\,D,\,\Omega_1$, having dimensions of a diffusion constant, should decrease with increasing density $c_0$ and particle size $a_0$. Further, $\Omega_2$ has the dimension of a hyperdiffusivity. Hence, { we set $\Omega_2\sim Da_0^2$. Then, $\lambda_0\tilde\Omega_0,\,\lambda_0\tilde\Omega_1\sim D/a_0^2$,} since both have the dimensions of a frequency. Thus, changing the number density and typical particle size, different regions of the phase space and hence different scaling behavior of the ordered states and the instabilities can be explored.

 

\section{Topological defects}\label{tp-def}

 The spin wave theory for the  classical XY model predicts QLRO, which in the absence of any spin wave - spin wave interactions should persist at all temperatures. This of course cannot be true, since at  high enough temperatures, the system must be in its paramagnetic phase with only a short-range order. This apparent contradiction is explained by the Kosterlitz-Thouless (KT) phase transition theory~\cite{kt}, which argues that there are localized excitations in the form of topological defects or vortices,  which arises due to the compactness of $\theta$. The KT theory predicts that the QLRO predicted by the spin wave theory with bound vortex pairs undergoes a defect unbinding transition to a high temperature paramagnetic phase  with free  vortices. While the KT theory makes predictions on the phase transition temperature $T_{KT}$ based on RG calculations, an estimate of $T_{KT}$ can be made heuristically by comparing the energy of a single vortex $E_v=\kappa\ln (L/a_0)$ in a system with linear size $L$, and the entropy of formation of a single vortex in the same area, $S=2\ln (L/a_0)$.  With a free energy of a single defect $F=E-TS$, we note that $T_{KT}=\kappa/2$~\cite{chaikin}.  If $\theta$ is interpreted as a single-valued height field, it is non-compact, and hence the question of topological defects does not arise.

The present hydrodynamic theory, which is a spin wave theory coupled with density fluctuations, does not contain any topological defects, for it is written assuming small phase and density fluctuations in all space and time. However, the possibility of formation of unbound defects and their proliferation cannot be ruled out {\em apriori}, given the general wisdom from studies on the $2d$ equilibrium XY model. It is then pertinent to ask whether the ordered states, i.e., SQLRO or WQLRO, survive these unbound defects at low enough noises (nonequilibrium analogs of temperature), or these defects {\em always} proliferate and destroy the order even for vanishingly low noises. For the present study, identifying renormalized, scale-dependent { ${\cal T}_\theta^R(k)$} in (\ref{red-tempR}) as the inverse of an effective reduced spin stiffness $\kappa_\text{eff}(k)$,  we get the vortex unbinding (reduced) temperature 
$T^\text{eff}_{KT}$ in a system of size $L$
\begin{equation}
 T^\text{eff}_{KT}\sim \kappa_\text{eff}(L)^{-1}\sim [\ln (L/a_0)]^\eta\times {\cal O}(1).
\end{equation}
We thus find that for SQLRO, $T^\text{eff}_{KT}(L)$ grows, whereas for WQLRO, $T^\text{eff}_{KT}(L)$ decays as $L$ gets bigger and bigger, eventually diverging and vanishing, respectively, in the thermodynamic limit. {This is however clearly unexpected}. In fact, the renormalized theory does not map to an effective equilibrium description, in part due to the non-vanishing cross-correlation function (see Appendix~\ref{fdt1}). In fact, even if it does map, it does not guarantee an equivalent mapping of the defect dynamics. This raises the question on the stability of bound vortices, necessary for stable ordered states~\cite{john-pvt} . Nonetheless, we can say with some degree of confidence that for SQLRO, the defect unbinding transition should take place at $T_{KT}^\text{eff}>T_{KT}$, whereas for WQLRO, $T_{KT}^\text{eff}<T_{KT}$ is expected. For the latter case, if $\eta$ is sufficiently negative, $T_{KT}^\text{eff}$ might even vanish in the thermodynamic limit, something which cannot be captured within our hydrodynamic theory.  Our numerical simulations of the agent-based model show that the stable ordered states that we study are stable against formation of these defects. Applications of more sophisticated theoretical approaches beyond heuristic arguments as above should be useful in this regard; see, e.g., Ref.~\cite{other-methods,john-defect}.

\section{Summary and outlook}\label{summ}

We have constructed an active XY model with a conservation law and thoroughly studied it in $2d$, exploring the stable phase-ordered states, their scaling properties, and also the instabilities and disordered states. In particular, we have constructed an agent-based model consisting of XY spins rigidly grafted on a lattice, which are in contact with a diffusively mobile conserved active species. We have simulated this system by usig Monte-Carlo methods, which detect stable phase-ordered and unstable phases in the system, controlled by the model parameters. In the phase-ordered states, we have numerically calculated $\langle \theta^2\rangle$ and $\sigma(N_0)$, which show scaling, respectively, with $\ln L$ and $N_0$. We find clear numerical evidence that these dependence on $\ln L$ and $N_0$ could be either simultaneously {\em stronger} or {\em weaker} than in equilibrium. This reveals a hitherto unknown, unique correspondence between the phase and density fluctuations in the stable phase-ordered states, demonstrating the simultaneous occurrence of SQLRO and MNF (WQLRO and GNF), which is usually not found in other orientationally ordered active matter systems. 
The agent-based model also admits   orientationally disordered states, which can be of two distinct types - no phase order with (i) no stable density clusters, and (ii) stable density clusters with preferred sizes. We name these disordered states as rough and aggregate phases, respectively.

In order to understand and provide firmer theoretical basis to the results from the agent-based model studies, we have also developed the hydrodynamic theory for a collection of nearly phase-ordered active mobile XY spins on a $2d$ substrate. Local phase and density fluctuations are the relevant hydrodynamic variables. Due to the presence of friction, any local advecting flow field is irrelevant in the RG sense, as shown in Appendix~\ref{irre-adv}. 
In this theory, the local phase fluctuations and density fluctuations are coupled both {\em linearly} and {\em nonlinearly}. The linear cross coupling terms are such that the nearly phase ordered states are either linearly stable with underdamped propagating waves with linear dispersion  that dominate over damping at small $k$, or linearly unstable.  In  the hydrodynamic theory, the time evolution  of phase fluctuations $\theta$ depends on the local number density fluctuations quadratically, in addition to a nonlinear term that is quadratic in the spatial gradient of $\theta$ itself. The latter nonlinearity is formally identical to that present in the KPZ equation for surface growth phenomena. The time evolution of the density fluctuation $\delta c$, a conserved density, is affected not only by conventional diffusion, but also by a current that depends on ${\boldsymbol\nabla}\theta$.  We have further argued that the hydrodynamic equations developed here hold also for a generalized-KPZ surface, e.g., inversion-asymmetric active biomembranes with a finite tension but without any momentum conservation, coupled with a conserved density on it. 

 By employing one-loop dynamic RG methods, we find that this model displays complex scaling behavior, hitherto unstudied. 
This theory predicts that the linearly stable states can be nonlinearly stable ordered states robust against noises for appropriately chosen model parameters.  Interestingly, the stable states that are robust against noises have essentially the same scaling laws as the linearly stable states, except for logarithmic modulations. We find that the phase-ordering could be generically ``logarithmically stronger'' (SQLRO) or ``logarithmically weaker'' (WQLRO) than QLRO that characterizes the equilibrium limit of the model, with  attendant ``miniscule number fluctuations'' (MNF) or ``giant number fluctuations'' (GNF) that characterize the density fluctuations. Nonlinear effects can also make phase-ordering unstable leading to loss of order or SRO. These are parametrized by $\mu_1,\mu_2$, the two dimensionless ratios of the three nonlinear coupling constants that appear in the hydrodynamic equations, and lead to the phase diagram (Fig.~\ref{phase-diag}). Our one-loop RG calculations predict that the relevant scaling exponents vary continuously with $\mu_1,\,\mu_2$, reflecting breakdown of the conventional notion of universality. We have defined a reduced temperature, that in the renormalized theory depends upon wavevector $k$ in a way such that for SQLRO, it decreases continuously and for WQLRO, it increases continuously as $k$ gets smaller and smaller.  Thus, the fluctuations about uniformly ordered stable states in the model can be either suppressed or enhanced vis-a-vis the equilibrium limit, a truly unique aspect of this model. These results from the hydrodynamic theory  are independent of the specific functional forms for $\Omega(c)$ and $\tilde \Omega(c)$, so long as the conditions of stability are generally met. 
The hydrodynamic theory predictions can be used to fit the numerical results on the scaling in the agent-based model, giving numerical estimations of the scaling exponents. 
Furthermore, our hydrodynamic theory, valid for small phase and density fluctuations, gets unstable either linearly, or due to nonlinear interactions. We have argued that these two routes to the disordered states correspond to the aggregate and rough phases, respectively, found in the agent-based model studies. Our hydrodynamic theory is clearly consistent with the results from the agent-based model studies. Our combined numerical and analytical studies, thus, firmly establish the complex scaling behavior of this model, distinguished by the SQLRO/MNF and WQLRO/GNF correspondence in the stable phase-ordered states, as well as the natures of the instabilities and the resulting disordered states. { Furthermore, we have considered zero noise crosscorrelations in our RG study. A non-zero noise crosscorrelation can be implemented in a straightforward way. This is not expected to change the qualitative features of our RG results.} 

We hope our theory will induce future experiments measuring $\langle \theta^2\rangle$ and $\sigma(N_0)$ in self-assembled spinners ~\cite{kotot}, rotor assemblies~\cite{shelley,rotor-transport}, programmable magnetic cilia carpets~\cite{mag-cilia,coll1}.


 As mentioned above, our agent-based model admits disordered states. In order to understand the disordered states, we have used our agent-based model to numerically explore the natures of these disordered steady states, lacking any orientational order.
 These disordered states are found be of two types, which are differentiated from each other by the respective density morphology or density distributions. One of them is the rough phase, which have fluctuations larger than the ordered phases, but without any long lived density clusters of any preferred sizes. In fact, the probability distribution $P(c)$ remains peaked about $c=c_0$, the mean density, similar to the phase-ordered states. However in this case, $P(c)$ is distinctly broader than that in the phase-ordered state. The second kind of the disordered states, named aggregate phase here and distinct from the rough phases, contain long-lived high density clusters of preferred sizes, with near-empty regions intervening these clusters. Thus the system separates into regions of high and low densities, i.e., it {\em phase separates}. In this case, $P(c)$ becomes double-peaked, with one peak at $c\gg c_0$ corresponding to the density clusters, and another larger peak near $c=0$, correspond to the near-empty sites between { the} density clusters. Clearly, this phase separation is a consequence of the interplay between the mobility and number conservation of the { diffusive species}. Thus these states form a novel example of ``motility induced phase separation''; see Ref.~\cite{cates11}. 
Can the existence of the disordered states be made related to the predictions from the hydrodynamic theory? Our hydrodynamic theory predicts that in some regions of the phase space, phase-ordered states are not possible, but breaks down for the disordered states. Nonetheless, it shows two distinct paths to disorder. Based on them, we speculate that those NESS of the disordered states could be of two types, one that has no long-lived density clusters of preferred sizes, and the others with clusters or patterns of a typical size (which is nonuniversal). In fact, the disordered  states of the agent-based model are analogous to the two different routes to instability in the hydrodynamic theory. The first kind, named rough phase, should correspond to instability of  linearly stable states that are rendered unstable by the nonlinear effects, or not robust against noises.  The second kind, named aggregate phase, should be linked  to the linearly unstable cases of the hydrodynamic theory.



 Our model forms a new class of ``polar'' active matter, known as {\em active spinners without directed walking}~\cite{kotot,vitelli}, different from the polar active matter models for flocking~\cite{2d-polar}.  Our model is fundamentally different from the celebrated ``moving XY model'' for polar ordered active fluids or flocking phenomena on substrates~\cite{2d-polar,john-flock,vicsek}. The latter model is characterized by ``self-propulsion'' of the particles. This is absent in our model; instead, the particles can rotate {\em actively}, resulting into stable phase-ordered NESS with novel parameter-dependent scaling behavior, or lack of phase-ordered states with distinct density morphologies. This model clearly has the vectorial symmetry, except that the spins here are ``vectors'' in the order parameter space, and not in the physical space. As a result, there is no ``spin-space'' coupling in this model, unlike the polar active matter models~\cite{2d-polar}. Connected to this is the absence of self-propulsion in this model. In contrast, self-propulsion is the crucial distinguishing feature of conventional polar active matter models. Lack of self-propulsion means we have no analog of the flocking phenomena in this model.  

At a technical level, considering the surface growth interpretation of the model, we have shown that the well-known thresholdless instability of the smooth phase in the $2d$ KPZ surface can be ``tamed'' or suppressed by coupling with additional degrees of freedom (in this case a conserved density). In fact, the taming of the instability is quite sensitive to the details - it occurs in parts of the phase space; in other parts of the phase space, the instability remains. Naturally this suppression of the instability implies the complete suppression of any putative roughening transition for $d>2$, that is well-known in the KPZ equation as also in the unstable sector of the phase space of the present theory for $d>2$. This is potentially significant from the point of view of nonequilibrium statistical mechanics and nonlinear dynamics, demonstrating how instabilities and phase transitions could be generically suppressed by adding extra degrees of freedom. It is known that $2d$ isotropic KPZ has only a rough phase, whereas $2d$ anisotropic KPZ equation can have a smooth phase~\cite{john-superfluid}. We have shown here that even $2d$ isotropic KPZ equation can have a ``smooth'' phase (different from the usual smooth phase of the $2d$ EW equation in having logarithmic corrections to the noninteracting theory), provided additional conserved degrees of freedom are coupled to the KPZ height field.  We expect this to give new understanding of generic order and instability in nonequilibrium systems.

We have limited our numerical studies using the agent-based model to just measuring $\langle\theta^2\rangle$ and $\sigma (N_0)$ in the steady states due to computational limitations. Although it would be a numerically challenging task, it would be useful in the future to measure the equal-time phase and density correlation functions as a function of separation $r$ in the agent-based model,  and  compare with the predictions from hydrodynamic theory. { It is also worthwhile to study and extract the universal aspects of the disordered phases (rough and aggregate), if any, by using different microscopic rules, all of which reduce to the same hydrodynamic equations in the long wavelength limit, and hence give the same universal scaling in the ordered phases. Whether or not microscopically different update rules affect the properties of the rough or the aggregate phase, and if so, to what degree, should be studied systematically.} 

Our study can be extended in various ways. For instance, one could consider a quenched disordered substrate. Quenched disorders are known to affect the universal behavior of pure systems, both equilibrium~\cite{weinrib,sudipON} and nonequilibrium~\cite{astik-prr}. It would be interesting to see how quenched disorder affects the results discussed above. How the  ordering found in this model is affected in the presence of both static and/or moving obstacles would be a worthwhile future work; see, e.g., Ref.~\cite{martin}  for a similar study on a flocking model. Secondly, noting that in the present study the density is a non-critical field, it would be of interest to study ordering and fluctuations when the density is {\em critical}, i.e., has a critical point,  separating a well-mixed and phase-separated state. At the critical point, the density fluctuations are long ranged. It will be interesting to study whether long range correlated density fluctuations at the critical point could induce long range orientational order. It would also be interesting to study the effects of multiple species and their mutual reactions including reproduction and death. This will have more direct biological implications. It should also be interesting to study the topological defects and the analogue of the KT transition in the model; see, e.g., Ref.~\cite{demian11} for a similar study in a model without mobility. The latter, if exists in our model, would be an example of a nonequilibrium KT transition. How that is influenced by the number conservation and mobility is a challenging theoretical question. We hope our studies will induce future research in these directions.

\section{Acknowledgement}
AH and AB thank T. Banerjee and N. Sarkar for critical reading and helpful suggestions. AS thanks University Grants Commission (UGC), India for a research fellowship. SC is financially supported by the German Research Foundation (DFG) within the Collaborative Research Center SFB 1027 and Indian Association for the Cultivation of Science, Kolkata.  AS and SC also acknowledges R. Paul (IACS, Kolkata) for some invaluable suggestions and discussions and are also thankful to him for providing the computational resources. AB thanks J. Toner for valuable  discussions in the early stages of this work, and comments concerning derivations of Eqs.~(\ref{ren-c-corr}) and (\ref{numb-fluc}), P. K. Mohanty, A. Maitra, and D. Levis for helpful comments, and the SERB, DST (India) for partial financial support through the MATRICS scheme [file no.: MTR/2020/000406]. AB has designed the problem, AH has contributed to the analytical part of the work, AS and SC have contributed equally to the numerical part of the work. All four authors jointly wrote the manuscript.

\appendix

\section{Glossary}

   Here we give rough definitions of some of the symbols used in different contexts in this paper.
 
\begin{enumerate}
   \item $L$: Linear system size [See Eq. (\ref{eta-intro})].
 \item $a_0$: Set as minimum lengthscale for hydrodynamic theory [See Eq. (\ref{eta-intro})]. 
 \item $\Delta_\theta^R$: Renormalized variance of the phase $\theta$ [See Eq. (\ref{eta-intro})]. 
  \item $R$: Stands for renormalised quantity [See Eq. (\ref{eta-intro}) and Eq. (\ref{eta})].
  \item $\eta$: A scaling exponent that determines { phase} ordering and density fluctuations [See Eq. (\ref{eta-intro})].
 \item $\eta_1$: A nonuniversal scaling exponent; it is positive definite [See Eq. (\ref{eta1})]. 
 \item $\eta_2$: A nonuniversal scaling exponent, also positive definite, determines the dynamics faster than diffusive [See Eq. (\ref{eta2})].
 \item $\sigma$:  Standard deviation of number fluctuations [see Eq.~(\ref{numb-fluc-lin}) for the linear theory, and Eq.~(\ref{numb-fluc}) for the nonlinear theory].
   \item $\Lambda$: Upper cutoff limit of wavevector i.e. $2\pi/a_0$.
 \item $l$: Renormalization group ``time'', which is the logarithm of a length-scale.
  \item $\mu_1$: Reduced dimensionless parameter: Ratio of two nonlinear term coefficients $\Omega_2$ and $\lambda$, the parameters present in hydrodynamic equations [see Eq.~(\ref{theta}) and Eq.~(\ref{C})].
 \item $\mu_2$: Reduced dimensionless parameter: Ratio of two nonlinear term coefficients $\lambda_1$ and $\lambda$, the parameters present in hydrodynamic equations [see Eq.~(\ref{theta}) and Eq.~(\ref{C})].
 \item $\chi_\theta$: Scaling exponent of the phase field $\theta$ [see Eq.~(\ref{basic-scale}), (\ref{field-scale})].
  \item $\chi_c$: Scaling exponent of the density fluctuation $\delta c$ [See Eq. (\ref{basic-scale}) and also Eq. (\ref{field-scale})].
 \item $\hat\chi_\theta$: Scaling exponent of the dynamic conjugate field $\hat\theta$ [see Eq.~(\ref{basic-scale}) and also Eq. (\ref{field-scale})].
 \item $\hat\chi_c$: Scaling exponent of the dynamic conjugate field $\hat{ c}$ [see Eq.~(\ref{basic-scale}) and also Eq. (\ref{field-scale})].
 \item $g:$ The dimensionless effective coupling constant in the present model [Eq.~(\ref{dim-coup-here}].
  \item $\tilde \gamma(r):$  Scaling exponent that describes the spatial decay of $C_{ZZ}^R(r)$ for large $r$ 
 [Eq.~(\ref{gamma-tilde}].
     \item $\omega_0$: Rotation frequency due to the nonlinear effects defined in the rotating frame of the system with uniform density [Eq.~(\ref{ang_freq} ].      
  \item $\gamma_1$: { Scaling exponent of $\Delta_\theta^R$ as a function of $\ln L$, calculated in the numerical simulations which determines the nature of phase-ordering of the active {XY spins (see Fig.~\ref{num-plot1-nofit}(a))}.}
 
 \item $\langle \cdots \rangle_{L^2}$: Averaging over $L^2$ lattice sites at a given time.
 
 \item $\langle \cdots \rangle_{t,L^2}$: Averaging over time and $L^2$ sites.
 
 \item $\langle \cdots \rangle_{nn}$: Averaging over the nearest neighbors.
 
 \item $g_2,g_1,c_s,c_{s2}$: Active coefficients used in the agent-based modeling.
 
 \item $\xi$: The amplitude of the noise in spin dynamics update rules in the agent-based model.
 
 \item $p_\theta$, $p_c$: Relative probabilities of the various update rules in the agent-based model.  
\end{enumerate}

\section{Details of Monte Carlo (MC) Update}
\label{MC-update}
{ In Section~\ref{modeling}, we have briefly discussed the central features of the numerical simulations performed on the agent-based model. 
Now, in this appendix, we present the particulars of the microscopic update rules which essentially govern the simulation procedure. At each MC step, we sequentially update both phase and density at each site $\ell$ of the square lattice. A single MC update can be described as follows$\colon$

\subsubsection{Update of $\theta_\ell$}
\label{phi_update}

(a) A random site $\ell$ out of $L^2$ lattice sites is chosen with $c_\ell$ number of particles and it has been assigned a unit spin vector ${Z}_\ell$ of angle $\theta_\ell$. \\

(b) The spin vector $Z_\ell$ then tend to align its direction with its local neighbors where the direction of $Z_\ell$ depends on the average direction of all spins (including $\ell$) in the nearest neighborhood. Similar to the Vicsek model~\cite{vicsek}, {the average spin orientation at site $\ell$ at the MC simulation time $t$ is computed in the following manner:}
\begin{equation} \label{theta_av}
 \bar{\theta}(t) = \arctan[\langle \sin(\theta(t)) \rangle_{nn}/\langle \cos(\theta(t)) \rangle_{nn}],
\end{equation} 
where $\langle ... \rangle_{nn}$ denotes the average over the nearest neighborhood including $\ell$. \\

(c) Next, we calculate $\langle \cos(\delta\theta) \rangle_{L^2}$ and $\langle \Omega(c) \rangle_{L^2}$ where $\langle \cos(\delta\theta) \rangle_{L^2}=\frac{1}{L^2}\sum_{\ell=1}^{L^2}\cos(\delta\theta_{\ell})$ and $\langle \Omega(c) \rangle_{L^2} = \frac{1}{L^2}\sum_{\ell=1}^{L^2}c_\ell^2$. For site $\ell$, $\cos(\delta\theta_\ell)$ is defined as$\colon$
\begin{equation}
\cos(\delta\theta_\ell)=\cos\big(\sum_{\langle \ell \ell^\prime \rangle}|\theta_\ell-\theta_{\ell^\prime}|/4\big),
\end{equation}
where ${\langle \ell \ell^\prime \rangle}$ denotes nearest-neighbor pairs with sums over $\ell^\prime$.\\

(d) Next we calculate the two parameters, $g_\ell^1$ and $g_\ell^2$ where $g_\ell^1$ is defined as the following:
\begin{equation} \label{g_1}
g_\ell^1=g_1(\cos(\delta\theta_\ell)-\langle \cos(\delta\theta) \rangle_{L^2}),
\end{equation}
where $g_1$ is the positive or negative coefficient of $g_\ell^1$. {Here, in Eq.~(\ref{g_1}), the mean rotation frequency, $g_1 \langle \cos(\delta\theta) \rangle_{L^2}$, is subtracted from the rotation frequency of all the individual spins, which is equivalent to being in the frame of reference of mean rotation, instead of the lab frame.} 

Likewise, we define $g_\ell^2$ as follows: \begin{equation} \label{g_2}
g_\ell^2=g_2(c_\ell^2-\langle \Omega(c) \rangle_{L^2}),
\end{equation}
where $g_2$ is defined as the {coefficient} of $g_\ell^2$ (could either be positive or negative). {Here also the mean rotation frequency, $g_2 \langle \Omega(c) \rangle_{L^2}$, is subtracted from the rotation frequency of all the individual spins to being in the frame of reference of mean rotation, instead of the lab frame.}
\\

(e) We then define a probability $p_\theta$ for phase update and choose a random number $r$. The updated phase $\theta_\ell^\prime(t)$ then takes the following form:
\begin{equation}
\theta_\ell^\prime(t) =
\begin{cases}
\bar{\theta}(t)+g_\ell^1+g_\ell^2+\Delta\theta_1 &  r \leqslant p_\theta, \\ \\
\bar{\theta}(t)+\Delta\theta_2  & \text{otherwise},
\end{cases}
\end{equation}
where $\Delta\theta_1$ and $\Delta\theta_2$ are random numbers chosen with a uniform probability from the interval $[-\xi/2,\xi/2]$, $\xi$ being the noise in the system similar to that in the Vicsek model~\cite{vicsek}. Thus, $p_\theta$ is the probability that the update of the phase includes contributions from local alignment as well as from the local density and phase fluctuations whereas phase update from the contribution of the local alignment  only  happens with probability $1-p_\theta$.

\subsubsection{Update of $c_\ell$}
\label{c_update}

(a) We choose a neighbor randomly out of the 4 possibilities and denote it by $v$.\\

(b) Analogous to our phase update scheme described above in Appendix~\ref{MC-update}, we define an active probability $p_c$ for the density update and compare it with a random number $r^\prime$.\\

(c) If $r^\prime \leqslant p_c$, we enter the active regime and calculate $\Delta\theta=\theta_\ell-\theta_v$ and the average local density $\langle c \rangle_v=(c_\ell+c_v)/2$. A new active parameter $c_{\rm sum}$ is then introduced and defined as follows:
\begin{equation}\label{c_sum}
c_{\rm sum}=\Omega^\prime(\langle c\rangle_v)\sin(\Delta\theta),
\end{equation}
where 
\begin{equation}\label{omega_prime_cv}
\Omega^\prime(\langle c\rangle_v)=c_s\langle c \rangle_v+c_{s2}\langle c \rangle_v^2
\end{equation}
{with $c_s$ and $c_{s2}$ being the positive or negative coefficients of $\langle c \rangle_v$ and $\langle c \rangle_v^2$ respectively}. Depending upon the sign of $c_{\rm sum}$, the following updates take place:
\begin{equation}
\label{c_update1}
\begin{cases}
c_\ell^\prime=c_\ell-1, c_v^\prime=c_v+1 &  {c_{\rm sum}>0 \;\& \;c_\ell \neq 0}, \\
c_\ell^\prime=c_\ell+1, c_v^\prime=c_v-1  & {c_{\rm sum}<0 \;\& \; c_v \neq 0},
\end{cases}
\end{equation}
where $c_\ell^\prime$ and $c_v^\prime$ are the updated densities at site $\ell$ and its neighboring site $v$.\\

(d) For $r^\prime > p_c$, pure diffusion happens from high density particle regime to low density particle regime where 
\begin{equation}
c_{\rm sum}=c_\ell-c_v,
\end{equation}
and the density update follows Eq.~\eqref{c_update1}.\\

One Monte Carlo (MC) step corresponds to $L^2$ such $\theta_\ell$ and $c_\ell$ updates.}


We define and calculate the angular frequency $\tilde\omega_0$ in NESS from the agent based model as follows:
\begin{equation} \label{ang_freq-sim}
\tilde\omega_0= g_1 \langle \cos(\delta\theta) \rangle_{L^2} + g_2 (\langle \Omega(c)\rangle_{L^2}-\Omega(c_0))
\end{equation}
where, $\Omega(c_0)=c^2_0$ is the rotation frequency in the uniform state;  $\tilde \omega_0$ gives the average rotation frequency of the spins in the laboratory frame.

Therefore, the mean rotation $\langle \theta \rangle$ yields 
\begin{align} \label{meanrotation-sim}
\langle \theta \rangle = \tilde\omega_0 t = \{g_1 \langle \cos(\delta\theta) \rangle_{L^2} + g_2 (\langle \Omega(c)\rangle_{L^2}-\Omega(c_0))\} t
\end{align}
where \begin{equation} \label{time-in-ss}
t=(t_{MC}-(t_{ss}+1))\Delta t,
\end{equation}
with $t_m \geqslant t_{MC} \geqslant t_{ss}+1$ and $\Delta t=10^{-3}$; $t_{MC}$ represents the number of Monte-Carlo step. {Here, $\Delta t$, a small number, is the time-increment after which $\langle \theta \rangle$ is measured, such that within a single Monte-Carlo step, $\langle \theta \rangle$ is measured many times. This ensures that $\langle \theta \rangle$ is measured near-continuously in time}.


\section{Results from the equilibrium limit of our model}
\label{equilibriumXYmodel}

\begin{figure*}[!hbt]
\includegraphics[width=\linewidth]{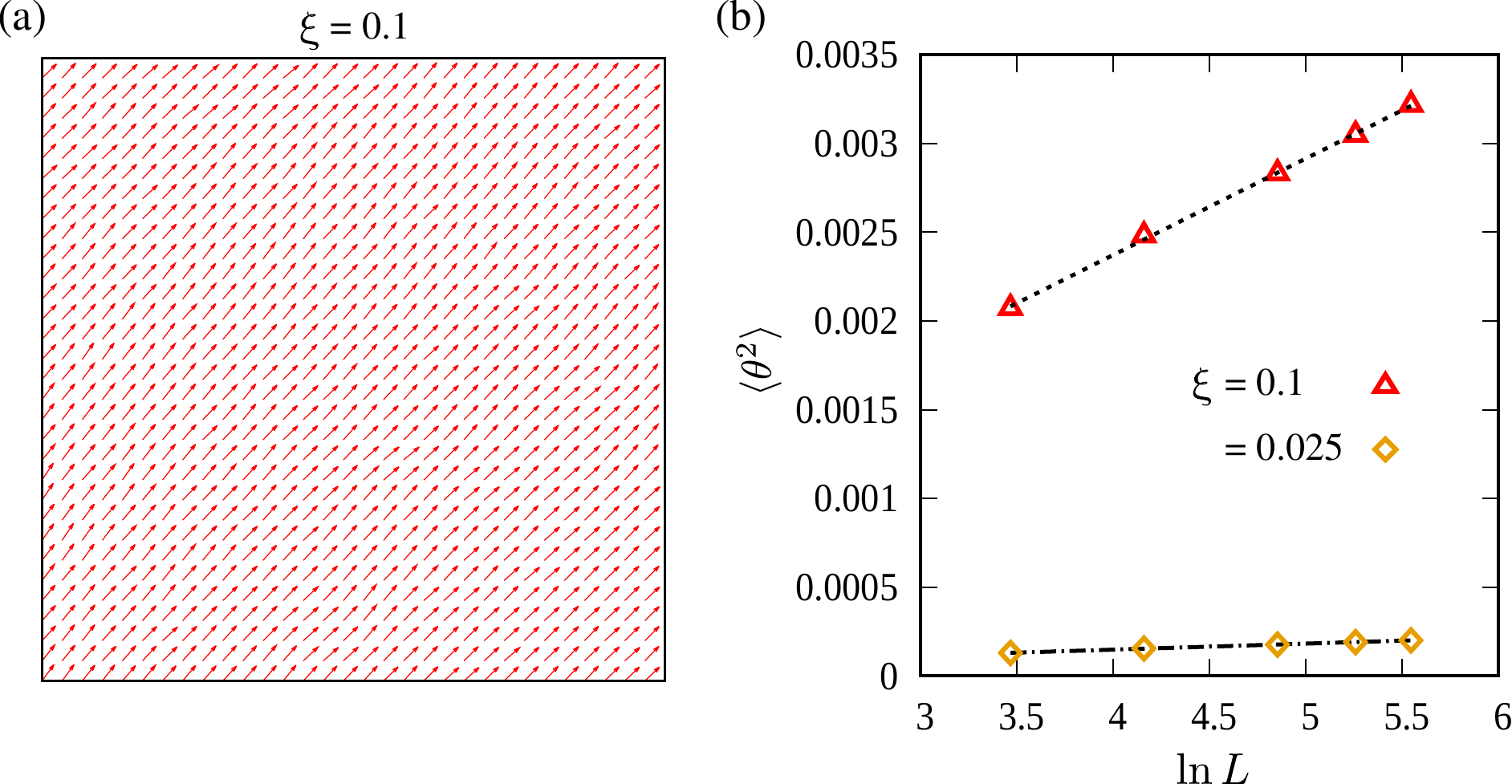}
\caption{(a) Steady-state snapshot of the spin orientations for $\xi=0.1$ ($32^2$ segment of the $256^2$ lattice is shown) in the pure 2d XY model without activity. (b) $\langle \theta^2 \rangle$ versus $\ln L$ for different noise amplitudes where a $n$ fold increase in noise corresponds to a $n^2$ fold increase in phase fluctuation. The dashed straight lines are guide to the eyes which signify the $\langle \theta^2 \rangle \sim \ln L$ behavior of the passive equilibrium XY model and corresponds to Eq.~\eqref{qlro}.}
\label{fig:PureXY}
\end{figure*}

The equilibrium picture of $2d$ XY model can be obtained from our model in the absence of any activity and uniform density. The dynamics of the phase in our agent-based model reduces to that of the $2d$ XY model in the limit $g_1=0=g_2$, i.e., when the activity is absent. We have simulated our model in this limit. The numerical results are presented in Fig.~\ref{fig:PureXY} where the steady-state spin configuration for $\xi=0.1$ is shown in Fig.~\ref{fig:PureXY}(a) and the phase fluctuation $\langle\theta^2\rangle$ as a function of system size $L$ is shown in Fig.~\ref{fig:PureXY}(b). The snapshot in (a) where a $32^2$ box of a $256^2$ lattice shows a nearly  phase-ordered state of the equilibrium  $2d$ XY model, expected  at a very low temperature ($T \ll T_{KT}$). The nature of the ordered phase is further investigated in  (b) where we see $\langle\theta^2\rangle \sim \ln L$, corresponding to QLRO and a $n^2$ fold increase in phase fluctuations for a $n$ fold increase in noise in agreement with the expectation $\langle\theta^2\rangle \propto \xi^2$. This picture is equivalent to Eq.~\eqref{qlro} with $T \equiv \xi^2$ which confirms that the nature of ordering in (a) is QLRO, equilibrium phase of the pure $2d$ XY model below the KT temperature.

At this point, we clarify a technical point. We note that unlike in Fig.~\ref{fig:PureXY}(a) obtained in the equilibrium XY limit of the model, the snapshot in Fig.~\ref{fig:DiffXYOrdering_zeta0_1}(a) (see main text) looks more noisy, although both are at the same $\xi=0.1$, and Fig.~\ref{fig:DiffXYOrdering_zeta0_1}(a) corresponds to SQLRO, an order {\em stronger than} QLRO; see Fig.~\ref{num-plot1-nofit}(a). We believe this is essentially due to the renormalization effects, because of which the effective noise in the simulation of Fig.~\ref{fig:DiffXYOrdering_zeta0_1}(a) is higher than $\xi=0.1$, which is borne out by the results $\overline D(l)\gg \overline D(0)$ for large $l$ from the hydrodynamic theory.  Nonetheless, the $\ln L$-dependence of $\langle \theta^2\rangle$ is {\em still} weaker in  SQLRO  than the pure equilibrium XY model, because the effective or renormalized damping too gets upwardly renormalized in a way to offset the upward renormalization of the noise, rendering a weaker $\ln L$-dependence of $\langle \theta^2\rangle$. Long wavelength divergence of the renormalized damping is however not reflected in a single snapshot.

\section{Insignificant noise sensitivity and  global rotation of spins in our active XY model}
\label{noise-effects-active-rot}
In order for the simulation results to correspond with the RG results obtained from the hydrodynamic equations, we need to ensure that the steady states obtained in the numerical simulations  correspond to the renormalized steady states of the hydrodynamic theory. In the equilibrium limit of our model, whose dynamics in the continuum limit is {\em linear} in the phase $\theta$, has only the noise $f_\theta$ (which represents $\xi$ in the agent-based model) in it. Obviously, doubling the amplitude of $f_\theta$ (or, $\xi$) enhances the amplitude of the phase correlator by a factor of 4; see Fig.~\ref{fig:PureXY}(b). On the other hand, in the full nonlinear model, the dynamics of $\delta c$ and $\theta$ are coupled; see Eqs.~(\ref{theta-full-eq}) and (\ref{c-full-eq}). In these equations, each of $\theta$ and $\delta c$ are affected by {\em both} $f_\theta$ and $f_c$. In addition, the nonlinearities affect the size of the fluctuations in $\theta$ and $\delta c$.  Equivalently, in the agent-based model the updates of each of $\theta$ and $ c$ at any site are affected by multiple sources of noises. Thus, doubling on $f_\theta$ or $\xi$ {\em does not} enhance the phase correlator by a factor of 4. Results from our numerical simulations are consistent with this physical picture. In Fig.~\ref{fig:DiffXYOrdering_zeta0_1Andzeta0_025}, $\langle \theta^2\rangle$ vs $\ln L$ is shown for two different noise amplitudes  $\xi = 0.025$ and $0.1$. We observe that $\langle\theta^2\rangle$ increases with $\xi$ only marginally, in contrast  to what we find in the equilibrium XY model as shown in Fig.~\ref{fig:PureXY}(b). As already discussed, this difference is attributed to the renormalization effects in a nonlinear theory as ours. Exponent $\gamma_1$ gives SQLRO for both $\xi=0.025$ and $\xi=0.1$ although further investigations with increasing $\xi$ show that the system becomes disorder beyond a certain value of $\xi$ (data not presented here).

\begin{figure}[!hbt]
 \includegraphics[width=\columnwidth]{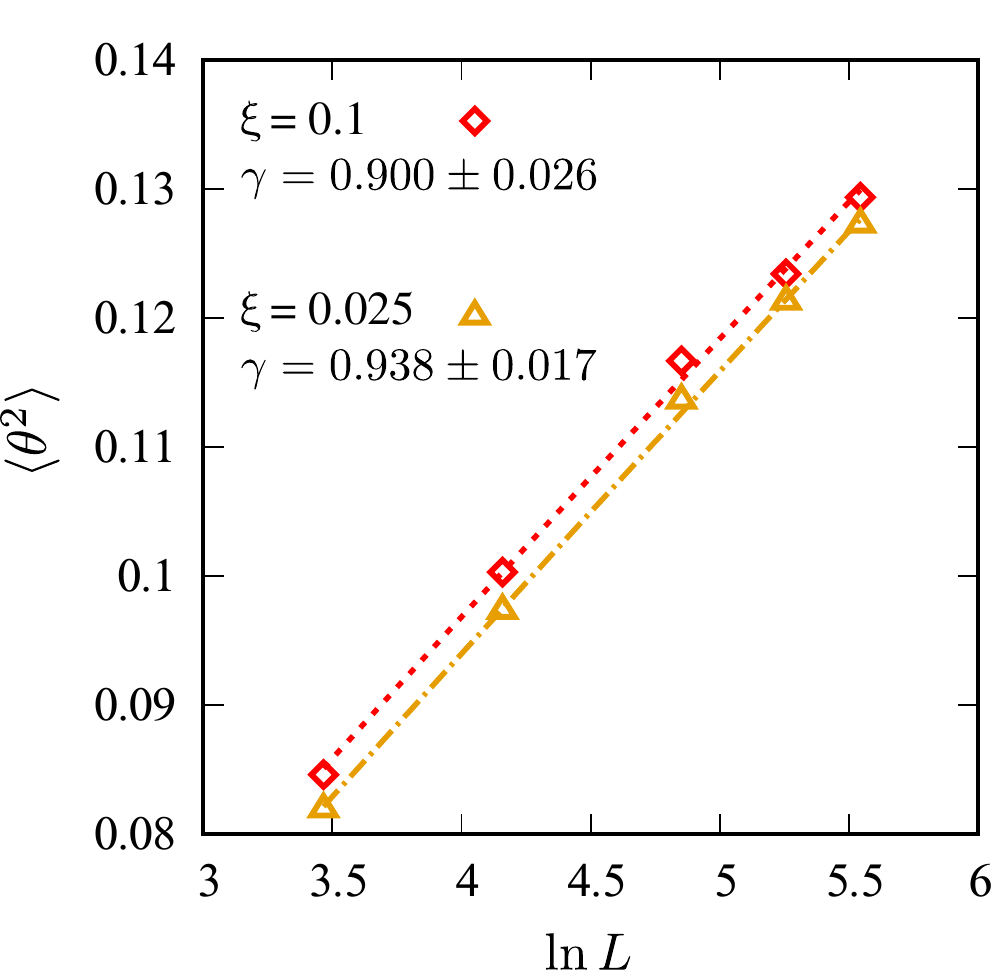}
 \caption{Variance of the phase fluctuations $\langle \theta^2 \rangle$ in NESS is plotted against $\ln L$ for two different values of the noise, $\xi=0.1$ and $0.025$ respectively. The dashed lines are the representation of the fitting function $f(x) \sim x^{\gamma_1}$ where $\gamma_1$ is the fitting exponent. Parameters: $c_0=5$, $p_\theta=p_c=0.5$, $g_1=1.0$, $g_2=0.02$, $c_s=1$, and $c_{s2}=1$.}
\label{fig:DiffXYOrdering_zeta0_1Andzeta0_025}
\end{figure}

Next, we calculate the mean rotation of the spins, $\langle \theta \rangle$, in the numerical simulations of our agent-based model with $ \langle \theta \rangle=\tilde \omega_0 t$ ($t$ is time defined in the NESS; see Eq.~\eqref{meanrotation-sim} and~\eqref{time-in-ss} in Appendix~\ref{MC-update} for precise definition). We find that in the parameter regime that exhibits the ordered NESS, the mean rotation of the spins exhibit a directional periodic rotation over the time, in agreement with the predictions made from the perturbative RG calculations in Eq.~(\ref{ang_freq}) (see main text).
Fig.~\ref{fig:rotation} shows the mean spin orientation $\langle\theta\rangle$ (in radian, $\langle\theta\rangle \in [0,2\pi], \pi \simeq 3.14$) versus $t$ for (a) anti-clockwise and (b) clockwise rotation of the spins with $\xi=0.1$ and $p_\theta=p_c=0.5$ in the ordered states. The ``phase wave'' plot in (a) and (b) are pictorial representations of the angular frequency $\tilde \omega_0\equiv  \frac{\partial\langle\theta\rangle}{\partial t}$ where a careful observation reveals that higher $g_1$ corresponds to smaller wavelength and thus larger angular frequency. Now comparing $\tilde \omega_0$ with Eq.~\eqref{ang_freq-sim} (see Appendix~\ref{MC-update}), we get $\tilde \omega_0$ to depend explicitly on $g_1$ and $g_2$, which justifies the simulation picture for a fixed magnitude of $g_2$ which shows that for a constant $\mid g_2\mid=0.01$, higher $\mid g_1 \mid$ is consistent with a larger $\tilde\omega_0$. Also note that, positive $g_1$ ($g_2$ positive) gives rise to an anti-clockwise rotation (Fig.~\ref{fig:rotation}(a)), whereas negative $g_1$ ($g_2$ negative) leads to clockwise rotation (Fig.~\ref{fig:rotation}(b)) (almost a mirror reflection of Fig.~\ref{fig:rotation}(a)) keeping the magnitude of $\langle\theta\rangle$ unaltered.  In view of Eq.~\eqref{ang_freq} (see main text) and Eq.~\eqref{ang_freq-sim} (see Appendix~\ref{MC-update}), our results are consistent with the fact that the change in the signs (or magnitudes) of the active coefficients can give rise to different values of the angular frequency $\tilde \omega_0$, hence both clockwise and anti-clockwise rotation of the spins. Here, we would like to emphasize that the clockwise and anti-clockwise rotations of the mean spin orientation which we have presented in Fig.~\ref{fig:rotation}, predominantly depend upon $g_1$ purely due to the higher magnitude of $g_1$ compared to $g_2$ in the control parameter space that defines ordered NESS. 

\begin{figure}[!t]
\includegraphics[width=\linewidth]{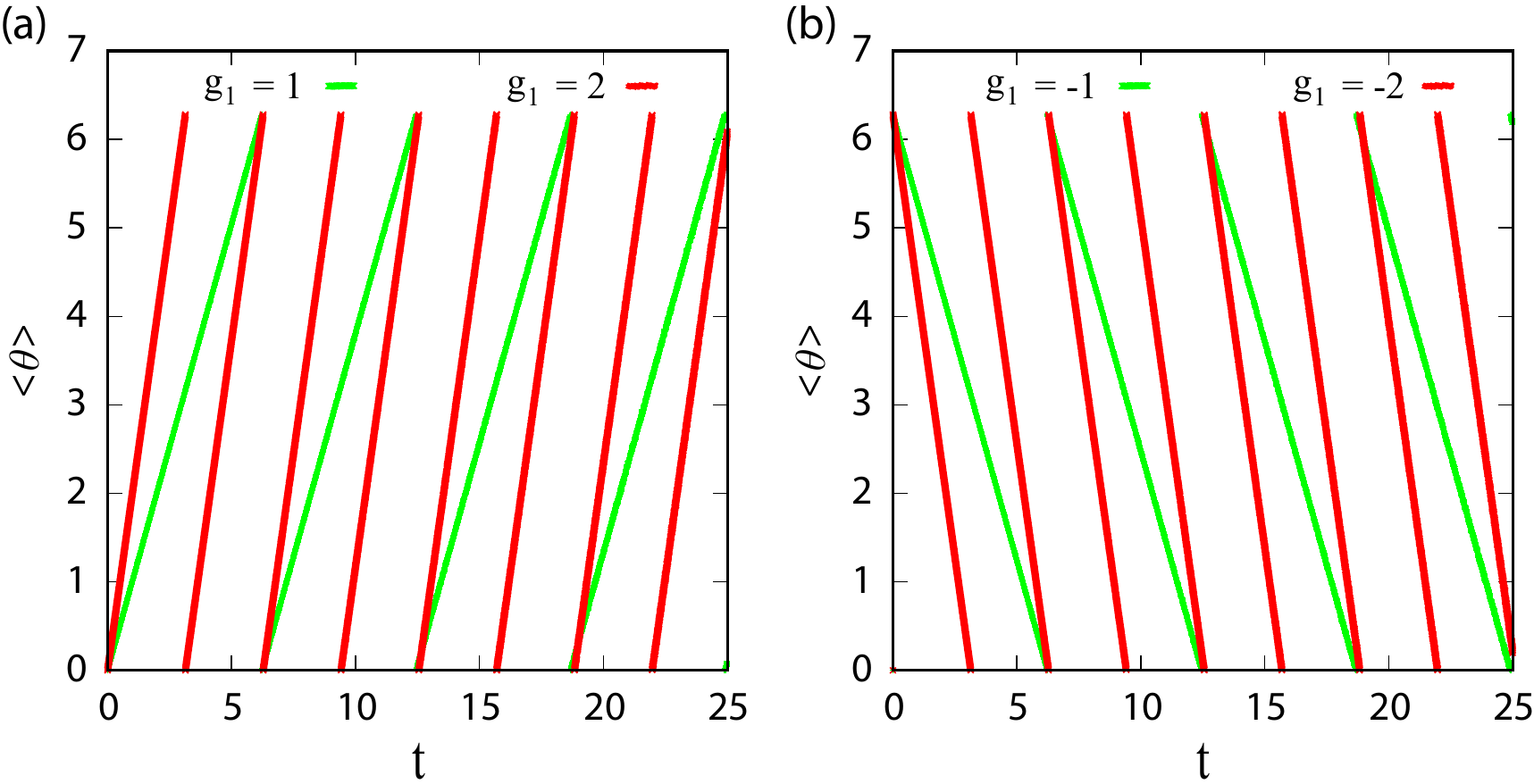}
\caption{Plots of $\langle\theta\rangle$ versus time $t$ for (a) anti-clockwise and (b) clockwise rotation of the spins in the ordered NESS, plotted for two different magnitudes of $g_1$: $|g_1|=1$, and $2$ respectively. Positive values of the coefficients $g_1$, $g_2$, $c_s$ and $c_{s2}$ lead to anti-clockwise rotation whereas negative values of the coefficients gives rise to the clockwise rotation. Parameters: $L =64$, $c_0=5$, $\xi = 0.1$, $p_\theta=p_c=0.5$, with $\mid g_2 \mid=0.01$, $\mid c_s \mid=1$, and $\mid c_{s2} \mid=1$.}
\label{fig:rotation}
\end{figure}

The global rotation of the spins is essentially a nonlinear effect, as is evident from (\ref{theta-full}) and (\ref{c-full}) in the main text. This, together with the effects of noise enhancement on the amplitude of the phase correlator clearly indicate that the ordered NESS obtained in our simulations indeed correspond to the renormalized steady states of the hydrodynamic theory. 

At the qualitative levels, we have also found from our simulations that in the ordered states, the rotation frequency $\tilde \omega_0$ that is calculated by averaging over all the $L^2$ site at a give time $t$ becomes less and less fluctuating in $t$ for bigger and bigger systems. This is a result of the fact that in the ordered states, the whole system behaves like a single ``domain'' of spins, such that the fluctuations in $\tilde \omega_0$ decreases as domain size increases. In contrast, we did not observe any reduction in the fluctuations in $\tilde \omega_0$ in the disordered aggregate states, for those states are composed of many finite size ``domains'', whose size do not increase with the system size.

\section{Irrelevance of advection in the hydrodynamic equations}\label{irre-adv}

In the above, we have assumed that the { particles} move only diffusively, ignoring any possibility of advection by a flow field on the substrate. We now show here that the effects of such an advection is irrelevant in the RG sense. To start with let us assume that the { particles} are advected by a local velocity $\bf v$. Due to the friction from the underlying substrate, there is no momentum conservation, and $\bf v$ satisfies a generalized Darcy's~\cite{darcy,darcy1} law, which is a statement of the balance of the frictional forces by the stresses arising from the phase fluctuations and density fluctuations.  Maintaining rotational symmetry and invariance under a constant shift of the phase $\theta$, and retaining up to the lowest order terms in gradients and fields, the equation of the velocity must have the form 
\begin{eqnarray}
&&\gamma^{-1} v_i({\bf x},t)=-\chi_1\nabla_ic +\chi_2\nabla_i\nabla_j(c\nabla_j\theta) \nonumber \\&+&w_1 \nabla_i(\nabla_j \theta)^2 + w_2 \nabla_j(\nabla_i\theta \nabla_j\theta) + f_{vi},\label{v-eq}
\end{eqnarray}
where stresses from the phase and number fluctuations (the $\chi_2,\,w_1,\,w_2$ terms) are included.
Here, $\chi_1,\,\chi_2,\,w_1$ and $w_2$ are coupling constants; $f_{vi}$ is a stochastic force, assumed to be a zero-mean, Gaussian white noise.

The coupled hydrodynamic equations of motion for $\theta$ and $c$ after inclusion of advection are
 \begin{eqnarray}
  &&\partial_t \theta = \lambda_\theta\left({\bf v}\cdot {\boldsymbol\nabla}\right)\theta+ \kappa \nabla^2 \theta + \frac{\lambda}{2} ({\boldsymbol\nabla} \theta)^2 + \Omega(c) + f_\theta,\nonumber\\ \label{theta-full1}\\
  &&\partial_t c = \lambda_c{\boldsymbol\nabla}\cdot\left({\bf v}c\right)+D \nabla^2 c +\lambda_0 {\boldsymbol\nabla}\cdot(\tilde \Omega(c){\boldsymbol\nabla}\theta) + f_c,\nonumber \\ \label{c-full1}
 \end{eqnarray}
  where
$\Omega(c)$ and $\tilde \Omega (c)$ are independent arbitrary functions of the number density $c$ introduced earlier in (\ref{theta-full}) and (\ref{c-full}) in the main text. Equation~(\ref{c-full1}) corresponds to a particle current 
\begin{equation}
 {\bf J}_c= - D{\boldsymbol\nabla c} - \lambda_0 \tilde \Omega (c) {\boldsymbol\nabla}\theta - \lambda_c {\bf v}c, \label{full-curr}
\end{equation}
which generalizes (\ref{c-curr}) in the presence of advection.

Notice that different coupling constants $\lambda_\theta$ and $\lambda_c$ are used in the advective terms of (\ref{theta-full1}) and (\ref{c-full1}), respectively, in an apparent violation of the Galilean invariance. However, truly there is {\em no} Galilean invariance in this model, since the frictional velocity field given by (\ref{v-eq}) has no invariance under the Galilean transformation. Furthermore, by substituting for $\bf v$ in (\ref{theta-full1}) and (\ref{c-full1}), we find that the advective nonlinear terms are actually irrelevant in the long wavelength limit. This can be easily seen by using the scaling of the fields defined in Eq.~(\ref{basic-scale}). Henceforth, we ignore these advective terms and set $\lambda_\theta=0=\lambda_c$. With this, velocity drops out of the problem completely and we get Eqs.~(\ref{theta-full}) and (\ref{c-full}) of the main text.

\section{Complex Ginzburg Landau equation}\label{app-cgle}

We here outline an alternative derivation of Eqs.~(\ref{theta-full}) and (\ref{c-full}). We start from the complex Ginzburg Landau equation (CGLE) for a non-conserved complex field or a two-component spin $Z$ coupled with a conserved density field. We define $Z\equiv Z_0({\bf r},t)\exp[i\theta({\bf r},t)]$. Amplitude $Z_0$ and phase $\theta$ are real functions of $\bf r$ and $t$. The CGLE is given by
\begin{equation}
 \partial_t Z=-\frac{\delta {\mathcal F}}{\delta Z^*} 
-i\Gamma_I(c)\frac{\delta 
{\mathcal F}}{\delta Z^*} +\Phi,\label{cgle1}
\end{equation}
where $\Gamma_I(c)$, a function of density $c$, is real, and
\begin{equation}
 {\mathcal F}=\int d^dx [\frac{\tilde r}{2}|Z|^2 +\frac{\hat g}{2}|{\boldsymbol 
\nabla}Z|^2 
+ u|Z|^4],\label{free}
\end{equation}
$\tilde r=0$ gives the mean field second order transition temperature, $\hat g>0$, $u>0$.
Here, $\Phi$ is a complex Gaussian noise with zero mean and a variance
\begin{equation}
 \langle\Phi({\bf x},t)\Phi^*(0,0)\rangle = 2D_\xi \delta ({\bf x})\delta 
(t),\,
 \langle\Phi({\bf x},t)\Phi(0,0)\rangle=0.
\end{equation}
The continuity equation for the density $c$ has the form
\begin{equation}
 \frac{\partial c}{\partial t}=-{\boldsymbol \nabla}\cdot {\bf \tilde J},
\end{equation}
where 
\begin{equation}
{\bf \tilde J}=-D{\boldsymbol\nabla}c -\frac{\lambda_0\tilde\Omega(c)}{2}{\cal I}m\left(Z{\boldsymbol \nabla} Z^* - Z^*{\boldsymbol\nabla}Z\right)+\tilde\lambda (c){\boldsymbol\nabla} \left(ZZ^*\right).
\end{equation}
We consider the ordered phase with $\tilde r<0$, i.e., $Z$ has a non-zero average magnitude. Fluctuations about this non-zero magnitude are {\em massive}, i.e., relax fast and hence these amplitude fluctuations are nonhydrodynamic variables. Thence, in the long time limit, we set $|Z|=Z_0=1$ (this can be ensured by appropriately choosing the parameters in  (\ref{free})). Setting $ZZ^*=Z_0^2=1$, we get (\ref{theta-full}) and (\ref{c-full}) above, by appropriately choosing the parameters.

\section{Field theory analysis}\label{generating-fn}

\subsection{Action functional and bare two point functions}\label{action-function}

It is convenient to recast Eqs.~(\ref{theta-full-eq}) and (\ref{c-full-eq}) as a path integral over configurations of $\theta({\bf x},t)$ and $\delta c({\bf x},t)$~\cite{janssen}. 
The corresponding generating functional is defined as 
\begin{equation}
\mathcal{Z}=\int \mathcal{D} \theta \mathcal{D} \hat\theta \mathcal{D} \delta c \mathcal{D} \hat{c} \exp \{-\mathcal{S}[\theta, \hat\theta, \delta c, \hat{c}]\},
\end{equation}
where $\mathcal{S}$ is the action functional, obtained from (\ref{theta}) and (\ref{C}):
\begin{align}
 &{\cal S} [\theta, \hat\theta, \delta c, \hat{c}]= \int_{{\bf x},t} \bigg{\{}-D_\theta \hat\theta \hat\theta + D_c \hat{c} \nabla^2 \hat{c} \nonumber \\
 & + \hat\theta \left[(\partial_t - \kappa \nabla^2) \theta - V_0 \delta c- \frac{\lambda}{2} (\nabla \theta)^2 - \Omega_2 (\delta c)^2 \right]\nonumber\\
 &+ \hat{c} \bigg{[}(\partial_t - D \nabla^2)\delta c - V_0 \nabla^2 \theta - \lambda_1 \nabla(\delta c \nabla \theta) \bigg{]}\bigg{\}}. \label{action}
\end{align}
Here, fields $\hat\theta ({\bf x},t)$ and $\hat c ({\bf x},t)$ are the response fields~\cite{janssen}. Action functional $\cal S$ contains both { harmonic and anharmonic terms}; the { anharmonic} terms are represented diagramatically in  Fig.~\ref{vertex}.

 \begin{figure}
  \includegraphics[width=0.9\columnwidth]{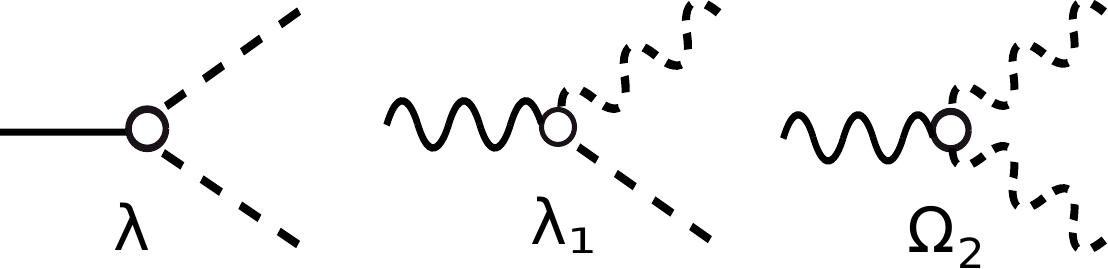}
  \caption{Diagramatic representations of the {\em anharmonic} terms in $\mathcal S$ (corresponding to the nonlinear terms in (\ref{theta-full-eq}) and (\ref{c-full-eq})).}
  \label{vertex}
 \end{figure}

\begin{figure}
  \includegraphics[width=\columnwidth]{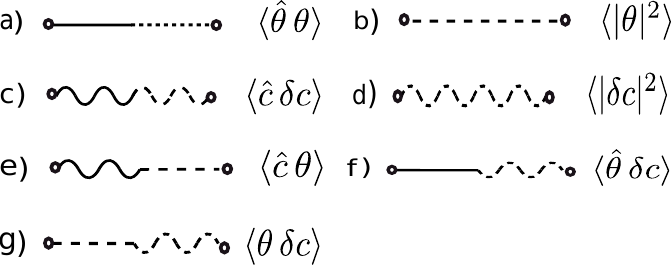}
  \caption{Diagramatic representation of two point functions.}
  \label{two_point}
 \end{figure}
 
 The two point functions are diagrammatically shown in Fig.~\ref{two_point}; the bare values  can be read off the harmonic part of \ref{action}: 
 {
 \begin{widetext}
 \begin{subequations}
 \begin{align}
    &\langle \hat\theta({\bf k},\omega)\hat\theta({\bf k^\prime},\omega^\prime) \rangle = \langle \hat c({\bf k},\omega)\hat c({\bf k^\prime},\omega^\prime) \rangle = \langle \hat\theta({\bf k},\omega) \hat c({\bf k^\prime},\omega^\prime) \rangle = 0, \\
  & \langle \hat\theta({\bf k},\omega) \theta({\bf k^\prime},\omega^\prime) \rangle = \frac{i\omega+Dk^2}{V_0^2k^2-\omega^2+i\omega k^2(\kappa +D)}~~\delta^d({\bf k+k^\prime})\delta(\omega+\omega^\prime),\\
  &\langle \hat\theta({\bf k},\omega) \delta c({\bf k^\prime},\omega^\prime) \rangle = \frac{-V_0k^2}{V_0^2k^2-\omega^2+i\omega k^2(\kappa +D)}~~\delta^d({\bf k+k^\prime})\delta(\omega+\omega^\prime).\\
 &\langle \hat c({\bf k},\omega) \theta({\bf k^\prime},\omega^\prime) \rangle = \frac{V_0}{V_0^2k^2-w^2+i\omega k^2(\kappa +D)}~~\delta^d({\bf k+k^\prime})\delta(\omega+\omega^\prime),\\
  &\langle \hat c({\bf k},\omega) \delta c({\bf k^\prime},\omega^\prime) \rangle = \frac{i\omega+\kappa k^2}{V_0^2k^2-\omega^2+i\omega k^2(\kappa +D)}~~\delta^d({\bf k+k^\prime})\delta(\omega+\omega^\prime),\\
  &\langle \theta({\bf k},\omega)\theta({\bf k^\prime},\omega^\prime) \rangle = \frac{2D_\theta(\omega^2+D^2k^4)+2V_0^2D_ck^2}{(V_0^2k^2-\omega^2)^2+\omega^2k^4(\kappa +D)^2}~~\delta^d({\bf k+k^\prime})\delta(\omega+\omega^\prime),\\
  &\langle \theta({\bf k},\omega) \delta c({\bf k^\prime},\omega^\prime) \rangle = \frac{2V_0k^2\left[-D_\theta(-i\omega+Dk^2)+D_c(i\omega+\kappa k^2)\right]}{(V_0^2k^2-\omega^2)^2+\omega^2k^4(\kappa +D)^2}~~\delta^d({\bf k+k^\prime})\delta(\omega+\omega^\prime) ,\\
 &\langle \delta c({\bf k},\omega)\delta c({\bf k^\prime},\omega^\prime) \rangle = \frac{2D_\theta V_0^2k^4+2D_ck^2(\omega^2+\kappa^2k^4)}{(V_0^2k^2-\omega^2)^2+\omega^2k^4(\kappa +D)^2}~~\delta^d({\bf k+k^\prime})\delta(\omega+\omega^\prime).
  \end{align}
  \end{subequations}
\end{widetext}

We can then define $ \langle \theta({\bf k},\omega)\theta({\bf k^\prime},\omega^\prime) \rangle \equiv \langle \theta(-{\bf k},-\omega) \theta({\bf k},\omega) \rangle = \langle| \theta({\bf k},\omega)^2| \rangle\,\delta^d({\bf k+k^\prime})\delta(\omega+\omega^\prime)$, which allows us to identify
\begin{equation}
 \langle | \theta({\bf k},\omega)^2| \rangle = \frac{2D_\theta(\omega^2+D^2k^4)+2V_0^2D_ck^2}{(V_0^2k^2-\omega^2)^2+\omega^2k^4(\kappa +D)^2}.
\end{equation}
 All the other two point functions $\langle \hat\theta ({\bf -k},-\omega)\theta({\bf k},\omega)\rangle$ etc can be obtained similarly.
}
  
 \subsection{Equal-time correlation functions in the Gaussian theory}\label{same-time-corr}
 
 { In this Section, we give the details of the results from the linearized equations of motion, which can equivalently be calculated from the harmonic part of the action functional (\ref{action}).}

{
\begin{eqnarray}
 &&\langle |\theta_{{\bf k},t}|^2\rangle = \int^\infty_{-\infty} \frac{d\omega}{2\pi} \langle |\theta_{{\bf k},\omega}|^2\rangle \nonumber\\
 &&= \int_{-\infty}^\infty \frac{d\omega}{2\pi} \frac{2D_\theta(\omega^2+D^2k^4)+2V_0^2D_ck^2}{(V_0^2k^2-\omega^2)^2+\omega^2k^4(\kappa +D)^2}\label{theta-cor-time}
 \end{eqnarray}
 
  \begin{figure}
  \includegraphics[width=0.8\columnwidth]{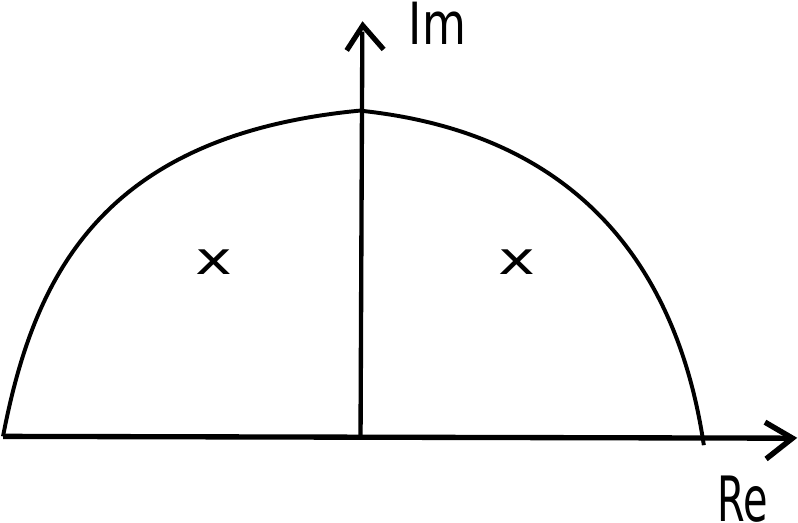}
  \caption{Contour chosen in the upper half plane for the frequency integral in the considered complex plane; two cross signs represent the two poles included in the contour (see text).}\label{omega-int}
 \end{figure}
 
 The denominator of (\ref{theta-cor-time}) can be written as $(\omega-V_0 k - i\Gamma k^2)(\omega +V_0 k - i\Gamma k^2)(\omega-V_0 k + i\Gamma k^2)(\omega +V_0 k + i\Gamma k^2)$ for small $k$. This gives four first order poles at $\omega=\pm V_0 k \pm i\Gamma k^2$. Considering a closed contour in the upper half of a complex plane, noting that the relevant poles included inside the contour are $\omega = \pm V_0 k + i\Gamma k^2$} and the contribution from the semicircle at infinity vanishes; see Fig.~\ref{omega-int}. Application of the residue theorem gives
 
 \begin{eqnarray}
 &&\langle |\theta_{{\bf k},t}|^2\rangle\nonumber \\&&=2\pi i[{\rm residue}(V_0 k + i\Gamma k^2) + {\rm residue}(-V_0 k + i\Gamma k^2)]\nonumber \\&&= 
2\pi \,\left[\frac{2V_0k^2(D_\theta+D_c)}{(2V_0k)^2 2\Gamma k^2}+\frac{2V_0k^2(D_\theta+D_c)}{(-2V_0k)^2 2\Gamma k^2}\right]\nonumber\\ &&=\frac{D_\theta+D_c}{2\Gamma k^2}.
\end{eqnarray}
in the limit of small $k$; we keep the term with highest power of $k$ in denominator.

 { Other equal-time correlators can be evaluated in the same manner, which are given by
\begin{eqnarray}
 \langle |\delta c_{{\bf k},t}|^2\rangle=  \frac{D_\theta+D_c}{2\Gamma }.
\end{eqnarray}
\begin{eqnarray}
 && \langle \theta_{-{\bf k},t} \delta c_{{\bf k},t}\rangle=  \frac{D_\theta+D_c}{2\Gamma V_0} - \frac{D_\theta+D_c}{2V_0}, \nonumber\\
 && \langle \theta_{{\bf k},t} \delta c_{-{\bf k},t}\rangle=  \frac{D_\theta+D_c}{2\Gamma V_0} + \frac{D_\theta+D_c}{2V_0}.
\end{eqnarray}
}
 

\section{Perturbation theory: Renormalization Group analysis}\label{perturbative}

We discuss the details of the perturbative RG on Eqs.~(\ref{theta}) and (\ref{C}) here.
 
\subsection{One-loop corrections}\label{one-loop}
  
  We integrate the higher wavevector parts of the fields (with wavevectors in the range $\Lambda/b$ and $\Lambda$). This cannot be done exactly, and, therefore, perturbative methods are needed. To that end,  we expand the anhamornic terms present in $\cal S$. The {\em fluctuation-corrected action} ${\cal S}^<[\theta^<, \hat\theta^<, \delta c^<, \hat c^<]$ contains the corrected model parameters. The fluctuation corrections to the model parameters are represented by the Feynman diagrams. We confine ourselves to obtain the fluctuation-corrections to the various model parameters in (\ref{action}) up to the one-loop order only. The one-loop Feynman diagrams are shown in Fig.~\ref{Dtheta-correction}, Fig.~\ref{Dc-correction}, Fig.~\ref{kappa-correction}, Fig. \ref{D-correction}, Fig.~\ref{V0-correction}, Fig.~\ref{V0_1-correction}, Fig.~\ref{lambda-correction}, Fig.~\ref{lambda_1-correction} and Fig.~\ref{Omega_2-correction}. We present the values of the one-loop diagrams below and write the {\em fluctuation-corrected} model parameters. 
 
 \begin{figure}
  \includegraphics[width=0.8\columnwidth]{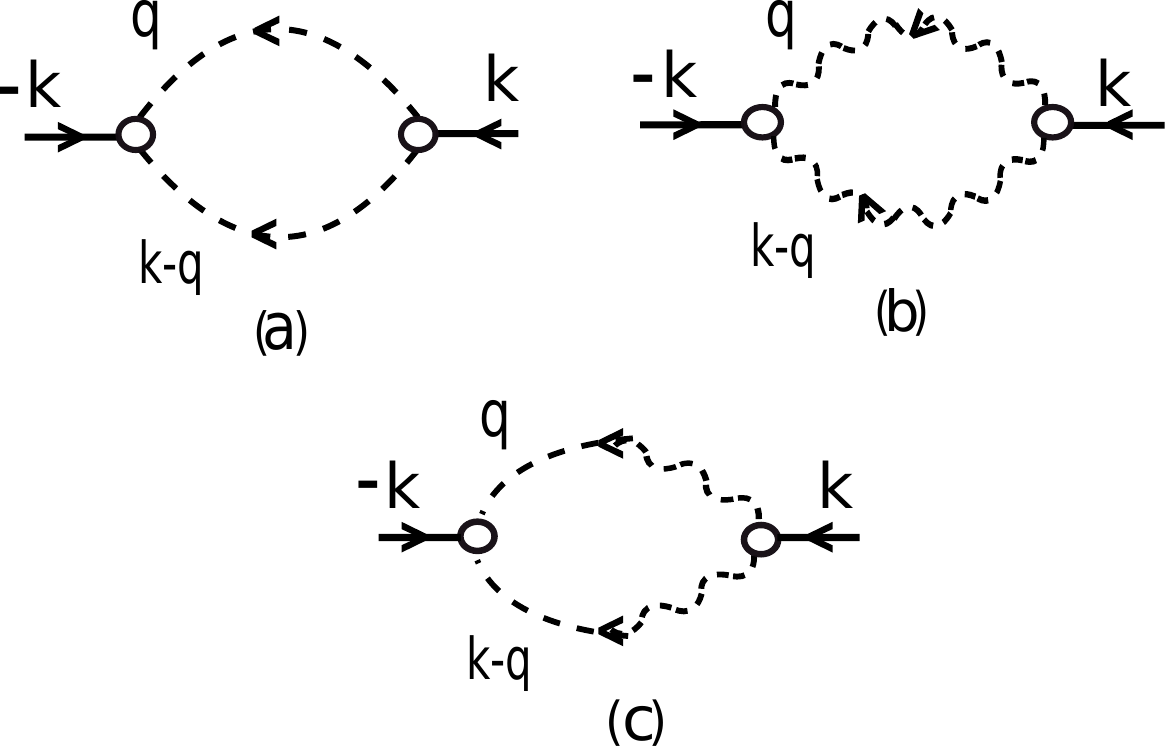}
  \caption{One-loop corrections for $D_\theta$ due to the anhamornic terms in $\cal S$.}
  \label{Dtheta-correction}

  \includegraphics[width=0.8\columnwidth]{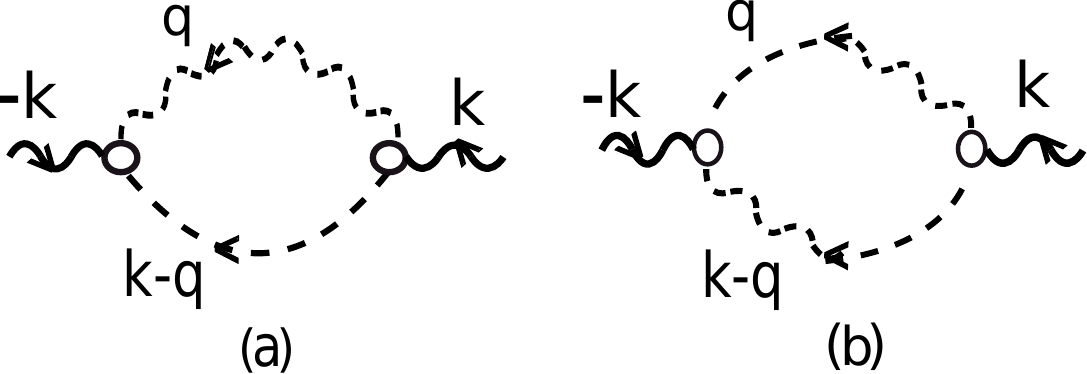}
  \caption{One-loop corrections for $D_c$ due to the anhamornic terms in $\cal S$.}
  \label{Dc-correction}
 \end{figure}

 \begin{subequations}
  \begin{align}
 &\text{Fig (\ref{Dtheta-correction}a)}=\frac{\lambda^2 (D_\theta +D_c)^2}{32\Gamma^3} \int^{\varLambda}_{\varLambda/b} \frac{d^2q}{(2\pi)^2} \frac{1}{q^2}.\\
 & \text{Fig (\ref{Dtheta-correction}b)}=\frac{\Omega_2^2 (D_\theta +D_c)^2}{8\Gamma^3} \int^{\varLambda}_{\varLambda/b} \frac{d^2q}{(2\pi)^2} \frac{1}{q^2}.\\
 & \text{Fig (\ref{Dtheta-correction}c)}=\frac{\lambda \Omega_2 (D_\theta +D_c)^2}{8\Gamma^3} \int^{\varLambda}_{\varLambda/b} \frac{d^2q}{(2\pi)^2} \frac{1}{q^2}.
  \end{align}

 \end{subequations}

 \begin{subequations}
  \begin{align}
    &\text{Fig (\ref{Dc-correction}a)}=\frac{\lambda_1^2 (D_\theta +D_c)^2}{32\Gamma^3} \int^{\varLambda}_{\varLambda/b} \frac{d^2q}{(2\pi)^2} \frac{1}{q^2}.\\
  &\text{Fig (\ref{Dc-correction}b)}=\frac{\lambda_1^2 (D_\theta +D_c)^2}{32\Gamma^3} \int^{\varLambda}_{\varLambda/b} \frac{d^2q}{(2\pi)^2} \frac{1}{q^2}.
  \end{align}

 \end{subequations}

  \begin{figure}
  \includegraphics[width=7cm]{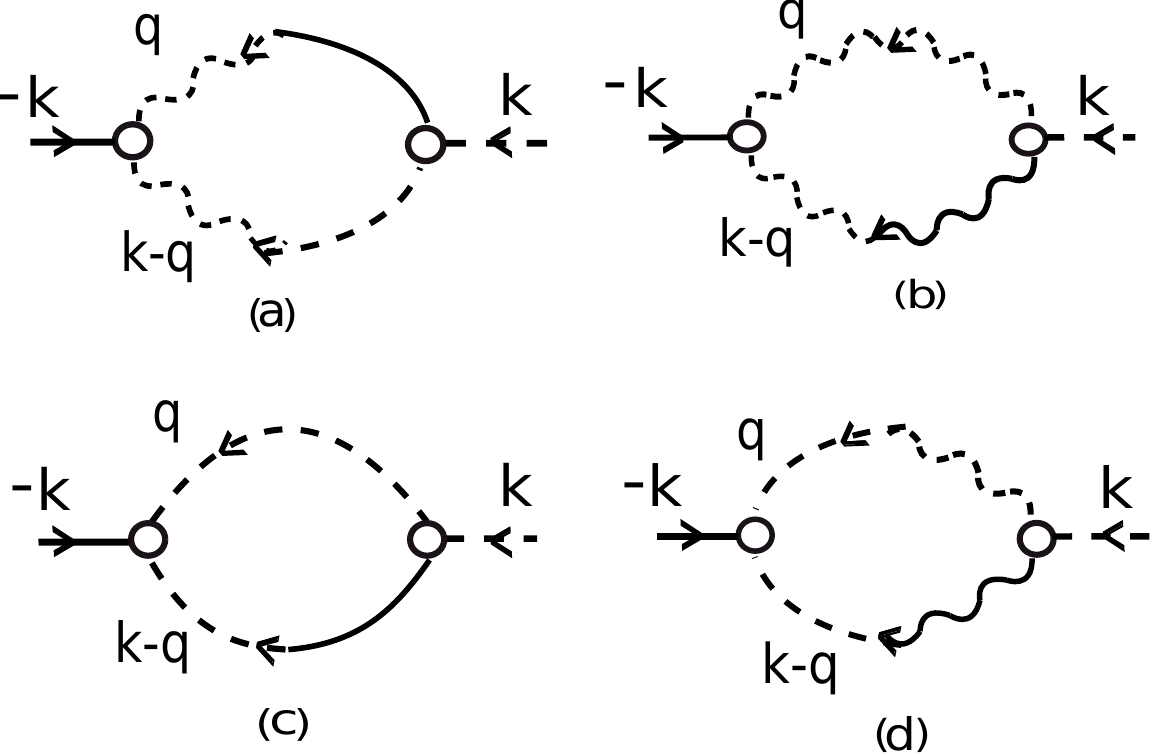} 
  \caption{One-loop corrections for $\kappa$ due to the anhamornic terms in $\cal S$.}  \label{kappa-correction}
   \end{figure}

   \begin{figure}
  \includegraphics[width=7cm]{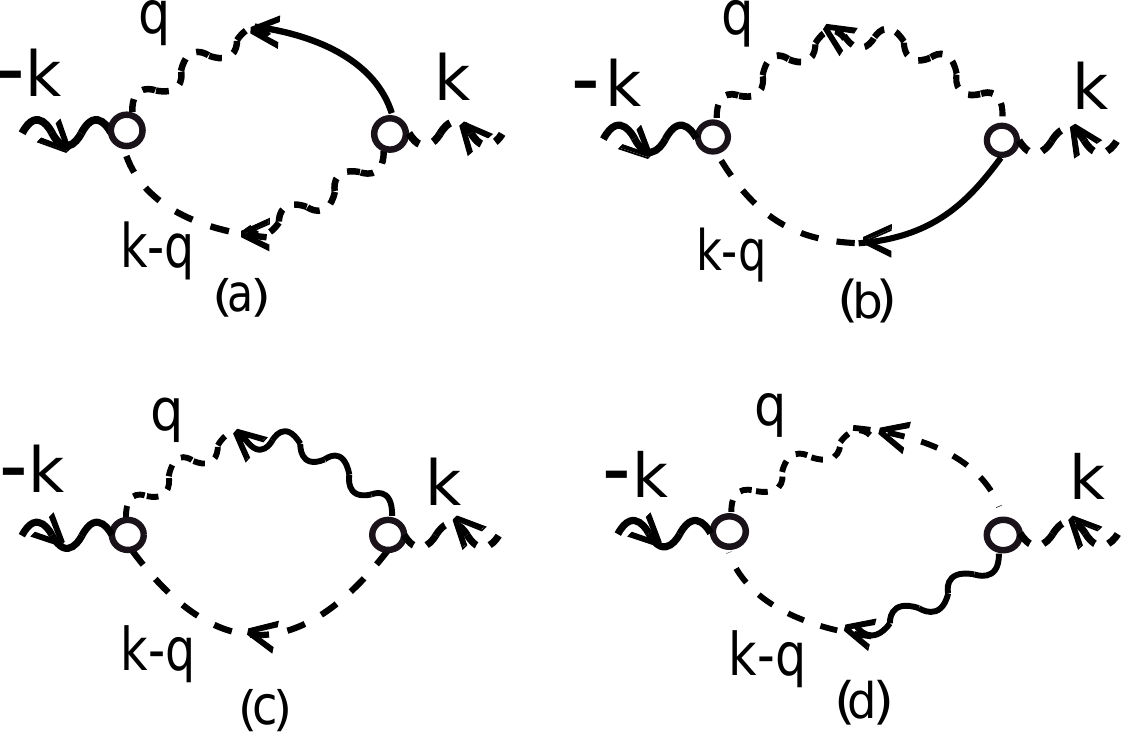}
  \caption{One-loop corrections for $D$ due to the anhamornic terms in $\cal S$.}
  \label{D-correction}
 \end{figure}

 \begin{subequations}
\begin{align}
 &\text{Fig (\ref{kappa-correction}a)}= \frac{\lambda \Omega_2}{8\Gamma^2}\left[-(D_\theta+D_c)+ \frac{(-D_\theta D + D_c \kappa)}{2\Gamma}\right]\nonumber\\
 &~~~~~~~~~~~~~~~~~~~~~~~\int^{\varLambda}_{\varLambda/b} \frac{d^2q}{(2\pi)^2} \frac{1}{q^2}.\\
 & \text{Fig (\ref{kappa-correction}b)}= -\frac{\lambda \Omega_2 (D_\theta +D_c)}{32\Gamma^2}\left[1+\frac{\Gamma+\kappa}{\Gamma}\right] \int^{\varLambda}_{\varLambda/b} \frac{d^2q}{(2\pi)^2} \frac{1}{q^2}.\\
  &\text{Fig (\ref{kappa-correction}c)}= \frac{\lambda^2 (D_\theta +D_c)}{32\Gamma^2}\left[1-\frac{\Gamma+D}{\Gamma}\right]\int^{\varLambda}_{\varLambda/b} \frac{d^2q}{(2\pi)^2} \frac{1}{q^2}.\\
  &\text{Fig (\ref{kappa-correction}d)}= -\frac{\lambda^2 }{16\Gamma^2}\left[(D_\theta +D_c)+\frac{-D_\theta D+ D_c\kappa}{4\Gamma}\right]\nonumber\\
  &~~~~~~~~~~~~~~~~~~~~~~~\int^{\varLambda}_{\varLambda/b} \frac{d^2q}{(2\pi)^2} \frac{1}{q^2}.
   \end{align}
   \end{subequations}

  \begin{subequations}
  \begin{align}
  &\text{Fig (\ref{D-correction}a)}= -\frac{\lambda \Omega_2}{16\Gamma^2}\left[(D_\theta+D_c)+ \frac{(-D_\theta D + D_c \kappa)}{2\Gamma}\right]\nonumber\\
  &~~~~~~~~~~~~~~~~~~~\int^{\varLambda}_{\varLambda/b} \frac{d^2q}{(2\pi)^2} \frac{1}{q^2}.\\
  &\text{Fig (\ref{D-correction}b)}= -\frac{\lambda \Omega_2 (D_\theta +D_c)}{32\Gamma^2}\left[1+\frac{\Gamma+D}{\Gamma}\right] \int^{\varLambda}_{\varLambda/b} \frac{d^2q}{(2\pi)^2} \frac{1}{q^2}.\\
  &\text{Fig (\ref{D-correction}c)}= \frac{\lambda^2 (D_\theta +D_c)}{128\Gamma^2}\left[1-\frac{\Gamma+\kappa}{\Gamma}\right]\int^{\varLambda}_{\varLambda/b} \frac{d^2q}{(2\pi)^2} \frac{1}{q^2}.\\
  &\text{Fig (\ref{D-correction}d)}= \frac{\lambda^2 }{32\Gamma^2}\left[-(D_\theta +D_c)+\frac{-D_\theta D+ D_c\kappa}{4\Gamma}\right]\nonumber\\
  &~~~~~~~~~~~~~~~~~~~~\int^{\varLambda}_{\varLambda/b} \frac{d^2q}{(2\pi)^2} \frac{1}{q^2}.
  \end{align}
  \end{subequations}
  
 
 \begin{figure}
  \includegraphics[width=8cm]{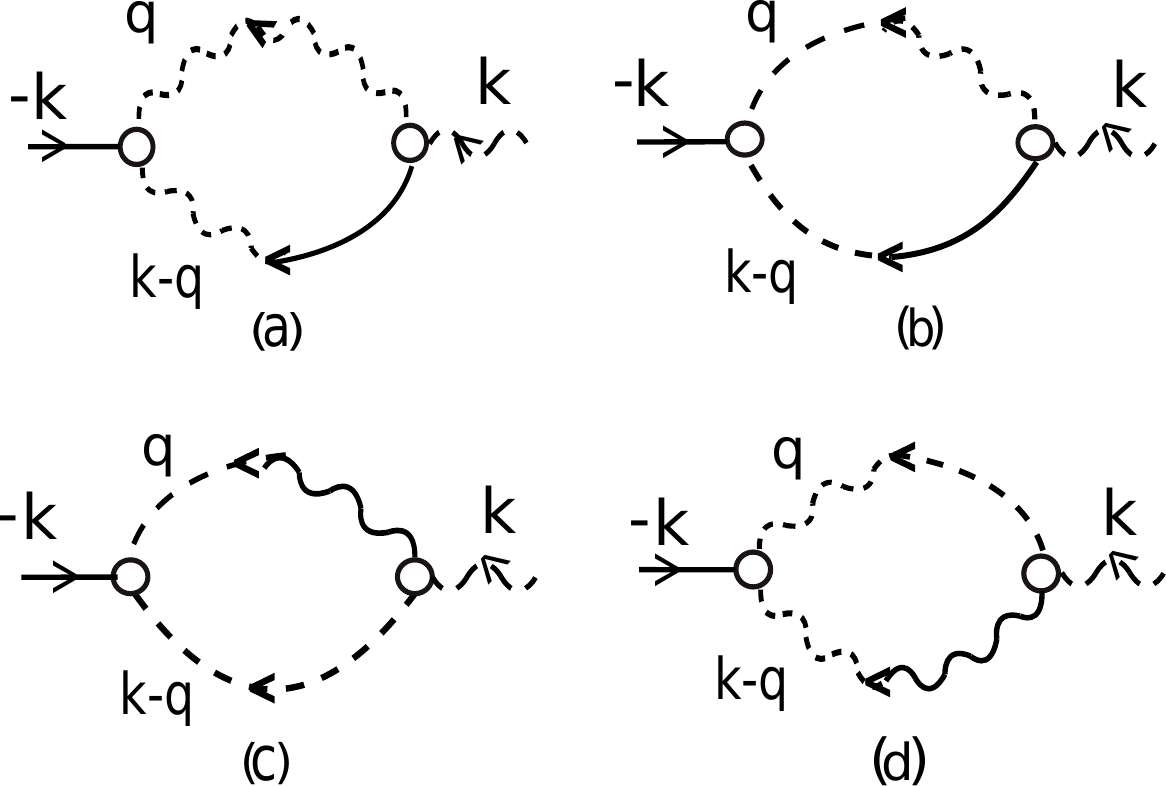}
  \caption{One loop diagrams for correction of $V_0$ parameter (the coefficient of the term $\int_{{\bf x},t}\hat\theta \delta c$ in $\cal S$) due to the anhamornic terms in $\cal S$.}
  \label{V0-correction}

  \includegraphics[width=8cm]{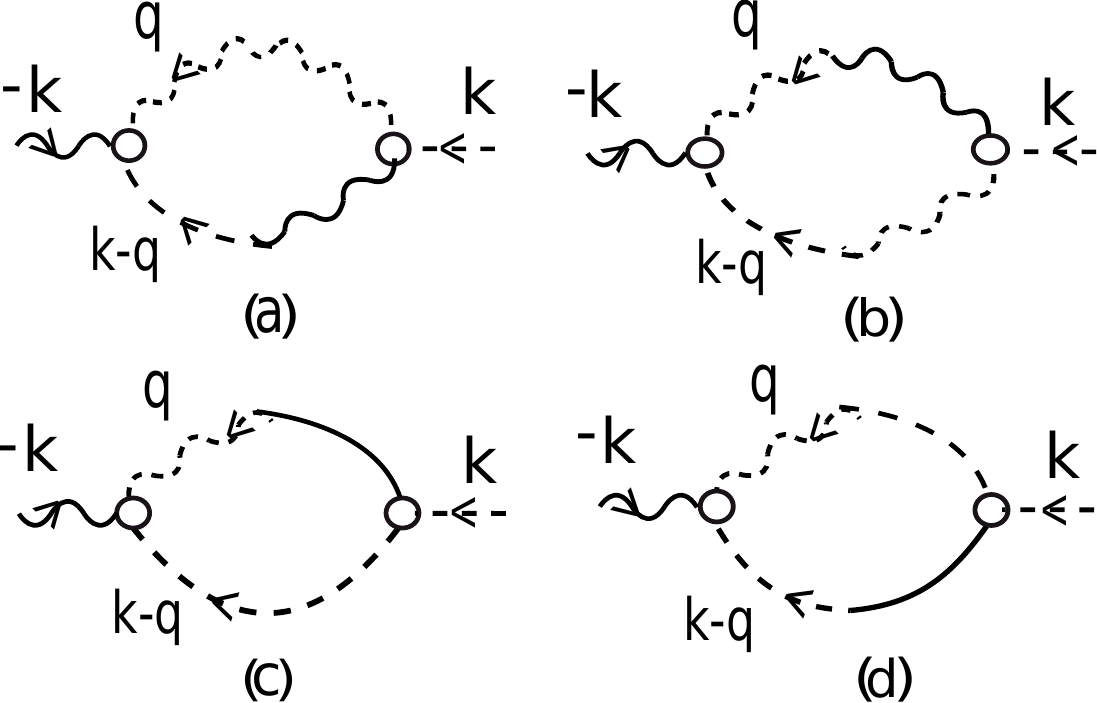}
  \caption{One loop diagrams for correction of $V_0$ parameter (which is the coefficient of term  $\int_{{\bf x},t} \hat c \nabla^2 \theta$) due to the anharmonic terms in $\cal S$.}
  \label{V0_1-correction}
 \end{figure}
 Each diagram in Fig.~\ref{V0-correction} and Fig.~\ref{V0_1-correction}  contribute individually as subleading corrections. Therefore, the parameter $V_0$ in (\ref{action}) will not be fluctuation-corrected.

\begin{figure}
  \includegraphics[width=0.9\columnwidth]{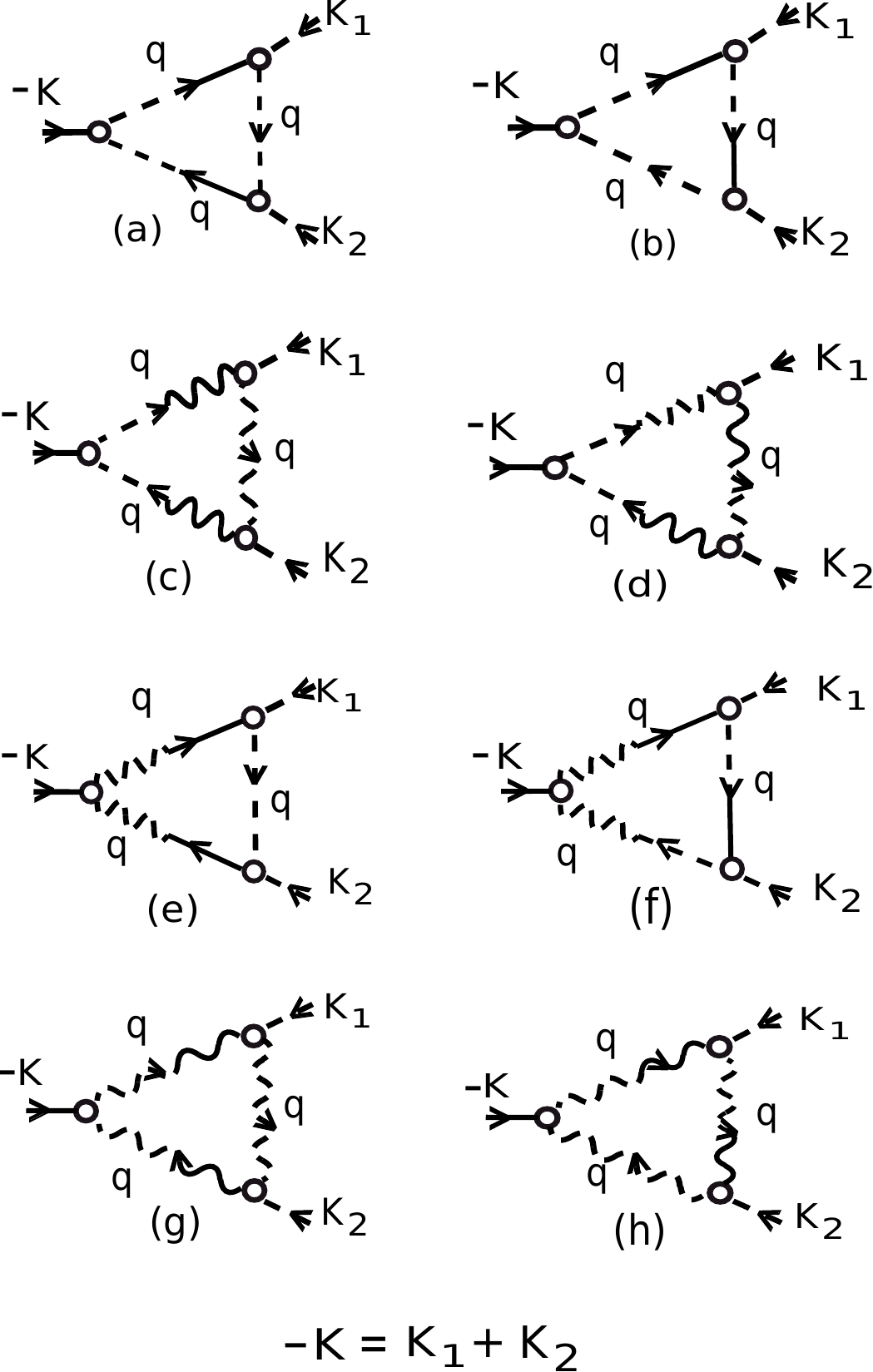}
  \caption{One loop corrections for $\lambda$ due to the anharmonic terms in (\ref{action}).}
  \label{lambda-correction}
 \end{figure}
 
 { The diagrams in Fig.~\ref{lambda-correction} formally correct $\lambda$, but their combined leading order contributions actually vanish. For instance:\\
 Diagrams with three $\lambda$ vertices $\implies$ $\text{Fig.~\ref{lambda-correction}(a)}+\text{Fig.~\ref{lambda-correction}(b)}\rightarrow$ no relevant correction.\\
 Diagrams with one $\lambda$ vertex and two $\lambda_1$ vertices $\implies$ $\text{Fig.~\ref{lambda-correction}(c)}+\text{Fig.~\ref{lambda-correction}(d)}\rightarrow$ no relevant correction.\\
 Diagrams with one $\Omega_2$ vertex and two $\lambda$ vertices $\implies$ $\text{Fig.~\ref{lambda-correction}(e)}+\text{Fig.~\ref{lambda-correction}(f)}\rightarrow $ no relevant correction.\\
 Diagrams with one $\Omega_2$ vertex and two $\lambda_1$ vertices $\implies$ $\text{Fig.~\ref{lambda-correction}(a)}+\text{Fig.~\ref{lambda-correction}(b)}\rightarrow $ no relevant correction.
 }
 
 \begin{figure}
  \includegraphics[width=0.9\columnwidth]{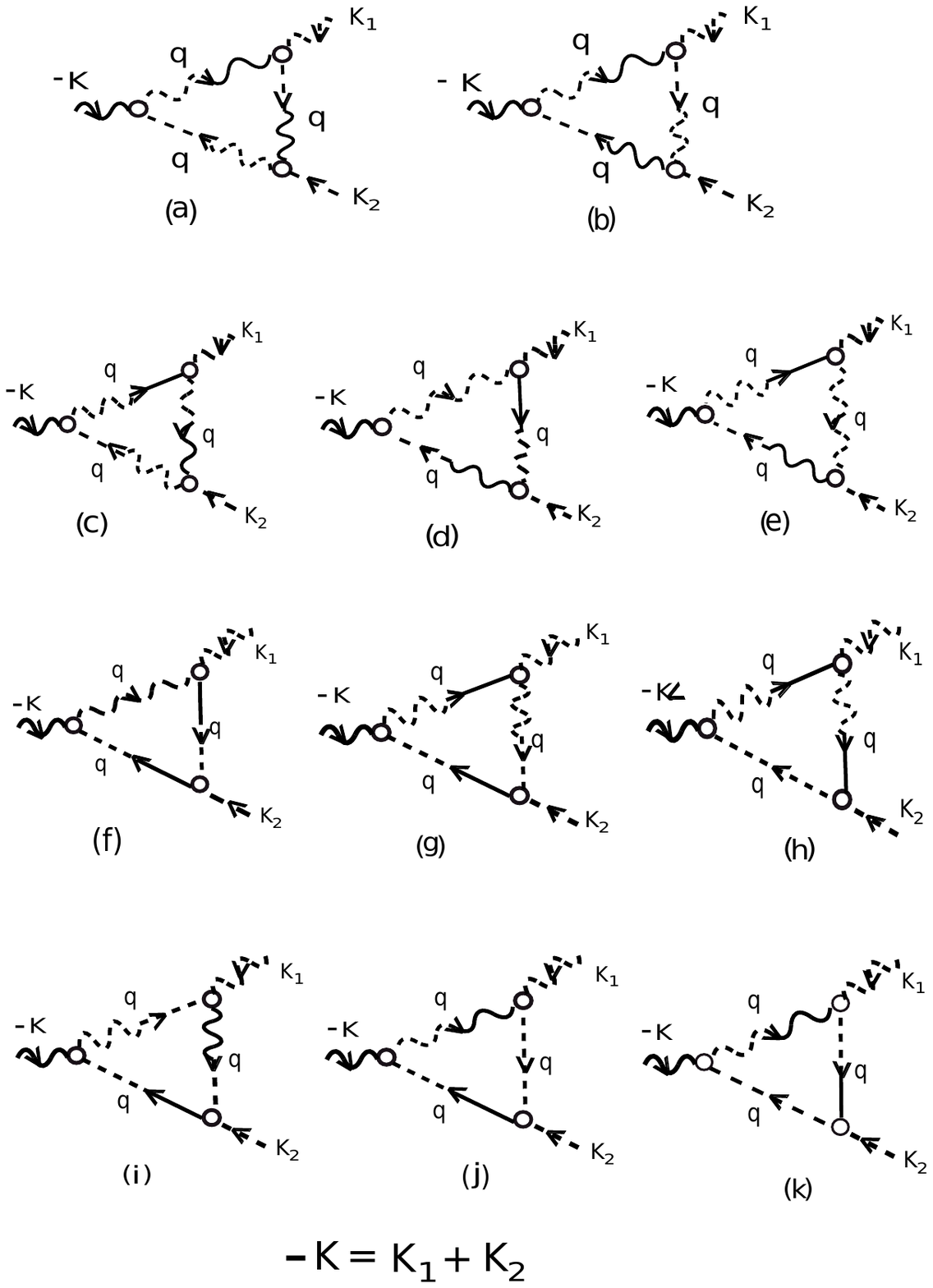}
  \caption{One loop corrections for $\lambda_1$ due to the anharmonic terms in (\ref{action}).}
  \label{lambda_1-correction}
 \end{figure}
 
   The diagrams in Fig.~\ref{lambda_1-correction} formally correct $\lambda_1$, but do not give any relevant correction. For instance:\\
 Diagrams with three $\lambda_1$ vertices $\implies$ $\text{Fig.~\ref{lambda_1-correction}(a)}+\text{Fig.~\ref{lambda_1-correction}(b)}\rightarrow$ no relevant correction.\\
 Diagrams with one $\Omega_2$ vertex and two $\lambda_1$ vertices $\implies$ $\text{Fig.~\ref{lambda_1-correction}(c)}+\text{Fig.~\ref{lambda_1-correction}(d)}+\text{Fig.~\ref{lambda_1-correction}(e)}\rightarrow$ no relevant correction.\\
 Diagrams with one $\Omega_2$, one $\lambda$, and one $\lambda$ vertices $\implies$ $\text{Fig.~\ref{lambda_1-correction}(f)}+\text{Fig.~\ref{lambda_1-correction}(g)}+\text{Fig.~\ref{lambda_1-correction}(h)}\rightarrow$ no relevant correction.\\
 Diagrams with one $\lambda$ vertex and two $\lambda_1$ vertices $\implies$ $\text{Fig.~\ref{lambda-correction}(i)}+\text{Fig.~\ref{lambda_1-correction}(j)}+\text{Fig.~\ref{lambda_1-correction}(k)}\rightarrow$ no relevant correction.

 \begin{figure}
  \includegraphics[width=0.9\columnwidth]{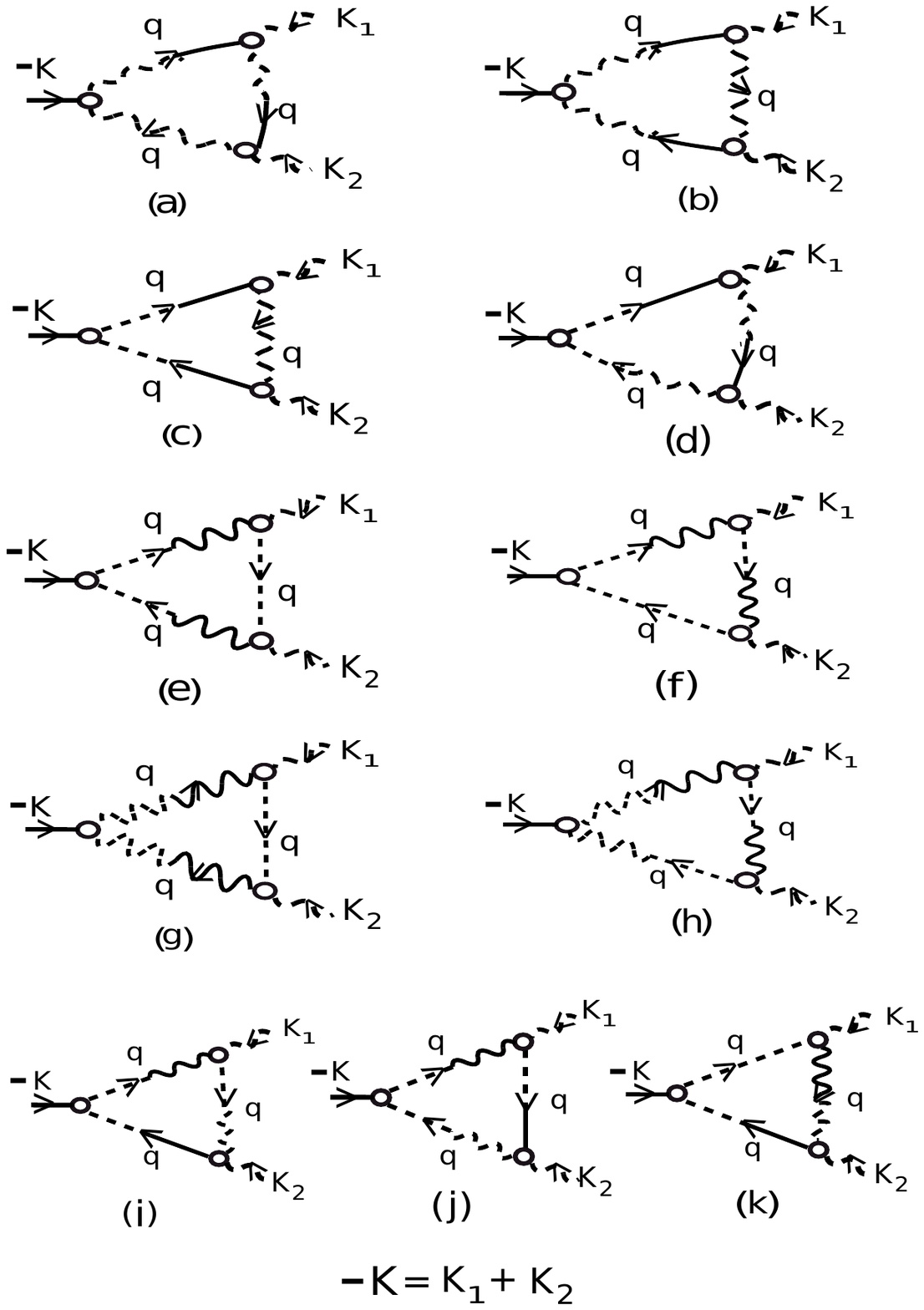}
  \caption{One loop corrections for $\Omega_2$ due to the anharmonic terms in (\ref{action}).}
  \label{Omega_2-correction}
 \end{figure}

   The diagrams in Fig.~\ref{Omega_2-correction} formally correct $\Omega_2$, but actually do not produce any relevant correction. For instance:\\
 Diagrams with three $\Omega_2$ vertices $\implies$ $\text{Fig.~\ref{Omega_2-correction}(a)}+\text{Fig.~\ref{Omega_2-correction}(b)}\rightarrow$ no relevant correction.\\
 Diagrams with one $\lambda$ vertex and two $\Omega_2$ vertices $\implies$ $\text{Fig.~\ref{Omega_2-correction}(c)}+\text{Fig.~\ref{Omega_2-correction}(d)}\rightarrow$ no relevant correction.\\
 Diagrams with one $\lambda$ vertex and two $\lambda_1$ vertices $\implies$ $\text{Fig.~\ref{Omega_2-correction}(e)}+\text{Fig.~\ref{Omega_2-correction}(f)}\rightarrow$ no relevant correction.\\
 Diagrams with one $\Omega_2$ vertex and two $\lambda_1$ vertices $\implies$ $\text{Fig.~\ref{Omega_2-correction}(g)}+\text{Fig.~\ref{Omega_2-correction}(h)}\rightarrow$ no relevant correction.\\
 Diagrams with one $\lambda$, one $\lambda_1$ and one $\Omega_2$ vertices $\implies$ $\text{Fig.~\ref{Omega_2-correction}(i)}+\text{Fig.~\ref{Omega_2-correction}(j)}+\text{Fig.~\ref{Omega_2-correction}(k)}\rightarrow$ no relevant correction.

The fluctuation-corrected model parameters are
\begin{subequations}
 \begin{align}
  V_0^<&=V_0,\\
  D_\theta^<&= D_\theta\bigg[ 1+\frac{\lambda^2(D_\theta+D_c)^2}{32D_\theta\Gamma^3}\bigg(1+4\mu_1^2+4\mu_1 \bigg)\nonumber\\
  &~~~~~~~~\times\int_{\varLambda/b}^\varLambda \frac{d^2q}{(2\pi)^2} \frac{1}{q^2} \bigg], \\
  D_c^<&= D_c \bigg[ 1+\frac{\lambda^2(D_\theta+D_c)^2}{16D_c\Gamma^3}\mu_2^2 \int_{\varLambda/b}^\varLambda \frac{d^2q}{(2\pi)^2} \frac{1}{q^2} \bigg],\\
  D^< &= D+\bigg[\frac{\lambda^2(D_\theta+D_c)}{8\Gamma^2}\bigg\{\frac{\mu_1\mu_2}{4}\bigg(9-\frac{\kappa-D}{\kappa+D}\bigg)\nonumber\\
  &+\frac{\mu_2^2}{8}\bigg(9+\frac{\kappa-D}{\kappa+D}\bigg)\bigg\} \nonumber\\
  & + \frac{\lambda^2(D_c \kappa-D_\theta D)}{16\Gamma^3}  \bigg(\mu_1\mu_2-\mu_2^2/2\bigg)\bigg]  \int_{\varLambda/b}^\varLambda \frac{d^2q}{(2\pi)^2} \frac{1}{q^2},\\
 \kappa^<&= \kappa+\bigg[\frac{\lambda^2(D_\theta+D_c)}{8\Gamma^2}\bigg\{ \frac{\mu_1\mu_2}{4}\bigg(5+\frac{\kappa-D}{\kappa+D}\bigg)\nonumber\\
 &+\mu_1+\mu_2 -\frac{\kappa+D}{8(\kappa+D)}+\frac{1}{8}\bigg\} \nonumber\\
 &+ \frac{D_c\kappa-D_\theta D}{16\Gamma^3}\bigg(\mu_2/2-\mu_1\bigg)\bigg] 
 \int_{\varLambda/b}^\varLambda \frac{d^2q}{(2\pi)^2} \frac{1}{q^2},\\
 \lambda^<&=\lambda,\\
 \Omega_2^<&=\Omega_2,\\
 \lambda_1^<&=\lambda_1.
 \end{align}
\end{subequations}



\subsection{Rescaling}\label{rescale}
 We scale momentum ${\bf k}\rightarrow b{\bf k}$ and frequency $\omega \rightarrow b^z\omega$ and then the momentum upper cutoff is reset to $\Lambda$. The lower wavevector parts of the fields are rescaled together with the rescaling of wavevector and frequency.  
 
\begin{align}
 &\theta({\bf k},\omega)=b^{\chi_\theta}\theta(b{\bf k},b^z\omega),\, \delta c({\bf k},\omega)=b^{\chi_c}\delta c(b{\bf k},b^z\omega),\nonumber \\
 &\hat\theta({\bf k},\omega)=b^{\hat\chi_\theta}\hat\theta(b{\bf k},b^z\omega),\, \hat c({\bf k},\omega)=b^{\hat\chi_c}\hat c(b{\bf k},b^z\omega),\label{field-scale}
\end{align}
  where $\chi_\theta,\, \chi_c$ are scaling exponents of the fields $\theta,\, \delta c$ respectively and $\hat\chi_\theta,\, \hat\chi_c$ are scaling exponents of the conjugate fields $\hat\theta,\, \hat\delta c$ respectively.
   { We calculate $\chi_\theta$ and $\chi_c$ by demanding that the following harmonic terms in (\ref{action}), written in the Fourier space,  $\int d^dk\,d\omega\,D_\theta |\hat\theta ({\bf k},\omega)|^2 $ , $\int d^2k\,d\omega\,D_c k^2|\hat{c} ({\bf k},\omega)|^2$, $\int d^dk\,d\omega\,\kappa \hat\theta({-\bf k},-\omega) k^2 \theta ({\bf k},\omega)$,  $\int d^dk\,d\omega\,D\hat{c}({-\bf k},-\omega) k^2 \delta c({\bf k},\omega)$,  do not scale under the rescaling of  wavevector and frequency. We also demand that  the terms $\int d^dk\,d\omega\,\omega \hat \theta ({-\bf k},-\omega) \theta ({\bf k},\omega)$ and $\int d^dk\,d\omega\,\omega \hat c({\bf -k},-\omega) \delta c({\bf k},\omega)$ are unchanged under the rescaling of wavevector and frequency. These gives the rescaling factors for the conjugate fields $\hat \theta,\, \hat c $.  These conditions give  the exponents  $z=2$, $\chi_{\theta }=3+d/2$, $\chi_{c}=2+d/2$, $\hat\chi_\theta=(d+2)/2$ and $\hat\chi_c= (d+4)/2$, giving $\chi_{\theta }=4$, $\chi_{c}=3$, $\hat\chi_\theta=2$ and $\hat\chi_c=3$ in $2d$ . 
 
 
 
 These conditions further give the rescaling factors of the model parameters:
  \begin{eqnarray}
  && \kappa'=b^{z-2}\kappa^<,\, D'=b^{z-2}D^<,\nonumber\\ && D'_\theta=b^{d+3z-2\chi_\theta}D^<_\theta,\, D'_c=b^{d+3z-2-2\chi_c}D^<_c,\nonumber\\
  && V'_0=b^{z-1}V^<_0, \lambda'=b^{-d-2+\chi_\theta}\lambda^<,\nonumber\\ 
  &&\Omega'_2=b^{-d-2+\chi_\theta}\Omega_2^<,\, \lambda'_1=b^{-d-2+\chi_\theta}\lambda^<_1.\label{resc-fact1}
 \end{eqnarray}
 These rescaling factors (\ref{resc-fact1}) can be used to show that the rescaling factor of the effective dimensionless constant $g$ is $b^{2-d}$; see Eq.~(\ref{scale-g}).
 
 \section{Calculation of the correlation functions in the real space}\label{real-corr}
 
 We now calculate the renormalized correlation functions of $\theta({\bf x})$, defined as
\begin{equation}
 C_{\theta\theta}^R( r)\equiv \langle [\theta({\bf x})-\theta({\bf x'})]^2\rangle_R.
\end{equation}
We start from
\begin{equation}
  \langle \theta({\bf k})\theta({\bf -k})\rangle_R \approx \frac{\overline D(0)}{2\Gamma(0) k^2[\ln (\Lambda/k)]^\eta},\label{anhar-k}
\end{equation}
Expression (\ref{anhar-k}) is no longer valid over the wavevector range from 0 to $\Lambda$, rather it is valid between 0 and $\tilde\Lambda\ll \Lambda$.
 We then obtain
 \begin{equation}
  C_{\theta\theta}(r)\approx\int_0^\Lambda \frac{d^2k}{(2\pi)^2}\left[1- \exp i{\bf k}\cdot ({\bf x-x}')\right]\frac{\overline D(0)}{ 2\Gamma(0) k^2 [\ln (\Lambda/k)]^{\eta}}.\label{anhar-r}
\end{equation}
Integrating over the angular variable, we get
\begin{eqnarray}
 C_{\theta\theta}(r)&\approx&\int_0^{\tilde\Lambda}\frac{dq\,\overline D(0)}{2\Gamma(0) q[\ln(\Lambda/q)]^{\eta}}\left[\frac{1}{2\pi} \int_0^{2\pi} d\theta (1-e^{iqr\cos\theta}\right]\nonumber \\
 &=& \int_0^{\tilde\Lambda}\frac{dq\,\overline D(0)}{2\Gamma(0)q[\ln(\Lambda/q)]^{\eta}}\left[1-J_0(qr)\right]\nonumber \\
 &=& \int_0^{\tilde\Lambda r}\frac{du\,\overline D(0)[1-J_0(u)]}{2\Gamma(0) u |\ln (\frac{r\Lambda}{u})|^{\eta}},
 \end{eqnarray}
where $J_0(u)$ is the Bessel function of order zero. Then

\begin{widetext}
 
 \begin{equation}
  C_{\theta\theta}(r)=\int_0^1\frac{du\,\overline D(0)[1-J_0(u)]}{2\Gamma(0)u[-\ln u + \ln (1/y)]^{\eta}}+\int_1^{\tilde\Lambda r}\frac{du\,\overline D(0)}{2\Gamma(0)u[-\ln u + \ln (\Lambda r)]^{\eta}} - \int_1^{\tilde\Lambda r} \frac{du\,\overline D(0)J_0(u)}{2\Gamma(0)u[-\ln u + \ln (\Lambda r)]^{\eta}}. \label{inter-appex}
 \end{equation}
 The first and the third terms on the rhs of (\ref{inter-appex}) are finite. Since $u_{max}=\tilde\Lambda r \ll \Lambda r$, the second contribution on the right may be integrated with the substitution $u=\exp(z)$ giving
 \begin{equation}
  \int_1^{\tilde\Lambda\, r}\frac{du}{u[\ln (\Lambda\,r)]^{\eta}}\approx 
  (1-\eta)[\ln(\Lambda\,r)]^{1-\eta}
 \end{equation}
 
 \end{widetext}
 in the limit of large $r$.
 We thus find $C_{\theta\theta}(r)\approx (1-\eta)\frac{\overline D(0)}{2\Gamma(0)}|\ln(\Lambda\,r)|^{1-\eta}$ in the limit of large $r$, with the remaining contributions on the right hand side of (\ref{inter-appex}) being finite or subleading for large $r$.
 \section{Fluctuations-Dissipation Theorem}\label{fdt1}
 
 The fluctuation-dissipation theorem (FDT) is a hallmark of all equilibrium systems that connects a correlation function with the associated response function. For a single-component linear system, this in the frequency domain essentially implies
 \begin{equation}
  \omega C(\omega)\propto 2{\cal I}m G(\omega),
 \end{equation}
where $G(\omega)$ is the propagator of the (linear) equation of motion of the single-component system. For a multi-component system this generalizes to
\begin{equation}
 i\omega C_{ij}(\omega)=2[G_{ij}(\omega) - G_{ji}(-\omega)].\label{fdt}
\end{equation}
Here, $i,j$  refer to the fields. In our case, $i,j$ are $\theta$ or $\delta c$. We can calculate the propagator matrix $\bf\underline  G$ from (\ref{lin1}) and (\ref{lin2}):

\begin{eqnarray}
{\bf \underline G}&=&\left(\begin{array}{cc}
-i\omega +Dk^2 & V_0\\
-V_0k^2 & -i\omega +\kappa k^2
\end{array}
\right)\nonumber \\&\times& \frac{1}{-\omega^2 -2i\omega\Gamma k^2 + V_0^2 k^2}.\end{eqnarray}

The correction function matrix $\bf \underline{C}$ is given by
\begin{eqnarray}
 {\bf \underline C} &=& {\bf \underline G}{\bf \underline N}{\bf \underline G^\dag},
\end{eqnarray}
where $\bf \underline N$ is noise correlation matrix given by
\begin{eqnarray}
 {\bf \underline N}&=&\left(\begin{array}{cc}
                             2D_\theta &0\\
                             0 & 2D_c k^2
                            \end{array}\right),
 \end{eqnarray}
which is diagonal. 
 Clearly, (\ref{fdt}) does not hold here. Thus, in spite of extracting the same effective temperature $T^\theta_{eff}= T^c_{eff}$, there is no FDT even at the linear level. This makes the linear theory out of equilibrium.  This can also be seen easily by going to the eigenbasis of (\ref{lin1}) and (\ref{lin2}), in which the propagator matrix is diagonal, but the noise matrix is not. Therefore, the correlation function matrix is too has off-diagonal entries, which manifestly breaks (\ref{fdt}).  It is straightforward to show that FDT remains broken even in the renormalized theory. { We briefly discuss if a non-zero noise crosscorrelation can restore FDT in the linear theory. If FDT could indeed be restored, then that would imply the existence of an effective (Gaussian) free energy $F_\text{eff}$, with an acceptable structure consistent with the basic symmetries of the problem, which in turn should give rise to (\ref{lin1}) and (\ref{lin2}) by taking appropriate functional derivatives of $F_\text{eff}$.  It is however impossible to generate the linear $\delta c$-term in (\ref{lin1}) from a free energy functional; see also Appendix~\ref{equi-hydro}. Thus, a non-zero noise crosscorrelation cannot restore FDT.}

 \section{Critical dimension of the model}\label{dc}
 
 We have found that $d=2$ is the critical dimension of the hydrodynamic equations (\ref{theta}) and (\ref{C}). This obviously raises the question whether this is the upper or lower critical dimension. A lower critical dimension is characterized by the possibility of a phase transition above it, whereas an upper critical dimension refers to the dimension above which the anhamornic terms in the action functional~(\ref{action})  are irrelevant, and the Gaussian theory suffices with no new phase transition appearing. This can be easily ascertained by considering the model at $d=2+\epsilon>2$ that we now discuss briefly. With this the flow equation for $g$ reads
 \begin{equation}
  \frac{dg}{dl}=-\epsilon g -\Delta_1 g^2.
 \end{equation}
For $\Delta_1>0$, which is the necessary condition for a stable phase, $g=0$ is the only fixed point which is linearly stable (as opposed to being marginally stable at $2d$). This implies that $d=2$ is the {\em upper critical dimension} of the model. On the other hand, if $\Delta_1<0$ (which corresponds to the marginally unstable situation at $2d$ with only short range order), then in addition to the stable fixed point at $g=0$, there is now an unstable fixed point at $g=\epsilon/|\Delta_1|$. This implies the existence of a phase transition at $d=2+\epsilon$, between a ``smooth phase'' controlled by the stable fixed point $g=0$ and a ``rough phase'' controlled by a fixed point at a finite $g$ inaccessible by perturbation theories. This is exactly analogous to the behavior of the KPZ equation at $d=2+\epsilon$ and implies that for the unstable sector in the phase diagram, $2d$ is actually the {\em lower critical dimension}. 
 
 \section{Coupled Burger-like model}\label{bmhd}
 
 If we write ${\bf v}_s\equiv {\boldsymbol\nabla}\theta$, where ${\bf v}_s$ is the ``superfluid velocity'', then the coupled equation for ${\bf v}_s$ and $\delta c$ to the lowest nonlinear order can be obtained from (\ref{theta}) and (\ref{C}):
 \begin{eqnarray}
  \frac{\partial {\bf v}_s}{\partial t}&=&\kappa \nabla^2 {\bf v}_s +V_0{\boldsymbol\nabla}\delta c+ \frac{\lambda}{2}{\boldsymbol\nabla} v_s^2 \nonumber \\&+& \Omega_2 {\boldsymbol\nabla}(\delta c)^2 + {\boldsymbol\nabla} f_\theta,\label{vs-eq}\\
  \partial_t c &=& D \nabla^2 \delta C + V_0 {\boldsymbol\nabla}\cdot {\bf v}_s +\lambda_1 {\boldsymbol\nabla}\cdot(\delta c{\bf v}_s) + f_c. \label{C-vs-eq}
 \end{eqnarray}
It is interesting to make a comparison of (\ref{vs-eq}) and (\ref{C-vs-eq}), respectively, with the one-dimensional coupled Burgers equation, discussed in Ref.~\cite{ab-jkb-sr} and subsequently generalized to $d$-dimensions in Ref.~\cite{abfrey,abfrey1}. These are (in $d$-dimension)
\begin{eqnarray}
\frac{\partial {\bf u}}{\partial t} &=& \left({\bf B}_0\cdot {\boldsymbol\nabla} \right) {\bf b} + \frac{\lambda}{2}{\boldsymbol\nabla}u^2 + \frac{1}{2} b^2 + \nu_u \nabla^2 {\bf u}+ {\bf f}_u,\label{burg_u}\\
\frac{\partial {\bf b}}{\partial t} &=& \left({\bf B}_0 \cdot {\boldsymbol\nabla}\right) {\bf u} + \lambda {\boldsymbol\nabla} ({\bf u\cdot b}) + \nu_b \nabla^2 {\bf b} + {\bf f}_b.\label{burg_b}
 \end{eqnarray}
Here, $\bf u$ and $\bf b$ are the Burgers velocity and magnetic fields, respectively; $\nu_u$ and $\nu_b$ are the corresponding diffusivities. Noises ${\bf f}_u$ and ${\bf f}_b$ are conserved noises. Parameter ${\bf B}_0$ is the ``mean-magnetic field''. 

The structural similarities between (\ref{theta}) and (\ref{C}), and (\ref{burg_u}) and (\ref{burg_b}) are quite apparent, with the understanding that the conserved scalar number density fluctuations $\delta c$ is now replaced by a conserved vector field $\bf b$ and ${\bf v}_s$ is replaced by the Burgers velocity field $\bf u$.
The dispersion relation of the linearized version of (\ref{burg_u}) and (\ref{burg_b}) are
\begin{equation}
 \omega = \pm B_0 k_x + ik^2(\nu_u +\nu_b)/2 \label{burg-disp}
\end{equation}
in the hydrodynamic limit, which is the analog of (\ref{lin-disp}) in the main text; here, ${\bf B}_0$ is assumed to be along the $x$-axis. Although (\ref{burg-disp}) is anisotropic, it has the same form (in a scaling sense) as (\ref{lin-disp}). Furthermore, the nonlinear coupling constants in (\ref{burg_u}) and (\ref{burg_b}) have 2 as their critical dimension as in (\ref{theta}) and (\ref{C}). In spite of these similarities at the scaling level, Eqs.~(\ref{theta}) and (\ref{C}) {\em do not} show any analog of the ordered states here, with the effective coupling constant that is the exact analog of $g$ above diverges as soon as the system size exceeds a (small) value at $2d$ in a manner similar to the $2d$ KPZ equation~\cite{abfrey,abfrey1}. Thus, the ordered states in the present study essentially are outcomes of the precise form of the fields (scalar versus vector) and the corresponding precise forms of the couplings, and not just the scaling dimensions of the coupling constants.

\section{Equilibrium limit of the hydrodynamic theory}\label{equi-hydro}

We briefly discuss the equilibrium limit of the theory developed here. We consider the relaxational dynamics, obtained from a free energy function ${\cal F}_\text{eq}$:

\begin{equation}
 {\mathcal F}_\text{eq} = \int d^2x [\frac{\kappa}{2}({\boldsymbol\nabla}\theta)^2 + 
A\delta c \nabla^2\theta + \frac{1}{2}(\delta c)^2].\label{free-eq}
\end{equation}
 Here, we have set a susceptibility to unity, parameter $A$ is a thermodynamic coefficient of arbitrary sign.
Free energy (\ref{free-eq}) yields (assuming simple relaxational dynamics, 
ignoring any advection for simplicity)
\begin{equation}
 \frac{\partial\theta}{\partial t}=-\frac{\delta {\mathcal F}}{\partial {\mathcal 
\theta}} 
+f_\theta = -[-\kappa\nabla^2\phi + A\nabla^2\delta c] +f_\theta,\label{eq1}
\end{equation}
and
\begin{equation}
 \frac{\partial \delta c}{\partial t} 
=D\nabla^2 \frac{\delta {\mathcal F}}{\delta (\delta c)} + { f}_c= D\nabla^2 \delta c + DA\nabla^4\theta + 
{ f}_c.\label{eq2}
\end{equation}
Equations~(\ref{eq1}) and (\ref{eq2}) do not have any underdamped propagating modes. In fact,
using the rescaling defined in Sec.~\ref{rescale} that keep $D$ and $\kappa$, unchanged under rescaling, it is easy to show that $A$ decreases under rescaling. Thus, in the long wavelength limit $A$ can be dropped, and therefore, $\theta$ and $\delta c$ effectively decouple, giving QLRO for the phase fluctuations in 2D, and NNF for the number density fluctuations. 

\section{Equations of motion and topological defects}\label{EquationLattice}

We note that the  hydrodynamic equations (\ref{theta-full}) and (\ref{c-curr}) are invariant under a constant shift of the phase variable $\theta({\bf x},t)$. However, the physical symmetry of the problem is nothing but the invariance of the dynamics under the transformation $\theta ({\bf x},t)\rightarrow \theta({\bf x},t) + 2\pi$ at all ${\bf x}$ and $t$ separately, which reflects the compactness of the phase variable $\theta({\bf x},t)$. As a consequence, the hydrodynamic theory is unable to capture the topological defects. It is possible to generalize the hydrodynamic equations that manifestly show the invariance under individual spin rotation by $2\pi$, and hence will in-principle include the topological defects. We use a discrete notation for space for convenience.

We first examine (\ref{theta-full}) and generalize its different terms:
\\\\
(i)$\nabla^2\theta \rightarrow -\sum_{\hat e}\sin(\theta_{\bf x} - \theta_{\bf x+\hat e})$,
\\\\
(ii) $({\boldsymbol\nabla}\theta)^2\rightarrow - \sum_{\hat e}\left[\cos(\theta_{\bf x} - \theta_{\bf x+\hat e})-1\right]$,
\\\\
where $\bf x+\hat e$ is the nearest neighbor of the lattice site $\bf x$, i.e., the sum is over ${\bf \hat e}\in \{\hat e_x,\,\hat e_y\}$, where lattice spacing $a=1$.

Next we consider the current ${\bf J}_c$ as given by (\ref{c-curr}). The first term corresponds to the usual diffusion process governed by the local spatial inhomogeneities of $c$. The second term may be generalized to make it manifestly $U(1)$ invariant as
\\\\
$\tilde\Omega(c){\boldsymbol\nabla}\theta \rightarrow \left[ \tilde\Omega(c ({\bf x},t)) +\tilde\Omega( c({\bf x+\hat e},t))\right]\sin (\theta_{\bf x} - \theta_{\bf x+\hat e})$.
\\\\

The generalized equations of motion then read

\begin{eqnarray}
 \frac{\partial\theta}{\partial t}&=&-\nu \sum_{\hat e}\sin(\theta_{\bf x} - \theta_{\bf x+\hat e}) -\frac{\lambda}{2} \sum_{\hat e}\left[\cos(\theta_{\bf x} - \theta_{\bf x+\hat e})-1\right] \nonumber \\&+&\Omega ( c) + f_\theta,\label{neweq1}\\
 \frac{\partial\delta c}{\partial t} &=& D\nabla^2 \delta c - \left[ \tilde\Omega(c ({\bf x},t)) +\tilde\Omega( c({\bf x+\hat e},t))\right]\sin (\theta_{\bf x} - \theta_{\bf x+\hat e}) \nonumber \\&+& f_c.\label{neweq2}
 \end{eqnarray}
In principle, Eqs.~(\ref{neweq1}) and (\ref{neweq2}) include spin configurations, which show topological defects in this model. Studies on the dynamics of such configurations are outside the scope of the present study.


}

\section{Additional Numerical Results}\label{add-numericalresults}
 
\subsection{Nature of the disordered states}\label{disd-st}

In this Section we discuss the nature of the disordered states. To this end, we rely on the numerical studies of the agent-based model, as the hydrodynamic theory is no longer valid in the disordered phase. We have studied the disordered states by varying the parameters $g_1,\,g_2,\,c_s$ and $c_{s2}$. For instance, the simulation snapshots in Fig.~\ref{fig:LinUns} show the spin and density configurations on a $64^2$ lattice (a $32^2$ corner is shown in the spin configuration snapshots for better visualization) when the system is in the aggregate phase. For different sign of $g_2$ and $c_{s2}$ we find the aggregate phase, where altering the signs of other parameters ($g_1$ and $c_s$) does not appear to have any effect on the stability scenario, it only activates a different disordered spin morphology (Fig.~\ref{fig:LinUns}(a)-(b)) and correspondingly different density profiles (Fig.~\ref{fig:LinUns}(c)-(d)) clearly demonstrating the aggregation or clustering of the particles in separate large bands with most of the sites left almost vacant. The different aggregate phases of the simulation results  obtained as a function of specific signs of the control parameters are tabulated in Table~\ref{tab5Simulation}. We generally note that the different signs of the parameters $g_2$ and $c_{s2}$ (with $|c_s|=|c_{s2}|=1$) control the existence of the aggregate phase in our simulations. However, both the value and the sign of the parameters are important for the ordered or rough phase in our simulations. Depending upon the values of active coefficients, we either obtain the ordered or rough phases in our simulations.  
As shown in Fig.~\ref{snapshot-visual} in the main text, increasing the magnitude of $g_1$, the system shows a transition from an {\em ordered} (Fig.~\ref{snapshot-visual}(a)-(b) and (d)-(e)) to a {\em rough} phase (Fig.~\ref{snapshot-visual}(c) and (f)).  The values of the control parameters noted in Table~\ref{tab4Simulation1} or \ref{tab4Simulation} corresponds to the ordered and rough phases of the simulation results.

\begin{figure*}[hbt!]
\includegraphics[width=\linewidth]{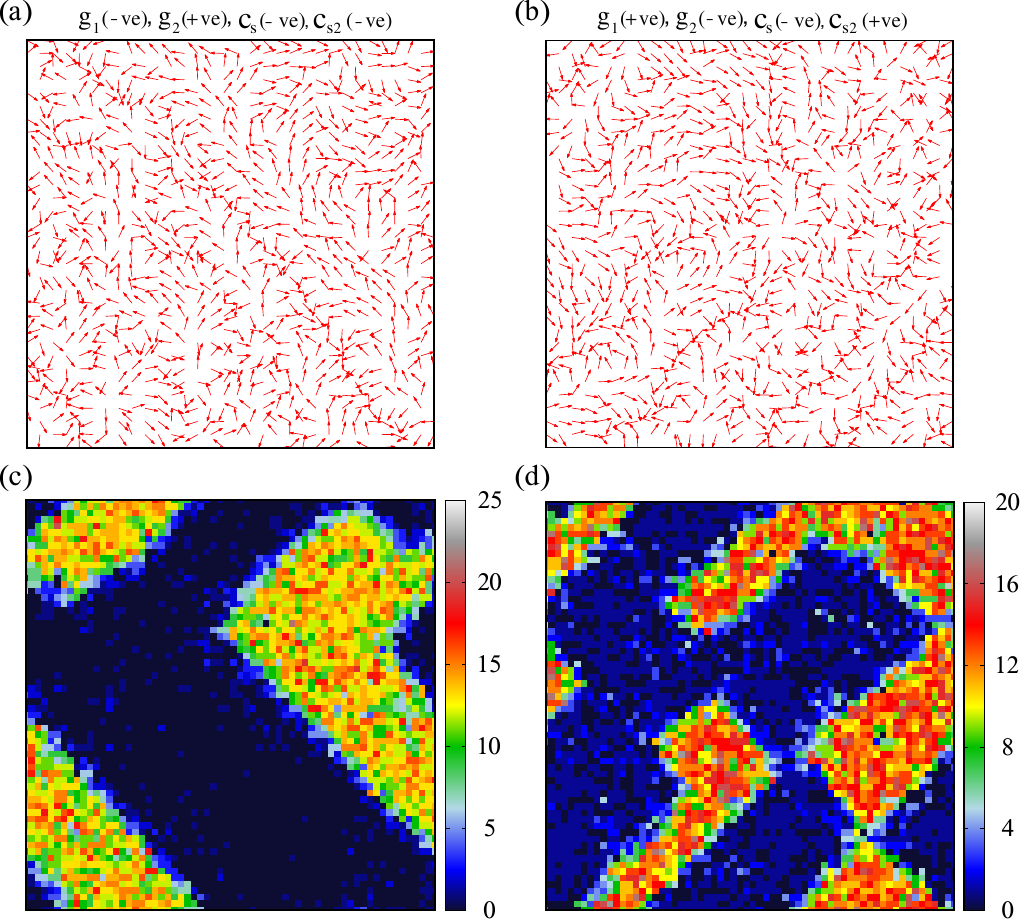}\\
\caption{Snapshots of NESS in the aggregate phase showing the (a-b) spin and (c-d) corresponding mobile species density configurations for two different sign combinations of the active coefficients (i.e., $g_1$, $g_2$, $c_s$ and $c_{s2}$) as mentioned on top of (a) and (b). Parameters: $L =64$ {(a $32^2$ corner is shown in the spin configuration snapshots)}, $c_0=5$, $\xi = 0.1$, $p_\theta=p_c=0.5$, $\mid g_1 \mid=1$, $\mid g_2 \mid=0.01$, $\mid c_s \mid=1$, and $\mid c_{s2} \mid = 1$. Visual inspections of the snapshots reveal bands and voids.}
\label{fig:LinUns}
\end{figure*}

In order to further distinguish between the rough and the aggregate phases, we numerically measure the average number of clusters and average cluster size as a function of $c$ in both rough and aggregate phases. The cluster size corresponding to a specific $c$ has been determined by estimating the total number of contiguous sites that contain the same number of particles. The time averages of the sizes of all such clusters for a given $c$ in the steady state formed within the whole lattice leads to the average cluster size. Likewise, the time averages of the number of all the clusters containing the same $c$ in the steady states give rise to the average number of clusters specific to that particular $c$. Remarkably, in the rough phase both these quantities are strongly peaked around $c=c_0$ ($=5$ here). In simple terms, this means most clusters have densities close to $c_0=5$. This is consistent with the fact that $P(c)$ too peaks around $c_0$ for rough phases; see  Fig.~\ref{rough-aggregate}(a)-(c) below.

We have a completely different picture in the aggregate phase, for which both the average cluster number and average cluster size as functions of $c$ show pronounced peaks close to $c\approx 0$ or $1$, which means the existence of a large regions  having no or very few {particles}, in other words voids or near empty regions. In addition, both in these plots a second smaller peak around $c\approx 12$ is observed, which implies the presence of a sizable number of clusters with density approximately $12$ or so. For all other ranges of $c$ both the average cluster number and average cluster size have comparatively very small values. This is clearly suggestive of a phase-separated state or the aggregate phase that typically contains either large voids or clusters with a density ($\approx 12$) much larger than $c_0=5$. Being consistent with these features, plotting $P(c)$ also shows two peaks, one large peak close to $c \sim 0$ (or $1$) and  a second one near $c\approx 12$ respectively. See Fig.~\ref{rough-aggregate}(d)-(f). It is also to be noted that, both in the aggregate and rough phases, the cluster size specific to a particular $c$ remains almost same for different system size, whereas the cluster number corresponding to some $c$ may increase with system size. 

As noted earlier when the hydrodynamic theory gets linearly unstable,  the eventual steady states, whose detailed properties cannot be found within the hydrodynamic theory, should be determined by the competition between the linear destablizing ${\cal O}(k)$ growth term  and the ${\cal O}(k^2)$ damping term, giving a typical cluster size that is independent of $L$. This strongly suggests that the aggregate phase with a preferred cluster size has a direct correspondence with the linear instability of the hydrodynamic theory.



\begin{figure*}[hbt!]
\includegraphics[width=\linewidth]{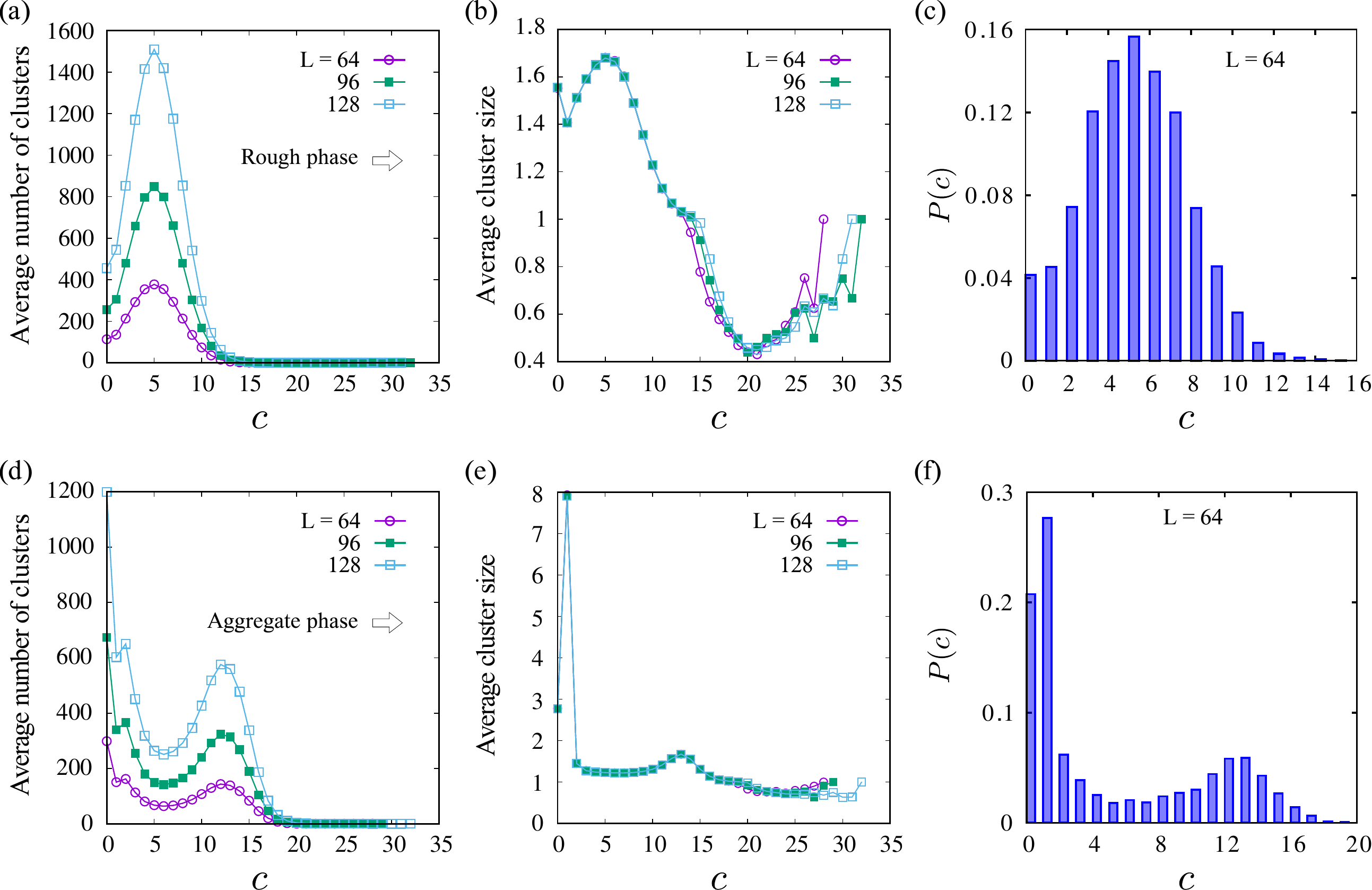}
 \caption{Plots of (a) average cluster number, (b) average cluster size and (c) the corresponding $P(c)$ for the `Rough' phase, plotted as functions of $c$ corresponding to the density snapshot and simulation parameters in Fig.~\ref{snapshot-visual}(f) of main text. (d-f) Similar plots but for the `Aggregate' phase, corresponding to the density snapshot and simulation parameters in Fig.~\ref{fig:LinUns}(d). 
}\label{rough-aggregate}
\end{figure*}

To further understand the nature of the cluster formation in the rough and aggregate phases, we have plotted the average number of clusters, and the corresponding average sizes of the clusters containing the number of particles $<3$, $3\leq c \leq 7$, and $>7$, respectively for $L=192$ in Fig.~\ref{aggregate3}. It turns out that in aggregate phase, the size of the clusters corresponding  to $3 \leq c \leq 7$ is very small as compared to the cluster's size for $c<3$ or $c>7$. This observation clearly tells that the system is dominated by the two distinct types of bigger sizes of clusters: one types of the clusters are made up of a smaller number of particles in the adjacent sites, while the others are formed by larger number of particles. However, many small sizes clusters with the intermediate range of particles ranging from 3 to 7 would present in between those large clusters. On the contrary, in the rough phases the clusters of particles (per site) ranging between $3-7$ are larger in size, relative to the cluster's size corresponding to $c<3$ or $c>7$. These observations are consistent with the prediction that $P(c)$ too peaks around $c_0=5$ for the rough phases, with the distribution being more wider than that for the ordered phase (see Fig.~\ref{histo1}(b)).

\begin{figure}[hbt!]
\includegraphics[width=\linewidth]{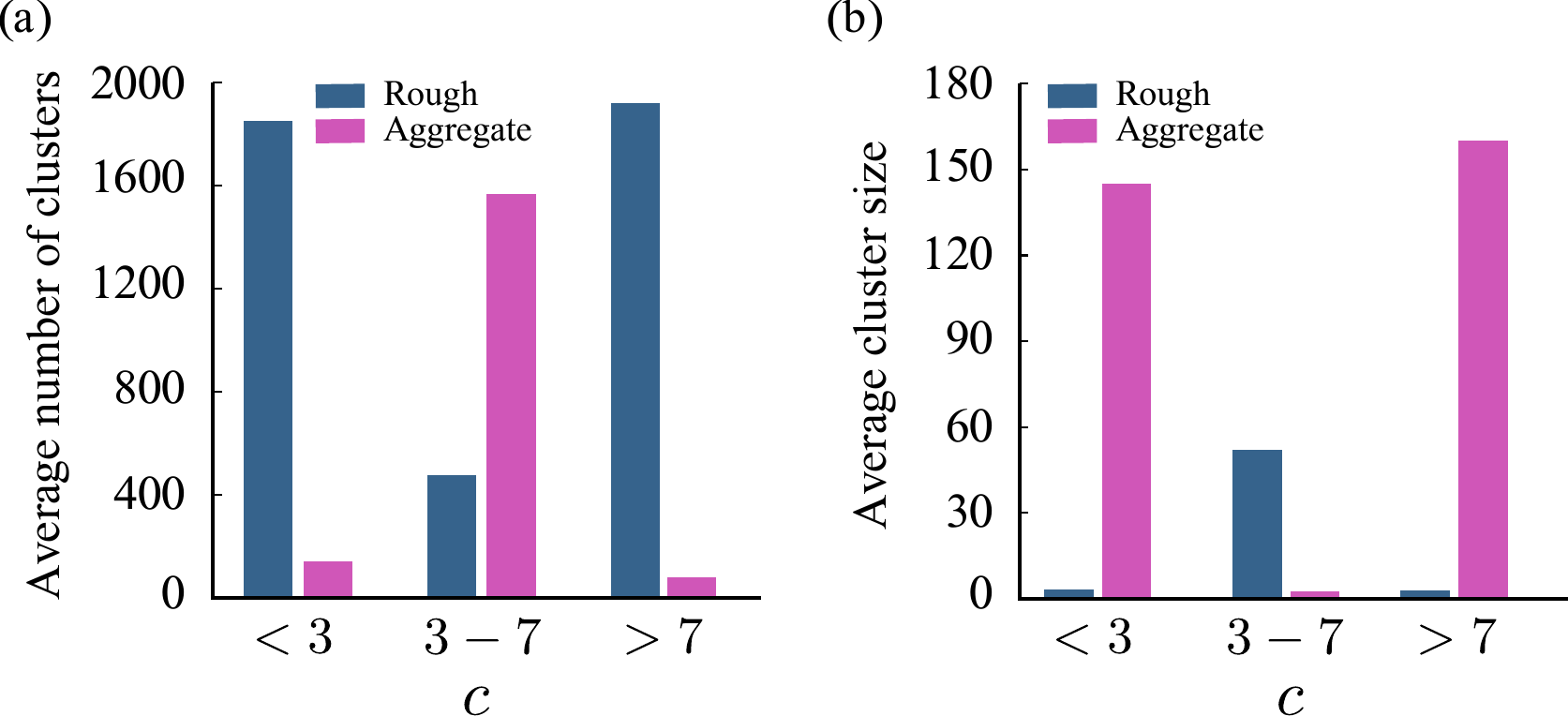}
 \caption{Plots of (a) average cluster number and (b) average cluster size corresponding to the per site number of particles, $c$, $<3$, $3-7$, and $>7$ respectively. Here $L=192$; the model parameters for the rough and aggregate phases are same as in Fig.~\ref{snapshot-visual}(f) and Fig.~\ref{fig:LinUns}(d) respectively.}      \label{aggregate3}
\end{figure}

Having established the existence of two distinct types of disordered states, {\em viz.}, the rough phase and the aggregate phase, we now illustrate how the disordered phase changes as model parameters are tuned. To this end, we study how the nature of the disordered states changes as we change the value of $c_s$. We plot the  probability distributions $P(\theta)$ and $P(c)$ for $c_s=-1$ and $-4$ in Fig.~\ref{crossover}, where the other active coefficients are $g_1=1.0$, $g_2=-0.01$, and $c_{s2}=1$. We noticed a smooth transition of the system towards the ordered phase from the aggregate phase as we increase the magnitude of $c_s$.

{ Furthermore, the snapshots corresponding to $c_s = -4$ in Fig.~\ref{crossover1} indicate that, although the system is exhibiting a disordered phase, it is more ordered as compared to Fig.~\ref{fig:LinUns}(b) and (d) which corresponded to $c_s = -1$. For $c_s = -6$, the system exhibits the ordered phase.
 While these transitions appear smooth in our simulations with finite values of $L$, it is possible that in the thermodynamic limit, these transitions are sharp and hence these are {\em not} crossovers, but genuine phase transitions, a class of structural transitions driven by some of the model parameters. Further studies will be helpful to analyze this more conclusively.}

\begin{figure}[hbt!]
\includegraphics[width=\linewidth]{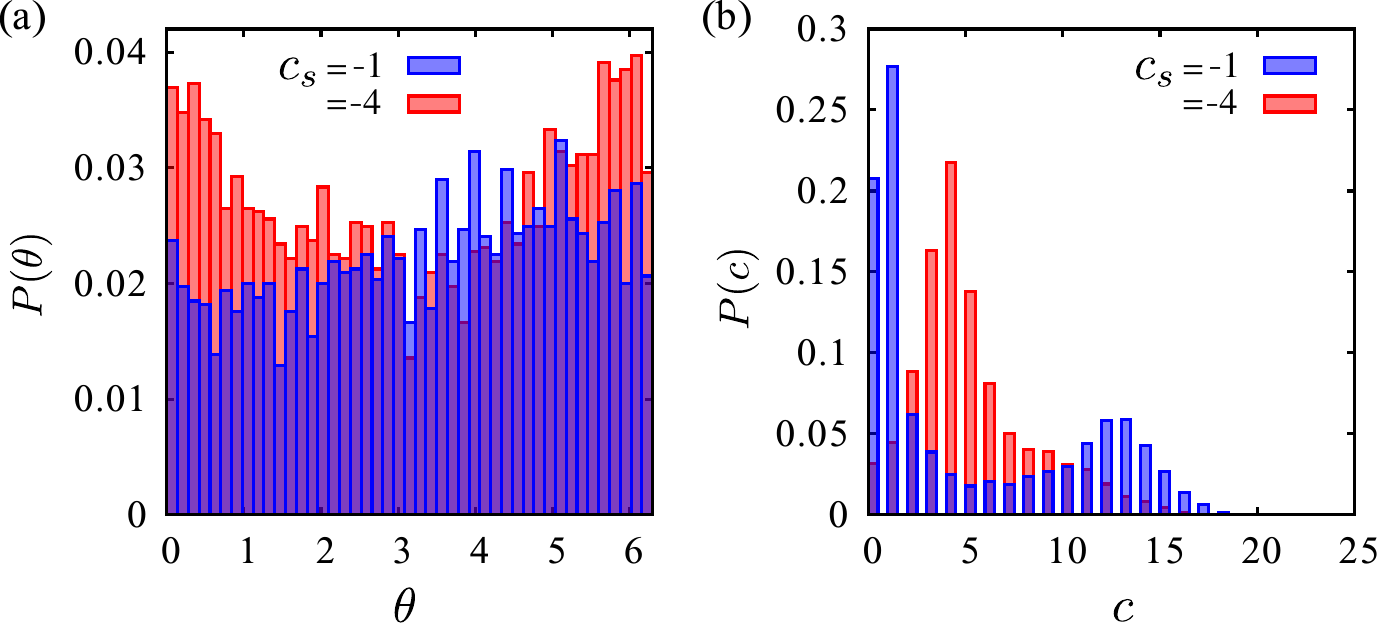}
\caption{{Plots of (a) $P(\theta)$ versus $\theta$ and (b) $P(c)$ versus $c$ for $c_s=-1$ (blue), $c_s=-4$ (red) and $c_s=-6$ (green). Other parameters are same as in Fig.~\ref{fig:LinUns}(b) or (d).}
}\label{crossover}
\end{figure}

\begin{figure*}[hbt!]
\hspace{10 mm}\includegraphics[width=0.9\linewidth]{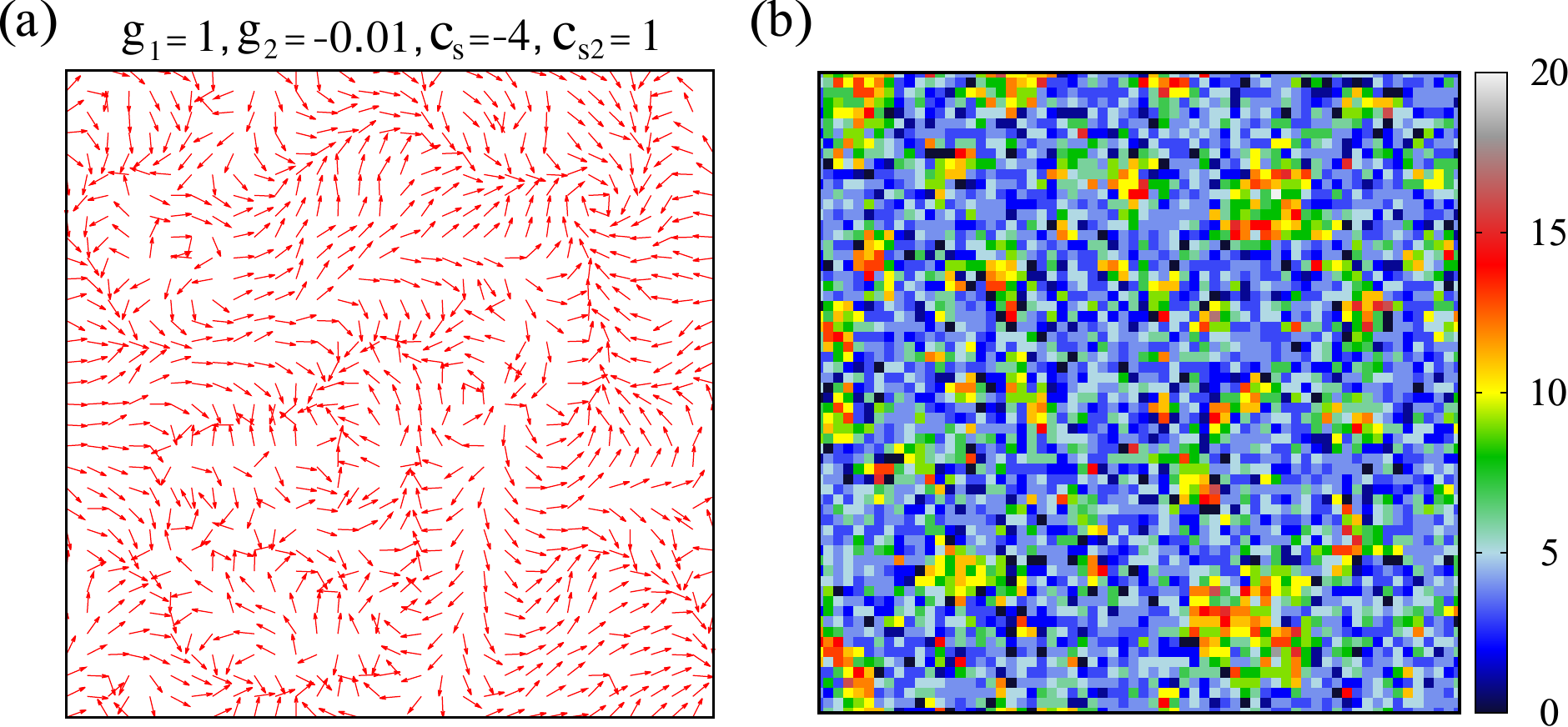}
\caption{{Phase and density configurations for $c_s=-4$ (a, b) and $c_s=-6$ (c, d) in Fig.~\ref{crossover}. For $c_s=-1$, plots are already shown in Fig.~\ref{fig:LinUns}(b) and Fig.~\ref{fig:LinUns}(d).}}\label{crossover1}
\end{figure*}


\subsection{Existence of rough phases and the restoration of order}\label{KPZ}
Fig.~\ref{fig:KPZ} demonstrates the restoration of order in a KPZ like system with the inclusion of activity. It is well known that the solutions to the $2d$ isotropic KPZ equation are known not to be smooth or regular but rather ``fractal'' or ``rough''. Fig.~\ref{fig:KPZ}(a) \& (c) represent such rough surfaces for $g_2=c_s=c_{s2}=0$ which signifies $\Omega(c)=0$ in Eq.~\eqref{theta-full} or $c=const$ (conditions at which Eq.~\eqref{theta-full} reduces to the KPZ equation). The $2d$ KPZ surface destabilizes and shows a rough phase for any nonzero $g_1$~\cite{stanley}. This however is true in the thermodynamic limit. For a finite system, it exhibits a rough phase when $g_1$ crosses a threshold that depends on $L$, the system size. This threshold is expected to be a decreasing function of $L$, ultimately vanishing in the thermodynamic limit. This is consistent with what we observe in our simulations: we find that the instability threshold $g_1(L=64)=2.826$ decreases to $g_1(128)=2.637$, in agreement with the discussions above. We also find that the ``rough'' or disordered phase  in the immobile limit (with $g_2=c_s=c_{s2}=0$) do get ordered upon making $g_2,c_s$ and $c_{s2}$ non-zero; see snapshots in Fig.~\ref{fig:KPZ}(b) \& (d) corresponding to ordered phases obtained by switching on the active parameters $g_2=0.001$, $c_s=1$ and $c_{s2}=1$ but for same values of $g_1$. The corresponding probability distributions $P(\theta)$ of phase $\theta$  are shown in Fig.~\ref{fig:KPZ1}. The sharply peaked nature of these probability distributions  in Fig.~\ref{fig:KPZ1}(b) \& (d) for the ordered state that persist for both $L=64$ and $L=128$, are to be contrasted with the uniform distributions of the phase for both $L=64$ and $L=128$ as shown in Fig.~\ref{fig:KPZ1}(a) and (c).

\begin{figure*}[hbt!]
\includegraphics[width=\linewidth]{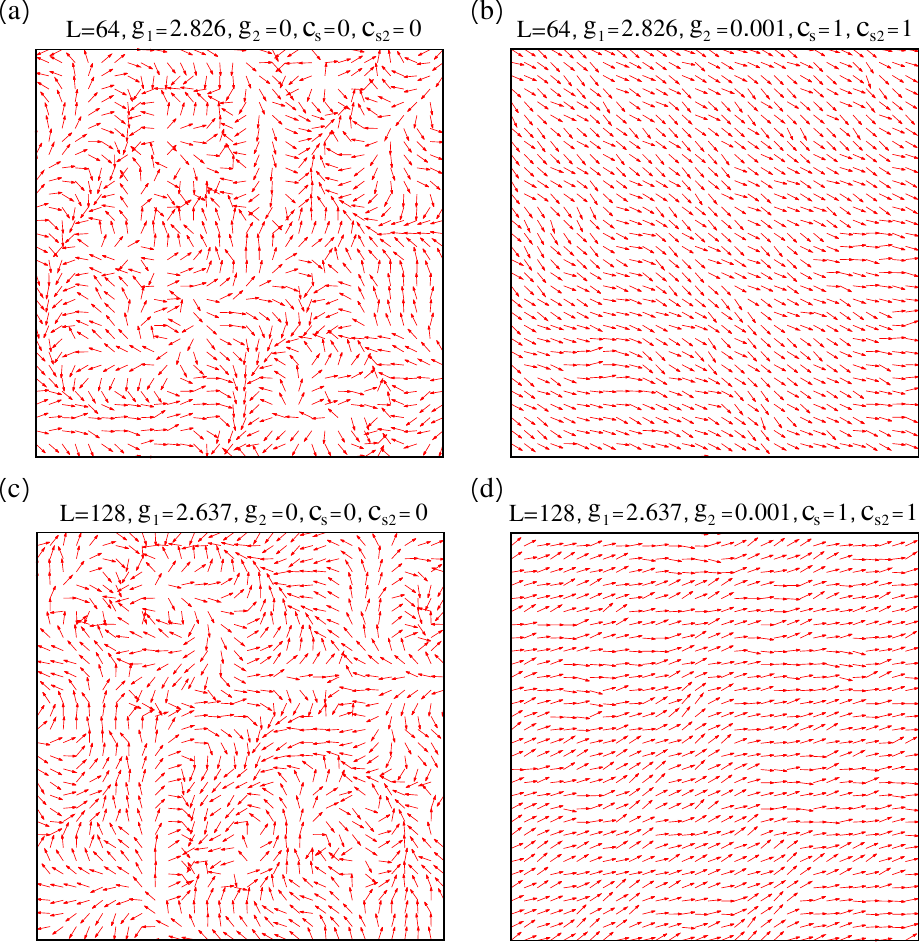}
\caption{(a \& c) Snapshots showing disordered rough phases in presence of active coefficient $g_1$ while the other active coefficients $g_2$, $c_s$ and $c_{s2}$ are zero. This is equivalent to the $2d$ KPZ equation for the hydrodynamic limit of the model in its immobile limit. { The system exhibits the disorder like phase at (and after) a critical value of $g_1$, before which it remains in ordered phase.} (b \& d) Ordering is restored following the inclusion of all the active coefficients in the system. The lattice dimensions are $64^2$ (a-b) and $128^2$ (c-d) ($32^2$ corner of the $128^2$ lattice is shown) respectively. The data also exhibit that the critical value of $g_1$ is a decreasing function of system size $L$. Parameters: $c_0=5$, $\xi = 0.5$, and $p_\theta=p_c=0.5$.}
\label{fig:KPZ}
\end{figure*}

\begin{figure*}[hbt!]
\includegraphics[width=\linewidth]{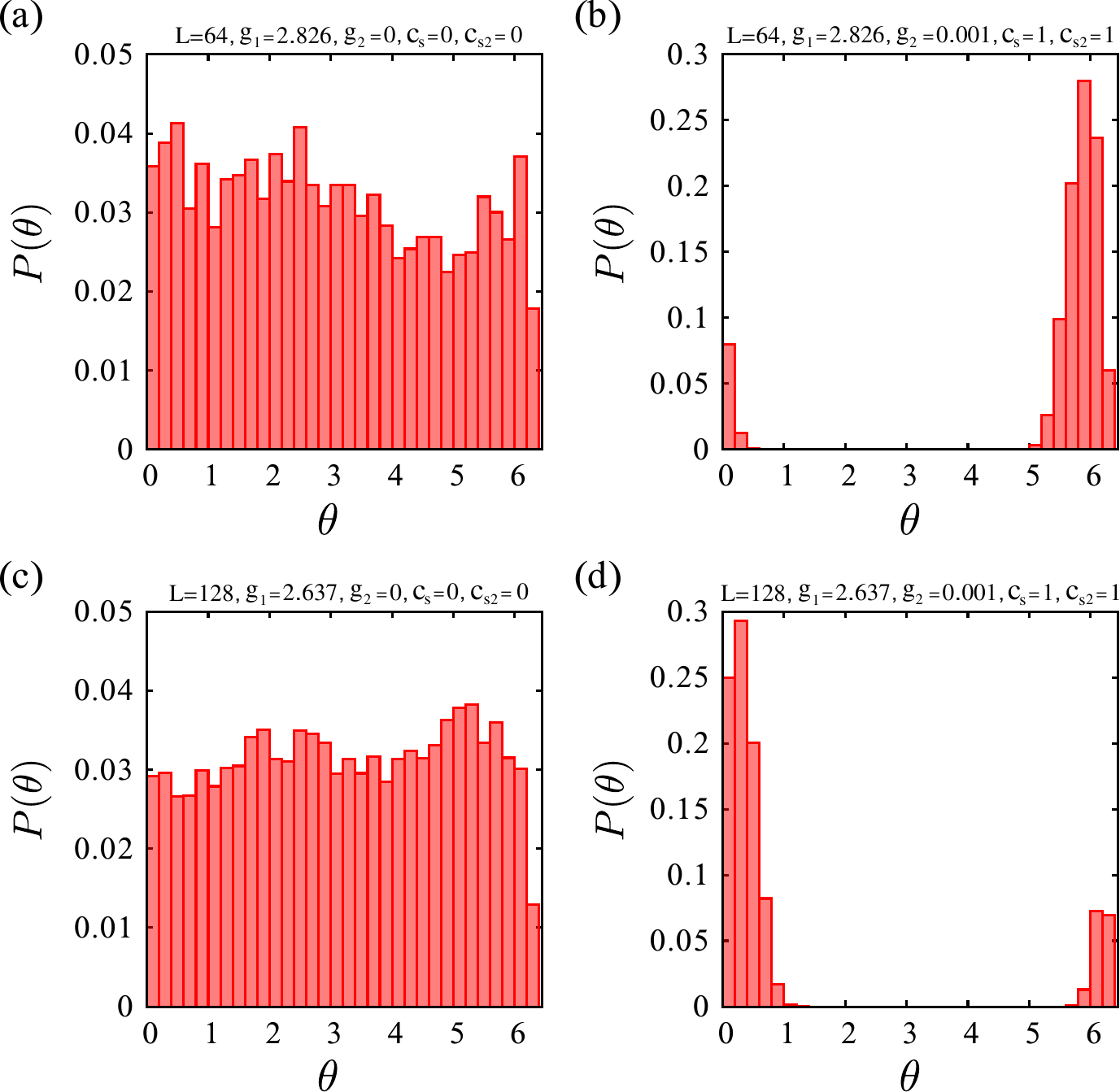}
\caption{(a), (b), (c) and (d) Plot of $P(\theta)$ versus $\theta$ corresponding to the snapshots in Fig.~\ref{fig:KPZ}(a), (b), (c) and (d). In the rough phases with snapshots in Fig.~\ref{fig:KPZ}(a) and (c), $P(\theta)$ is nearly flat, implying lack of orientational order, whereas in the ordered phases with snapshots in Fig.~\ref{fig:KPZ}(b) and (d), $P(\theta)$ is sharply peaked, as is expected in an orientationally ordered state. Clearly, non-zero values of $g_2,c_s,c_{s2}$ can suppress the disorder associated with the $2d$ KPZ equation, and introduce order in the system (see text).}
\label{fig:KPZ1}
\end{figure*}

\subsection{Results with $g_1=0$}
\label{limitingcase-ordered}

We further note that the hydrodynamic theory predicts ordered states even in the limit of $\mu_1,\mu_2\rightarrow \infty$  with appropriate choices of $\tilde \mu = \mu_1/\mu_2$ {(see Table~\ref{tab2})}. 

What is the nature of order if one or the other phase update rules are suppressed {in the agent-based model}? Since the conserved density fluctuations are expected to be crucial in stable phase order, setting $g_2$ to zero should lead to only phase disordered states. To see if any kind of phase order can survive with $g_2$, we perform simulation with $g_1=0$ along with finite values of $g_2,\,c_s$ and $c_{s2}$ where visual inspection of the snapshot for $L=64$ confirms the existence of an ordered state; see Fig.~\ref{infinite}. Also, the narrow distribution of $P(\theta)$ and $P(c)$ peaked around the respective means of $\theta$ and $c$ found in the studies on the agent-based model further signifies the ordered phase. Further detailed numerical investigations for various values of $L$ should reveal whether the order is SQLRO or WQLRO.

\begin{figure*}[!hbt]
\includegraphics[width=\linewidth]{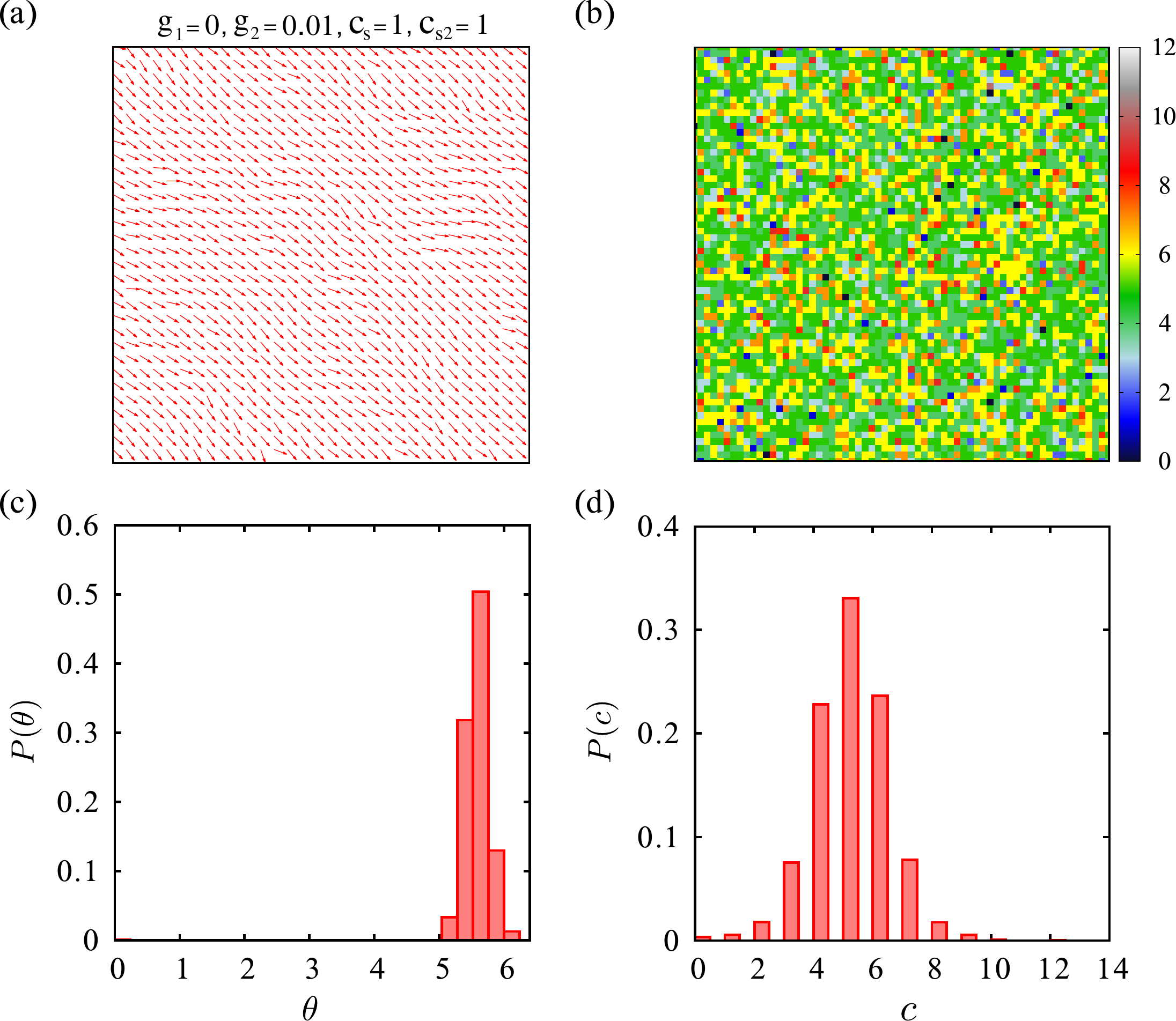}
 \caption{NESS  with $g_1=0$ and $c_0=5$, $\xi = 0.1$, $p_\theta=p_c=0.5$: (a) Spin configuration snapshot, (b) density configuration snapshot, (c) plot of $P(\theta)$ versus $\theta$, and (d) plot of $P(c)$ versus $c$. These clearly reveal the existence of an ordered state (see text). }\label{infinite}
 \end{figure*}

\clearpage

\bibliography{diffxyref}

\end{document}